\documentclass{aa}

\usepackage{graphicx}
\usepackage{txfonts}
\usepackage{hyperref}
\usepackage{natbib}
\usepackage{graphicx}
\usepackage{subcaption}
\usepackage[flushleft]{threeparttable}
\usepackage{epstopdf}
\usepackage{longtable}
\usepackage{graphicx}
\usepackage{multicol}
\usepackage{siunitx}
\usepackage{array,multirow}
\usepackage{float}
\usepackage{pdflscape}
\usepackage{todonotes}
\usepackage{color,soul}
\usepackage{array}
\usepackage{amsmath}

\newcolumntype{$}{>{\%global\let\currentrowstyle\relax}}
\newcolumntype{^}{>{\currentrowstyle}}

\newcommand{\comment}[1]{}

\begin{document}


\title{Oxygen line in fireball spectra and its application to satellite observations}
\titlerunning{The oxygen line in fireball spectra}

\author{V. Voj\'{a}\v{c}ek
\inst{1}
\and
J. Borovi\v{c}ka\inst{1}
\and
P. Spurn\'{y}\inst{1}
}

\institute{Astronomical Institute of the Czech Academy of Sciences,
Fri\v{c}ova 298, 251 65 Ond\v{r}ejov, Czech Republic\\
\email{vojacek@asu.cas.cz}
}

\date{}

\abstract
{}
{Lightning mapper sensors on board weather satellites can be successfully used to observe fireballs. These sensors use a very narrow spectral band at $777$nm, which is only a small fraction of the total fireball radiation. In this spectral band, the oxygen O I-1 triplet is dominant for fast meteors and the Planck continuum can prevail in slow meteors. It is possible to estimate the meteor brightness in the visible spectral range from this narrowband radiation, but it is vital to first study the dependence of this radiation on the meteor velocity.}
{We used observations from the well-established European Fireball Network with newly developed digital spectral cameras that allowed us to study the oxygen triplet in meteor spectra and its relation to the meteor velocity and altitude. In addition, we studied strong magnesium and sodium lines.}
{We developed a method for calibration of fireball observation reported by Geostationary Lightning Mapper (GLM) sensors on board the Geostationary Operational Environmental Satellite (GOES) weather satellites. We confirm that in slow meteors, the radiation of the Planck continuum dominates, but for faster meteors, a correction on velocity is needed. We observe that the altitude where the oxygen line was recorded  can also affect the radiation at $777$ nm. In addition, determining whether or not the meteor showed a bright
flare could also lead to a similar effect. Thus, the meteor brightness estimate may be impacted by these characteristics. We derived simple corrections on the altitude and on the meteor brightness that helped to improve the overall precision of the magnitude estimate of our sample. This allowed us to estimate the magnitude of meteors observed by GLM with an accuracy of $\approx$ 1 in magnitude. The Na/Mg line intensity ratio was found to be constant for velocities above $25$ km.s$^{-1}$ and increasing toward lower velocities. }
{}


\maketitle

\section{Introduction}
The atomic emission lines of elements evaporated from the meteoroid provide insights into the meteoritic elemental composition and physical conditions during meteor flights in the atmosphere \citep{Borovicka1994c}. The lines and bands of oxygen and nitrogen are a regular part of the meteor spectrum, which can have either meteoroid and atmospheric origins. The oxygen is abundant in the minerals of stony meteoroids, but most of the radiation usually originates in the atmospheric elements and molecules excited during the meteoroid flight.

The meteoritic lines in meteor spectra can have their origin in different parts of the meteor. There are low-temperature lines with the origin in meteor head plasma ($\approx 4500$K) and high-temperature lines ($\approx 10 000$K) that originate in the shockwave \citep{Borovicka1994b}. The lines of a meteor wake are naturally part of the spectrum \citep{Halliday1958}. They originate in the low-energy excitation intercombination region behind the meteor and last for a fraction of a second. The intensity of spectral lines of different elements can depend on the overall abundances of these elements in the meteoroid (i. e., the chemical composition). The plasma temperature also affects the intensity of spectral lines and bands due to different excitation and ionization potentials of meteoritic and atmospheric species. Other factors such as the meteoroid mass, the altitude, and the optical thickness of formed plasma can also affect the shape of the meteor spectrum. Lastly, the velocity of the meteor in the atmosphere significantly changes the spectrum as the mass and the relative luminosity of the high-temperature component rises in line with the meteor velocity \citep{Borovicka1994b}.

The relative intensity of atmospheric lines depends on the velocity of the meteor in the atmosphere \citep{Vojacek2015, Segon2018} and other parameters such as the altitude. The atmospheric lines are part of every meteor spectrum. In slow meteors, the excitation of atmospheric elements can be low and the brightness of atmospheric lines is often under the observation limit. On the other hand, in the spectra of some small and fast meteors, only the atmospheric lines have been detected \citep{Vojacek2015}.

The excited atmospheric atoms can also interact with elements of meteoritic material and form metallic oxides. The iron oxide is then observed in the spectrum in the form of molecular bands. These bands can be detected in spectra of very slow meteoroids with high content of iron \citep{Vojacek2020}.

Current lightning detection instruments on weather satellites are observing in a narrow spectral band of oxygen triplet at $777$ nm and can continuously detect bright bolides in the atmosphere as a byproduct of its original designation. These instruments can constantly observe a significant part of the Earth's globe. Observations from the Geostationary Lightning Mapper (GLM) are now filtered for fireball events and are regularly used in meteor research. Due to the observation in a narrow spectral band, calibrating the lightning detector data is not trivial. \cite{Jenniskens2018} successfully calibrated the radiation of slow fireballs observed by GLM assuming continuum radiation at $777$ nm. \cite{Brown2019} calibrated the GLM data of the Hamburg 2018 fireball using results of \cite{Jenniskens2018}, whereby  the threshold signal for GLM corresponds to an absolute magnitude brightness of -14  \cite{Vida2022} derived a simple calibration for fast fireballs by comparing GLM observation and ground observations of a fireball over Alberta in 2020. A better understanding of the behavior of oxygen lines in meteor spectra is thus more important than ever before. However, considering all aforementioned effects influencing oxygen brightness, establishing fireball properties using only a very narrow spectral window can be challenging. In this work, we used precise spectral measurements to calibrate narrow spectral satellite observations of bolides in the full range of their velocities.

All the optical data in this work were obtained by the newly modernized Czech part of the European Fireball Network (EN) \citep{Spurny2017}. This network consists of a battery of meteor observing instruments. The main instrument is the Digital Automated Fireball Observatory (DAFO). As of 2021, there were $20$ DAFOs placed in the Czech Republic (15), Slovakia (4), and Austria (1). The Spectral Digital Automated Fireball Observatories (SDAFO) are a new part of the European Fireball Network \citep{Borovicka2019IMC}. The first SDAFO was installed in 2015, and in 2021, eight cameras were running in the Czech Republic:\ one in Slovakia, and one in Germany. SDAFOs are located at the stations next to the DAFOs, with the exception of the SDAFO located at the Tautenburg Observatory as the only DAFO type instrument in Germany.

Although we focus mainly on the intensity of the oxygen line in this work, as this is the only region observed by GLM, we also analyzed two other spectral regions in spectra obtained by SDAFO. The spectral areas dominated by magnesium and sodium were measured and the results were compared to other works as a test of our calibration methods.

Most of the meteor spectra research projects such as those of the All-Sky
Meteor Orbit System (AMOS, \cite{Matlovic2019}) and Canary Island Long-Baseline Observatory (CILBO, \cite{RUDAWSKA2020}), as well as the observations of \cite{Abe2020} are focused on middle-sized meteoroids and their spectra, bright fireballs are usually saturated in these systems. As a result, their observations are not well suited for the calibration of bright fireballs.

\section{Observations of fireballs using optical and lightning detection instruments}

\subsection{Satellite observations of fireballs using lightning detection instruments}

To  adequately observe lightning on the daylight hemisphere, the weather satellites on the geostationary orbits use a very narrow spectral band with the center at the near-infrared oxygen line at $777.4$ nm.
Geostationary Lightning Mapper on board the GOES-16 and GOES-17 carry out observations in a $1.1$ nm wide spectral band with a time resolution of $2$ ms and spatial resolution of about $10$ km \citep{GOODMAN2013}. The GOES–16 satellite was initially placed in a non-operational test position at $89.5^{\circ}$W. In December 2017, the satellite was moved to its operational, geostationary position at $75.2^{\circ}$W. The GOES-17 satellite was launched on March 1, 2018 and orbits at $137.2^{\circ}$W longitude and has been in full operation since $2019$. The GLM detectors are equipped with $134$ mm lenses and with FOV $\pm8^{\circ}$. They cover the globe roughly from $16^{\circ}$ W to $165^{\circ}$ W longitude and from $55^{\circ}$ N to $55^{\circ}$ S latitude, especially including both the Americas and most of both the Atlantic and the Pacific Oceans \citep{Edgington2019}. These observations are freely accessible and they can be filtered for events lasting longer than lightning flashes. These potential fireball data are published by NASA \footnote{ \url{https://neo-bolide.ndc.nasa.gov/} \label{footnote_1} }. According to \cite{Jenniskens2018} the GLM detectors can detect fireballs of absolute magnitude of $\approx-14$ mag and brighter. This corresponds to sizes from several decimeters to bodies that are several meters in length. To block the sunlight, the GLM sensor uses a sun-blocking filter on top of the narrow $1.1$ nm filter, which changes the observed bandwidth and also the center wavelength according to the angle of incidence. The correction to the flux is less than $20 \%$ for observation less than $7^{\circ}$ from the nadir and it increases with the increasing angle from the nadir.

The Chinese satellite of the Fengyun series FY-4A, launched in December 2016, is China's second-generation geostationary meteorological satellite. It carries the Lightning Mapping Imaging (LMI), whose parameters are the same as those of the GLM detector: a $1.1$ nm narrow band with center at $777.4$ nm and a time resolution is $2$ ms. The satellite orbits at $86.5^{\circ}$E longitude and the LMI covers the surface of China \citep{Cao2021}. The data are provided by the National Satellite Meteorological Center, China Meteorological Administration.

The Lightning Imager (LI) on the EUMETSAT's (European Organisation for the Exploitation of Meteorological Satellites) Meteosat Third Generation (MTG-I) satellite will provide real-time data on the location and intensity of lightning, covering Europe and Africa. It  will be observing at $777.4$ nm with $1.9$ nm wide narrow band with time resolution $1$ ms.  The first satellite MTG-I is planned to be launched in the autumn of 2022 \citep{Holmlund2021}.

The requirement of narrowband observation (i. e., the ability of lightning observation on the daylight side) makes it challenging for meteor observations. To be able to estimate such parameters  as the absolute magnitude, some assumptions and simplifications are inevitable -- which may certainly can lead to uncertainties and discrepancies.

This work compares the meteor radiation at $777.4$ nm with other spectral regions and with the radiation of the spectrum as a whole. Because the velocity of a meteoroid in the atmosphere can significantly affect the intensity of the oxygen line \citep{Vojacek2015}, a representative sample of meteor velocities was used. Fireball observations from the European Fireball Network were used in the current analysis. These results can be useful for interpreting satellite observations of fireballs from the lightning imaging instruments that are observed in this narrowband spectral region.

\subsection{Optical observations using the European Fireball Network}

The spectral camera is a modification of the normal non-spectral cameras (DAFO). Two Canon 6D digital cameras were used. The IR cut filter was removed to allow observation of a broader spectral range. 

\begin{figure}[ht]\centering
{ \includegraphics[width=\hsize]{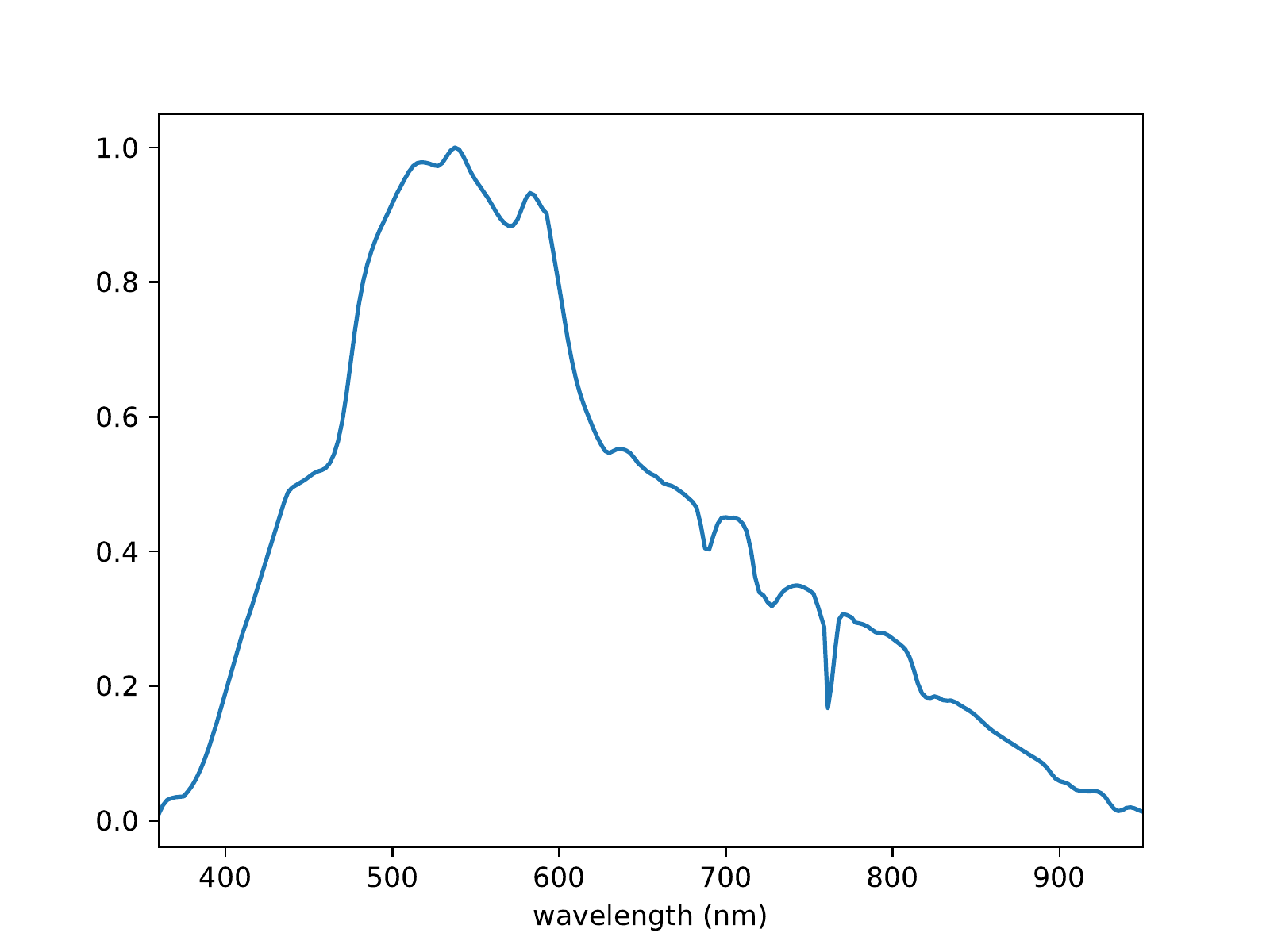}}
\caption{Spectral sensitivity of the system with the atmospheric absorption included. The curve was normalized at the maximum ($538$ nm).}
\label{Fig:sensitivity}
\end{figure}

The cameras are equipped with 15mm lenses. This gives the camera almost all-sky coverage. Exposures $30$ sec long are taken for the whole night. For spectroscopy, there is a spectral grating with $1000$ grooves/mm in front of the lens. The system covers the spectral range between approximately $360$ nm and $950$ nm (see Figure \ref{Fig:sensitivity}). The sensitivity curve in Figure \ref{Fig:sensitivity} was obtained using laboratory measurements and the spectrum of solar light reflected by the Moon and Venus. The spectral resolution is somewhere of about $1.2$ nm between a low-resolution video spectrum and a high-resolution analog film spectrum. The system can produce non-saturated spectra for fireballs between magnitudes of $-6$ and $-15$.

\section{Data reduction}

\subsection{Observations and calibrations of optical spectra}
\begin{figure}[ht]\centering
{ \includegraphics[width=\hsize]{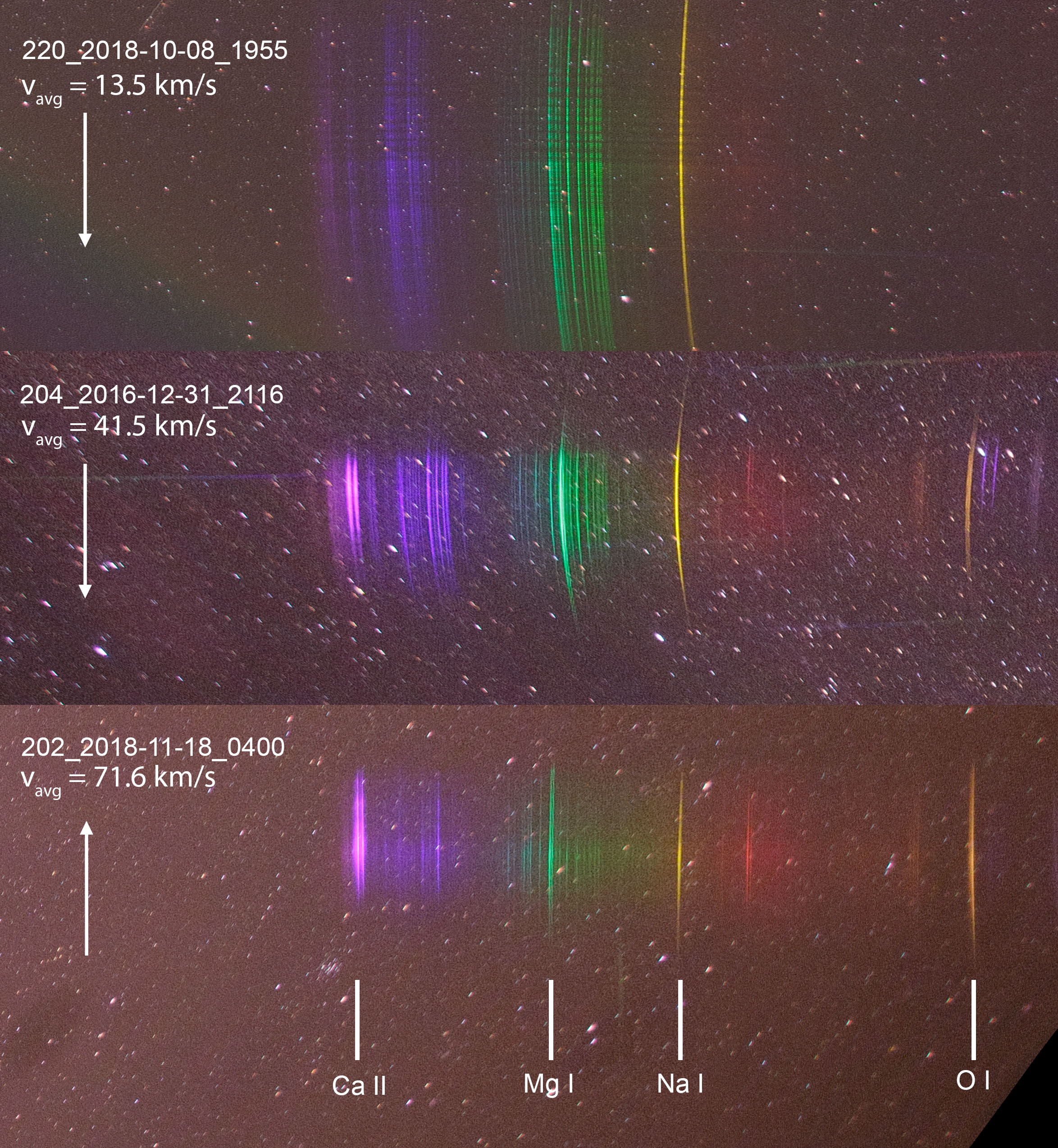}}
\caption{Examples of images of spectra of meteors from low to high velocity. The arrows indicate the direction of the meteor flight. Individual spectra were cropped from original images. The spectral lines on images were aligned in the same direction and scaled for better comparison.}
\label{Fig:examplePic}
\end{figure}

The spectra in this work were observed by the EN between December 2015 and April 2021. We selected $43$ meteors with representative speeds to cover the whole range of entry velocities. The absolute brightness of these selected meteors was in a range between $-8$ and $-15$ mag. Since multiple spectral cameras are in operation, some spectra were captured with multiple cameras from different observatory sites. In these cases, we chose the spectrum with the best combination of brightness and resolution for further measurements. Typical spectra from SDAFOs can be seen in Figure \ref{Fig:examplePic}. In this figure, we selected three spectra with slow, medium, and high entry velocities. In particular, we note the difference in oxygen intensity. The names of fireballs given in Figure \ref{Fig:examplePic} are in the date time format: SSS\_YYYY--MM--DD\_HHMM, where SSS is the EN station number of the SDAFO camera YYYY is the year, while MM and DD are the month and the day, respectively, and HHMM is the time in UT of the start of the $30$s exposure.

Observations from non-spectral cameras were used to determine the trajectory in the atmosphere for all $43$ meteors. The important parameter for this work is the velocity of the meteoroid in the atmosphere. For further analysis, we used the average velocity on the trajectory. We note that the velocity of the meteoroid at the altitude where the spectrum is captured is aptly represented by this average velocity.

The images from SDAFOs were processed by our self-developed software. All RAW RGB images were converted into grayscale images using the weighted method: $I = 0.299*R + 0.587*G + 0.114*B$. The images were dark frame-subtracted and flat-fielded, and star photometry was also performed for each image. The background image created from the previous photograph, taken just $30$ seconds before, was subtracted to remove stars and sky radiation from the image. The spectrum was scanned only in its brightest part for long meteors. For short meteors, the entire length of the spectrum was scanned. The distortion of the $15$ mm was taken into account when scanning the curved spectrum. The photometry of stars was used to calculate the energy emitted by the spectrum.

\begin{figure}[ht]\centering
{ \includegraphics[width=\hsize]{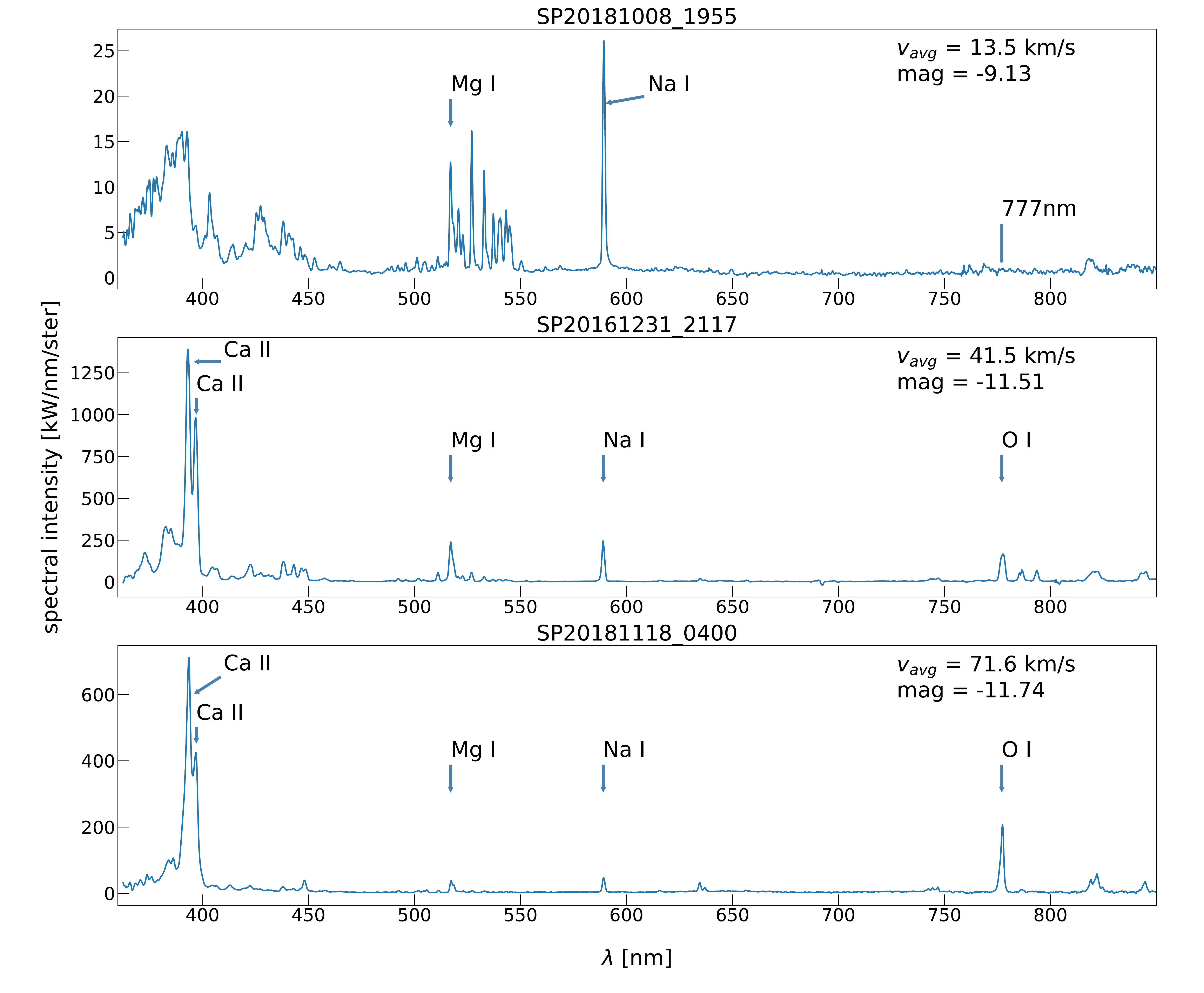}}
\caption{Examples of calibrated meteor spectra from Figure \ref{Fig:examplePic}. The prominent lines are marked. $v_{avg}$ is the average velocity on the trajectory and mag is the fireball magnitude computed from the DAFO photometry.}
\label{Fig:exampleSpec}
\end{figure}

All spectra were calibrated for wavelength using known wavelengths of lines identified in the spectrum. All spectra were then calibrated for the spectral sensitivity of our system using the normalized curve of spectral sensitivity. Examples of three calibrated spectra are shown in Figure \ref{Fig:exampleSpec}. These three spectra are the same as those shown in Figure \ref{Fig:examplePic}. Calibrated spectra names are given in the format: SPYYYYMMDD\_HHMM, where SP indicates the spectrum.

\subsection{Measuring the energy radiated in spectrum} \label{section:energyMeasure}

Knowing the distance of the meteor from the observatory that captured the given spectrum, we measured the intensity in the units of energy radiated by the given meteor per unit wavelength as the spectral intensity $I_{e,\Omega, \lambda}$ in W.ster$^{-1}$.nm$^{-1}$.

As a main region of interest, we measured the region at $777$ nm, where the oxygen line triplet O I -- 1 can be observed. To compare the oxygen line region with meteoritic lines, we measured spectra at $517$ nm, where lines of magnesium Mg I -- 2 are observed (also different multiplets of iron are overlapping here), and the region at $589$ nm, where sodium doublet Na I -- 1 dominates. We further refer to these regions in Figure \ref{Fig:exampleSpec}.

\begin{figure}[htb]
\centering
\begin{subfigure}{0.5\textwidth}
\includegraphics[width=\hsize]{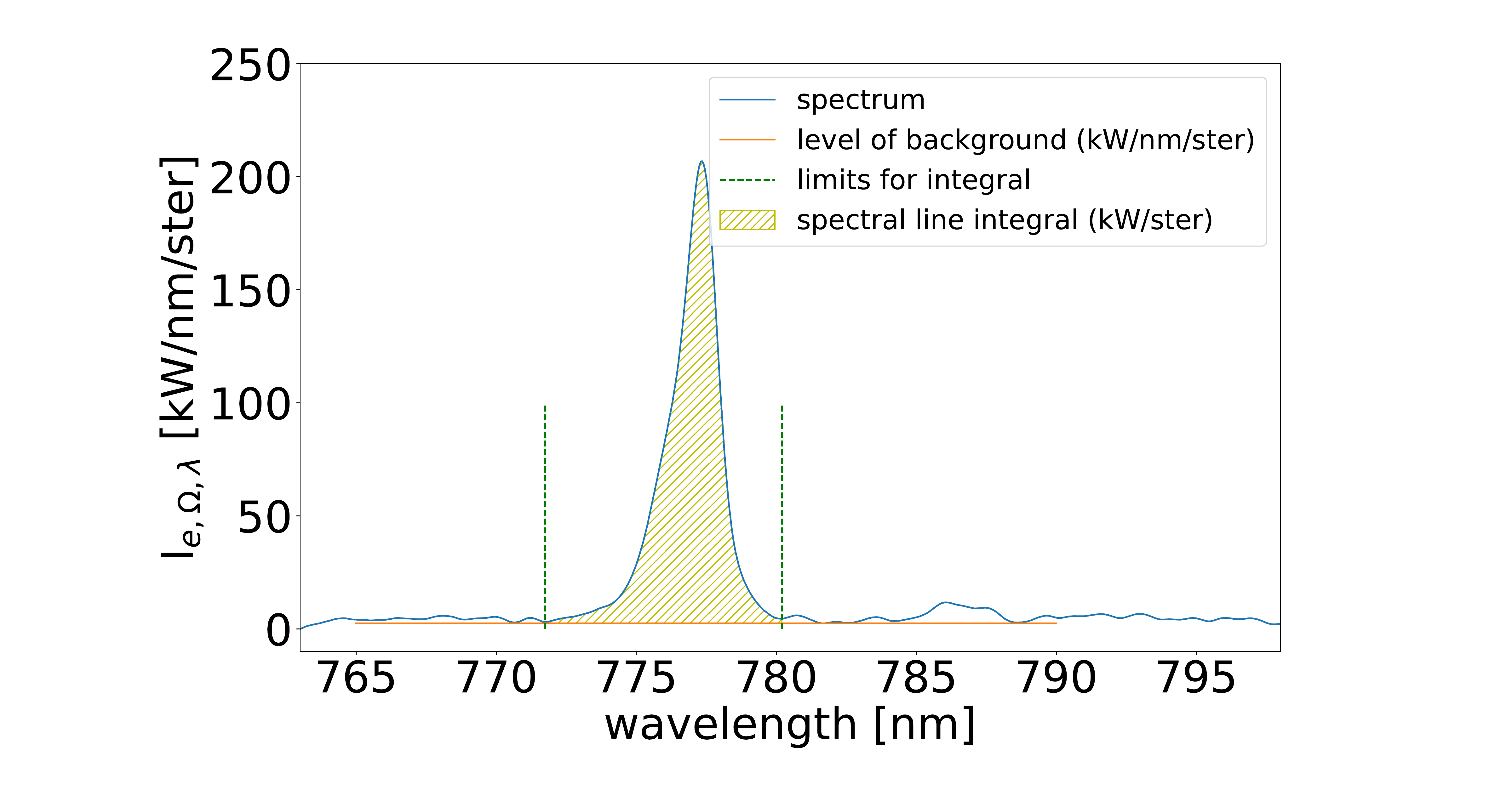}
\caption{ }
\label{Fig:LineMeasure_b}
\end{subfigure}
\begin{subfigure}{0.5\textwidth}
\includegraphics[width=\hsize]{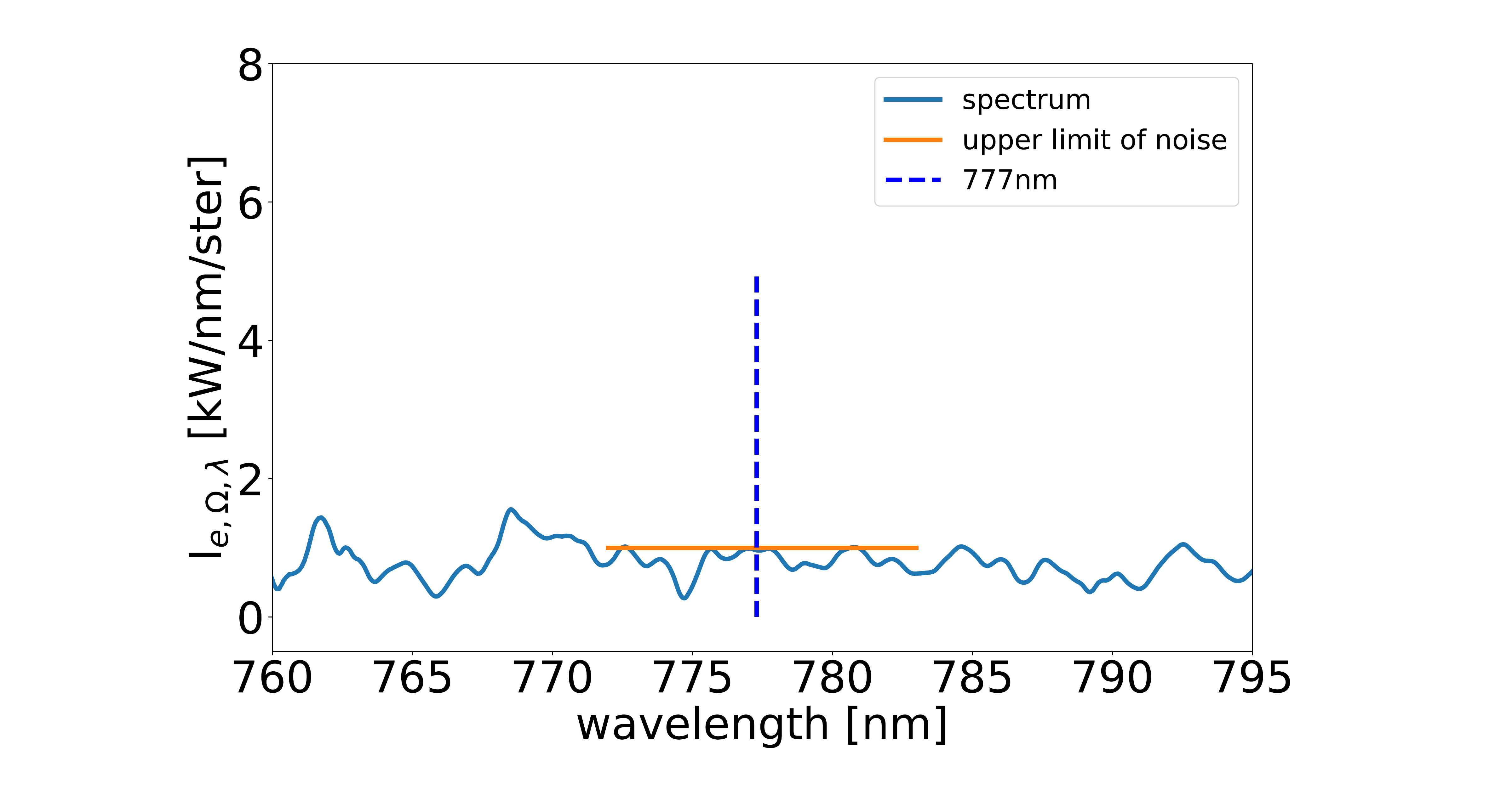}
\caption{}
\label{Fig:LineMeasure_a}
\end{subfigure}

\caption{Explanation of the method of measurement of the radiant intensity. (a) Measurement of the background as an estimate of local continuum and a measurement of the spectral line as an integral. (b) the case when only noise is observed. }
\label{Fig:LineMeasure}
\end{figure}

To measure the energy radiated by each region in EN meteor spectra, we simulated the measurement of narrowband filter observation. We measured the radiated energy in two steps. We measured the level of the radiation from the background in kW/nm/ster. This measured spectral intensity $I_{e,\Omega, \lambda}$ was then simply multiplied with $1.1$ nm narrow band filter spectral range to simulate the lightning imaging instrument measurement and we received the estimation of the radiant intensity $I_{e,\Omega}$ of the given $1.1$ nm spectral band. Then we measured radiation from the spectral line. The observed lines were instrumentally broadened due to the low spectral resolution of our cameras and thus wider than $1.1$ nm. Therefore, we integrated the whole broadened spectral line since the broadening happened in the camera and a high spectral-resolution camera would measure the same radiant intensity. An illustration of the measured line integral and level of the background can be seen in Figure \ref{Fig:LineMeasure_b}. The level of the continuum and the integrated line intensity were then summed up to obtain the radiation in the given region.
If there was only a continuum and no visible spectral line in the given region, we only measured  the level of the continuum in the same manner as the level of the background: measuring the level of radiation in kW/nm/ster and multiplying it with $1.1$ nm spectral band (see Figure \ref{Fig:LineMeasure_a}). These cases were then marked with crosses in Figure \ref{Fig:Energy}. In the case when no radiation was observed and only noise was detected, we measured the level of the noise as the upper value of the noise in a given spectral region (see Figure \ref{Fig:LineMeasure_a}. In other words, it is the upper limit of possible radiation that SDAFO was able to detect. Measurements of the noise are then shown as triangles in subsequent figures.

For further analysis, we measured the energy emitted by the spectral range between $380$ nm and $850$ nm (the limits were chosen mainly due to low sensitivity below and above these limits); hereafter, this energy marked as $I_{total}$. We also simulated measurements in the standard UBV system using the filter passbands of \cite{Bessell1990}. Specifically, we measured the absolute magnitude in the V filter (between $\approx 480$ nm and $\approx 650$ nm) and the B filter (between $\approx 380$ nm and $\approx 550$ nm).

To estimate the uncertainty of the energy measurement, we manually estimated the level of the noise near the measured region. This value was then used to estimate the uncertainty for all integrated energy radiated in a given region.

\section{Results}

\subsection{Oxygen at $777nm$ }

\begin{figure}[htb]
\centering
\includegraphics[width=\hsize]{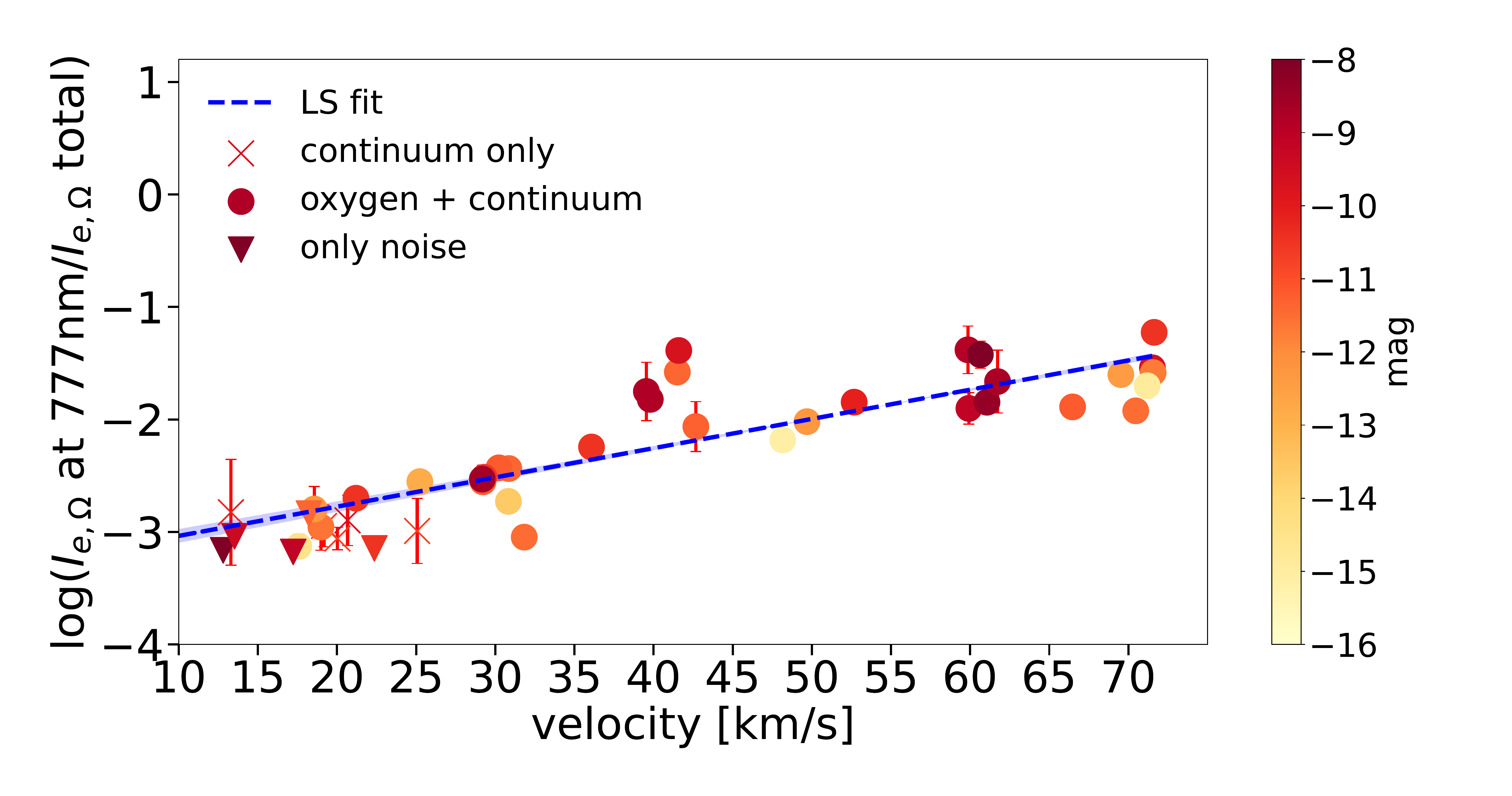}

\caption{Radiation at O I -- 1 triplet region (777 nm). Relative radiant intensities as a function of meteor velocity. Symbol colors mark the meteor absolute magnitude. Symbol shapes mark the presence or absence of the oxygen line.}
\label{Fig:Energy}
\end{figure}

The key analysis for this work is the behavior of the oxygen line O I at $777$ nm. We computed the relative intensity of the given region to the total radiant intensity $I_{total}$. Then we plotted this relative intensity as a function of the average velocity of the meteor in the atmosphere. The results can be seen in Figure \ref{Fig:Energy}. The dependence of intensities of atmospheric lines in meteor spectra on velocity is well known \citep{Millman&Halliday1961, Vojacek2015}. As expected, we observed this dependence in our data (see Figure \ref{Fig:Energy}). In seven cases, we detected only a continuum. In four cases, there was only noise without any detectable radiation. In Figure \ref{Fig:Energy}, we show the meteors colored according to the absolute magnitude in the V filter computed from their spectra.

\begin{figure}[htb]
\centering
{ \includegraphics[width=\hsize]{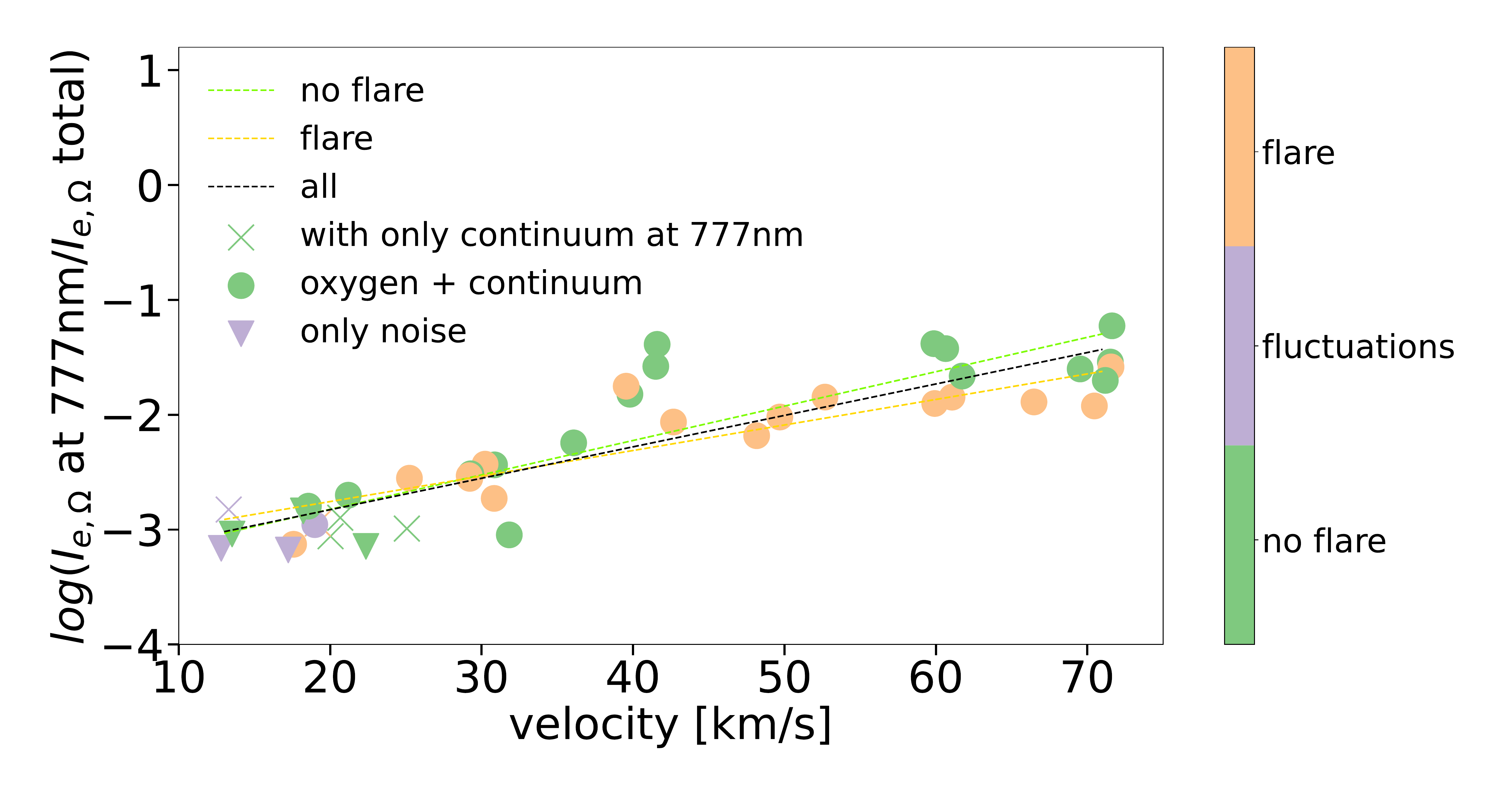}}
\caption{Radiation at O I -- 1 triplet region (777 nm). Relative radiant intensities as a function of meteor velocity. Symbol colors mark meteors with flares or fluctuations and showing fits for meteors with or without flare. Symbol shapes mark the presence or absence of the oxygen line.}
\label{Fig:O_Flare}
\end{figure}

The least-squares fit of all meteors, except those with only noise detected, gives us the dependence of the ratio between the radiation at $777$ nm and the radiation from the whole observed spectrum $I_{777}/I_{total}$ on the velocity $v$ in km/s as follows:

\begin{equation} \label{eqnO_vel}
log(I_{777}/I_{total})=0.026(\pm0.001) \times v - 3.294(\pm0.077)
.\end{equation}

To obtain an uncertainty estimate for least-squares fits, we used the Monte Carlo approach and we generated 10000 clones of each meteor point with normal distribution. The measurement error of the given point was used as the standard deviation of the normal distribution. Obtained sets of Monte Carlo clones were then fitted using the least-squares method. The standard deviation of parameters of all least-squares fits were then used as the uncertainty for the slope and the y-intercept of the final fit.

We note that meteors faster than $\approx 40$ km/s with flare on the light curve show lower relative radiation at $777$ nm. This is displayed in Figure \ref{Fig:O_Flare}, where meteors with flare and without flare in their light curves are fitted separately. The possible explanation is that a large amount of meteoric material is released during the flare and thus meteoric lines brighten more than atmospheric lines. In Figure \ref{Fig:O_Flare}, we also marked meteors that showed fluctuations on their light curve. Light curves of these meteors showed periodical brightness changes that were not as intense as flares, they occurred only in very slow meteors. We did not observe any significant difference for relative radiation at $777$ nm between meteors, with or without flares or with fluctuation for slow meteors.

\subsection{Estimating GLM detector response to meteors}\label{Caption_EstimateGLMtoMeteors}

With the known dependence of the radiation at $777$ nm on velocity, it is possible to estimate the absolute magnitude of the meteor observed by the GLM detectors if meteor velocity is known.
 From the GLM data, the radiant intensity $I_{777}$ at $777$ nm in $W.ster^1$ can be computed. We measured the same quantity for the sample of EN fireballs and we also computed the meteor V-band magnitude, $m_V,$ (see Section \ref{section:energyMeasure}). In the simplest case, the relation would be expressed as:

\begin{equation} \label{eqnMAGsimple}
m_V = -2.5 \times log_{10}(I_{777}) + b
,\end{equation}

where $b$ is a constant. However, since we know that $I_{777}$ depends strongly on velocity, we can expect the following dependency:

\begin{equation} \label{eqnMAG_first}
m_V = -2.5 \times log_{10}(I_{777}) + a \times v + b
.\end{equation}

\begin{figure}[htb]
\centering
{ \includegraphics[width=\hsize]{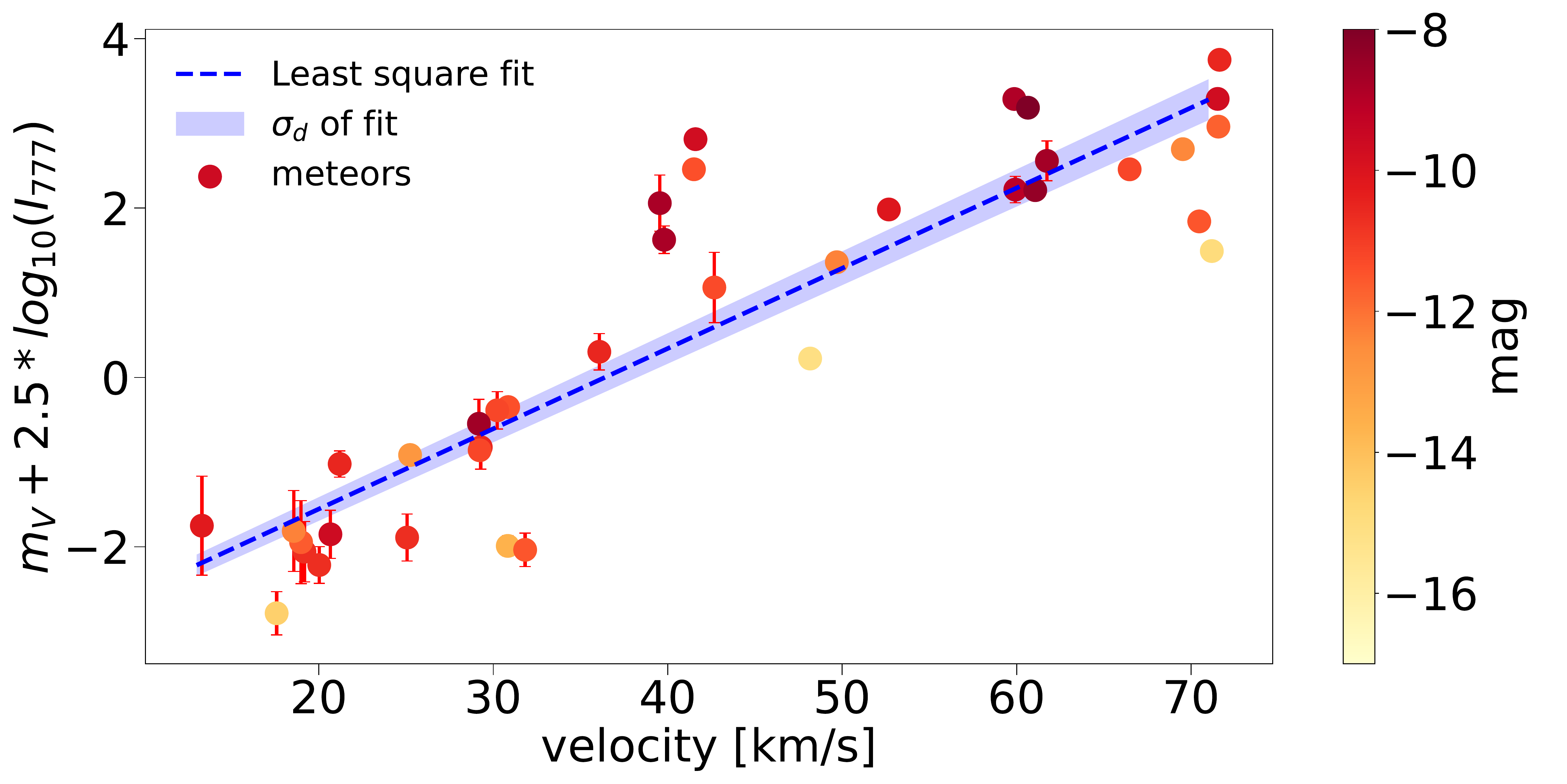}}
\caption{Magnitude in V filter and radiation at $777nm$ compared to the velocity of meteoroids in the atmosphere. The least-squares fit of all meteors and the Monte Carlo uncertainty of the fit are shown.}
\label{Fig:O_Mag}
\end{figure}

To find constants $a$, $b,$ we computed the sum of the absolute magnitude of the meteor, $m_V$, and the measured radiation, $I_{777}$, at $777$ nm as $m_V + 2.5 \times log_{10}(I_{777})$. The dependence of this quantity on the velocity is shown in Figure \ref{Fig:O_Mag}. Using a least-squares fit of the data, we obtained parameters $a$ and $b$ as follows:

\begin{equation} \label{eqnMAG}
m_V = - 2.5 \times log_{10}(I_{777}) + 0.0948(\pm0.002) \times v - 3.45(\pm0.1),
\end{equation}

where $v$ is in $km.s^{-1}$.

To be able to use Eq. (\ref{eqnMAG}) for the GLM data, we need to convert the energy measured by GLM detectors $E_{GLM}$ and reported in joules to radiant intensity $I_{777}$ emitted per unit solid angle in $W.ster^1$. The light curves of fireballs are reported by NASA as energy measured directly at the satellite's detector. On the other hand, the radiant intensity of $I_{777}$ is computed at the source. The conversion must then be expressed as:

\begin{equation} \label{eqnE_GLM}
I_{777} = \frac{E_{GLM} \times R^2 } {\Delta t \times A},
\end{equation}

where $\Delta t$ is the exposure time of the GLM detector $0.002$s, $A$ is the effective lens aperture, $0.0098 m^2$ \citep{Jenniskens2018}, and $R$ is the distance in meters between the fireball and the GOES satellite. The distance $R$ can be estimated for any geostationary GOES satellite. We know the position of the geostationary GOES satellites and the longitude and latitude of the fireball is provided along with the measured energy at the NASA website. The provided coordinates assume a "lightning ellipsoid" with an imaginary surface above $16$ km altitude at the equator and $6$ km at the poles \citep{Jenniskens2018}. The normal fireball altitude is, of course, several times higher and this creates noticeable parallax and adds some uncertainty to the magnitude estimation, which is, however, small in comparison with other uncertainties.

\subsection{Sodium and magnesium}\label{NaMg}

The spectral regions of sodium at $589$ nm and magnesium at $517$ nm contain relatively well-studied  spectral lines for meteors. To test our spectral measurements, we analyzed them and compared these results with the known behavior of lines in these regions and compared them with previous works.

\begin{figure}[htb]
\centering
\begin{subfigure}{0.5\textwidth}
\includegraphics[width=\hsize]{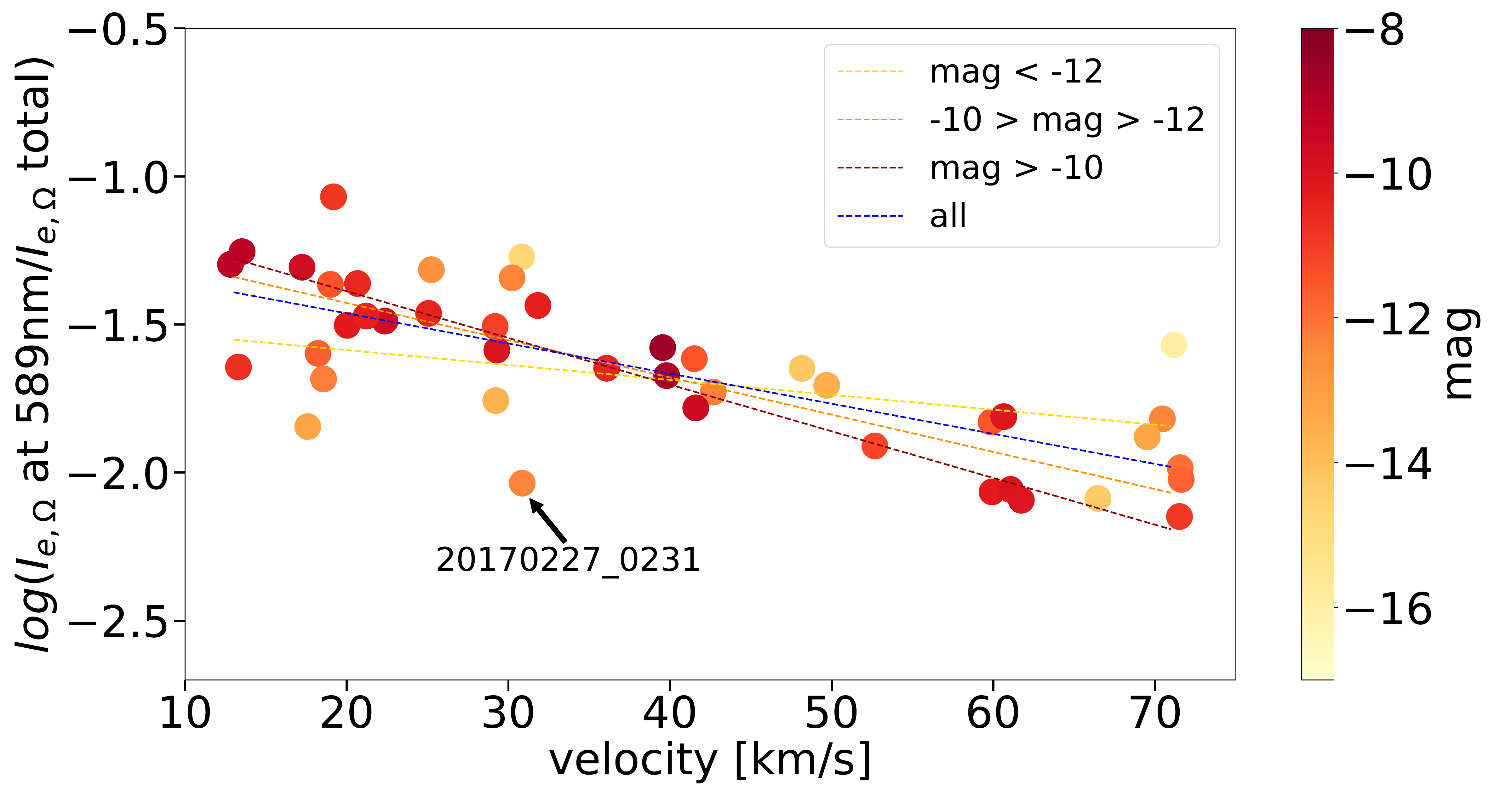}
\caption{}
\label{Fig:Energy_Na}
\end{subfigure}
\begin{subfigure}{0.5\textwidth}
\includegraphics[width=\hsize]{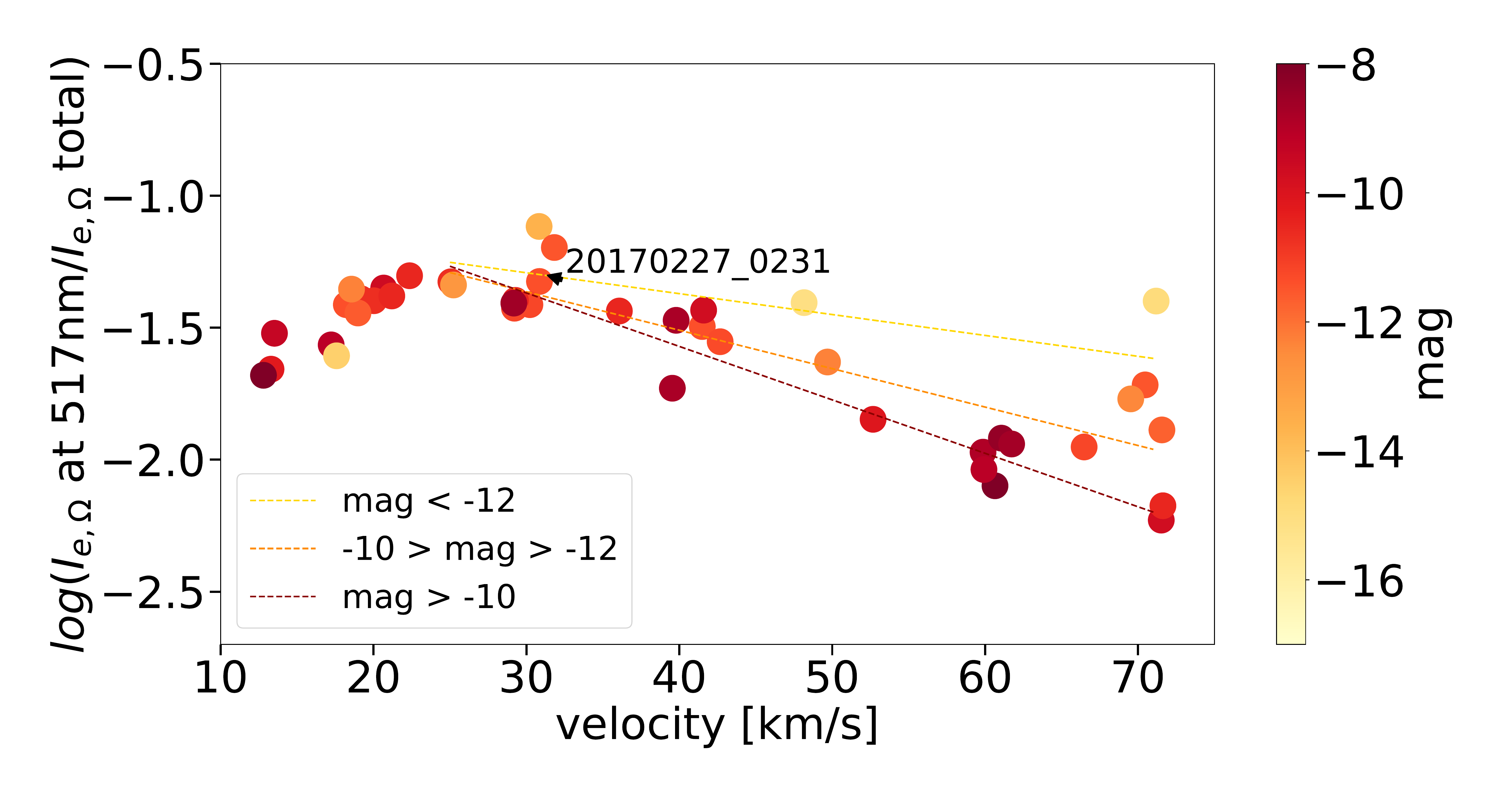}
\caption{}
\label{Fig:Energy_Mg}
\end{subfigure}
\caption{Relative radiant intensities in regions of magnesium and sodium as a function of meteor velocity. Symbol colors mark the meteor's absolute magnitude. Fits for three different magnitude intervals are shown. Figure \ref{Fig:Energy_Na}: Radiation at Na I -- 1 region (589 nm). Figure \ref{Fig:Energy_Mg}: Radiation at Mg I -- 2 region (517 nm).}
\label{Fig:EnergyNaMg}
\end{figure}

\begin{figure}[ht]\centering
{ \includegraphics[width=\hsize]{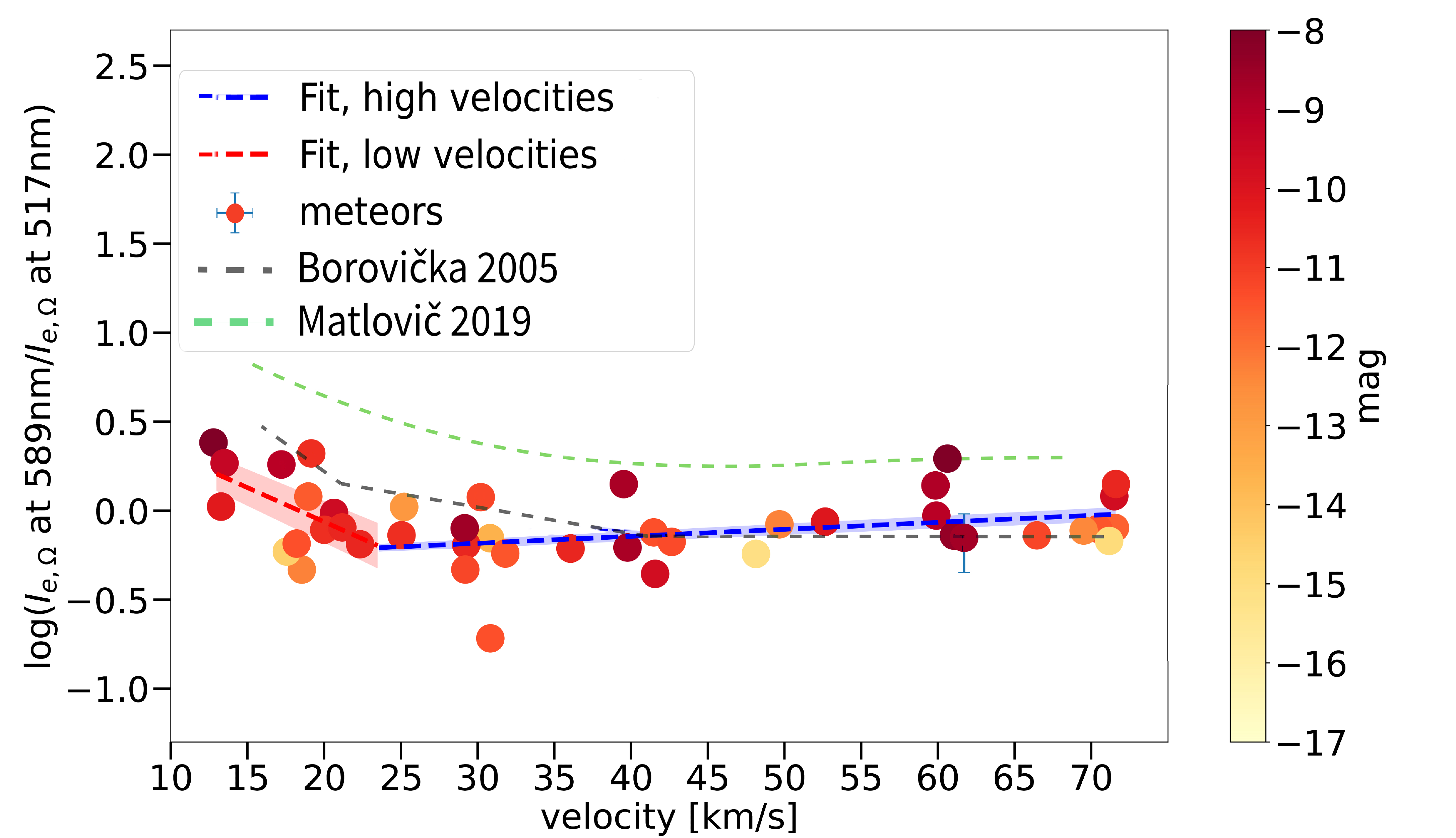}}
\caption{Dependence of the ratio of the radiant intensity at $589nm$ (Na I -- 1) and radiant intensity at $517nm$ (Mg I -- 2) on the velocity. All meteors are marked according to their absolute magnitudes. Monte Carlo fits of meteors slower than $25$km/s (red dashed line) and faster than $25$km/s (blue dashed line) are shown. Uncertainties of these fits are shown as red and blue colored regions. }
\label{Fig:NaMg}
\end{figure}

In Figure \ref{Fig:EnergyNaMg}, the relative radiation in regions at $589$ nm and $517$ nm, where sodium and respectively magnesium dominates, to the broad spectral radiation $I_{total}$ can be seen. Also, three least-squares fits of three brightness bins (meteors brighter than $-12$ mag, meteors with a brightness between $-12$ mag and $-10$ mag, and meteors weaker than $-10$ mag) are shown. For magnesium, only meteors faster than $25$ km/s were used for the brightness bin fits. In Figure \ref{Fig:NaMg} the ratio of intensities in the magnesium and sodium region to each other is shown and it is also compared to previous findings in the works of \cite{borovicka2005} and \cite{Matlovic2019}.

\section{Application of empirical parameters on real GLM data and comparison with known fireballs}

To test the derived parameters in Eq. (\ref{eqnMAG}) we chose several fireball events observed both by the GLM sensor by ground-based cameras. Since our European Fireball Network does not overlap with the coverage of the GOES satellites, we had to use other available ground-based observations.

We developed a Python script that directly calculates the absolute magnitude from GLM data using the method described in Section \ref{Caption_EstimateGLMtoMeteors}. The GLM detector parameters are already included in the script. To calculate the distance between a fireball and a geostationary satellite, it is necessary to know the latitude and the longitude of the fireball. If the altitude is known, it can be added to the script. If the event was recorded simultaneously from GLM--16 and GLM--17, NASA's website provides a corrected approximate altitude calculated from this stereo observation. Alternatively, some typical fireball altitude can be used, because the altitude is negligible compared to the total distance between the fireball and the geostationary satellite. The longitude and latitude of the meteor are provided along with the measured energy by the NASA website in a CSV file. The script does not take into account the parallax of the lightning ellipsoid and uses longitude and latitude as they are. As a result, a figure and an output table file with the light curve are created. If necessary, the light curve can be corrected for frame gaps caused by data overflow using interpolation. The script is available on the GitHub server \footnote{ \url{https://github.com/vojacekasu/GLM_Fireball_Magnitude} \label{footnote_2} }.

\subsection{Comparison with ground-based observations}\label{FireballsCompare}

\subsubsection{British Columbia fireball, September 5, 2017}
The British Columbia fireball and meteorite fall that occurred on September 5, 2017, at Crawford Bay in British Columbia, Canada, is a well-documented event. It was recorded by many observers, ground-based video cameras, and United States Government (USG) sensors \citep{Hildebrand2018}.

\begin{figure*}
\sidecaption
\includegraphics[width = 12cm]{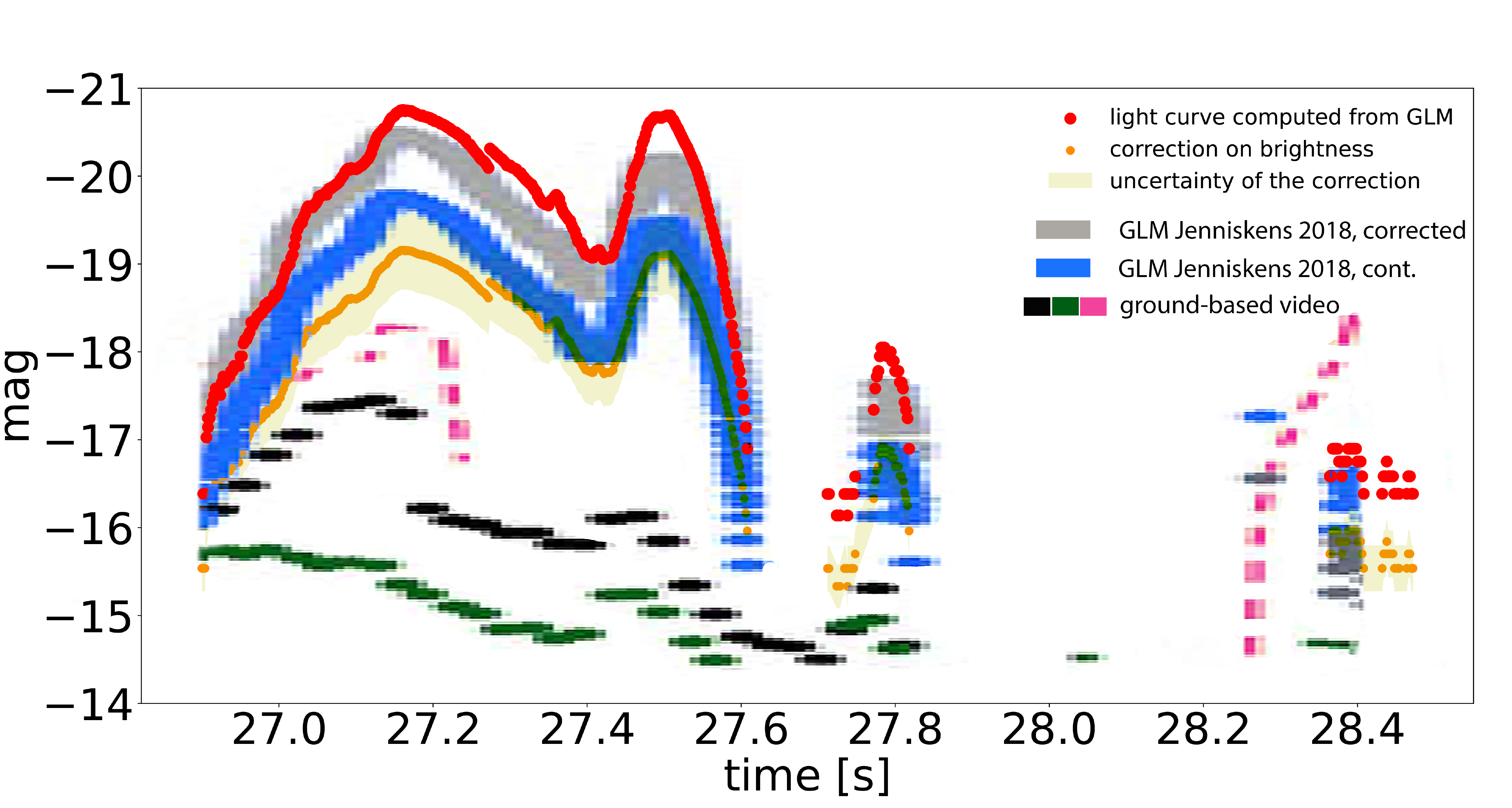}
\caption{British Columbia fireball, with the GLM-computed  light curves shown as: red (this work), gray (\cite{Jenniskens2018} radiation of oxygen lines at $777$ nm), and blue (\cite{Jenniskens2018} only blackbody radiation at $777$ nm). The black, green, and pink are light curves from the ground-based video observations in terms of visual magnitude (source \cite{Jenniskens2018}).}
\label{Fig:GLM_British}
\end{figure*}

The GLM sensors detected the four brightest flares. The magnitudes in the peaks derived from the ground-based videos were in the range from $\approx -15$ to $-18$. We estimated the absolute magnitude using Equations (\ref{eqnMAG}) and (\ref{eqnE_GLM}). The value for the velocity ($\approx 16.5$ km/s) and for the altitude ($\approx 35$ km) were used as reported in \cite{Hildebrand2018}. The estimated light curve, with a peak magnitude of $\approx -20.75$, can be seen in Figure \ref{Fig:GLM_British}. Our result is in good agreement with the light curve of \cite{Jenniskens2018} obtained also from the GLM using their calibration and assuming radiation from oxygen triplet at $777$ nm (gray color in Figure \ref{Fig:GLM_British}). With that assumption, the light curve was corrected for the Sun-blocking filter (the bandpass is not covering the O I triplet completely) and the peak magnitude was $\approx -20.5$. The difference was mostly less than $1$ mag. The blue-colored light curve in Figure \ref{Fig:GLM_British} is also from \cite{Jenniskens2018}, but without the correction on the Sun-blocking filter, that is, assuming only the continuum radiation at $777$ nm. This light curve differs a bit more from our estimate, namely, by more than one magnitude in brigtness. The peak magnitude computed in this case was $\approx -19.5$. The difference between GLM light curve and ground-based observation is assumed to be due to the saturation in video data.

\subsubsection{Arizona fireball, November 15, 2017}
\begin{figure*}[htb]
\sidecaption
{ \includegraphics[width=12cm]{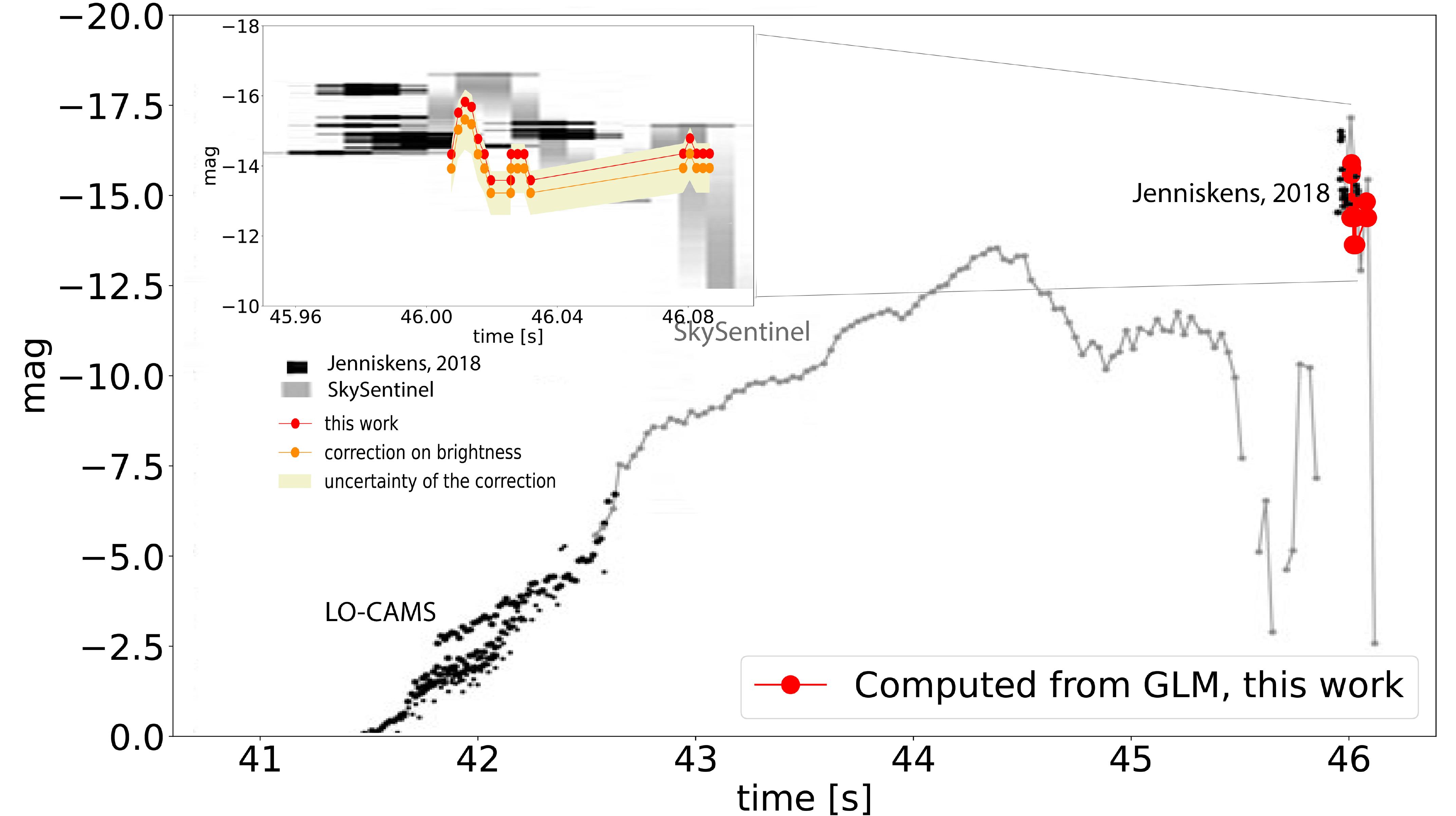}}
\caption{Light curve of the Arizona fireball. Comparison of the estimate from our method and GLM light curve from \cite{Jenniskens2018} and observations from LO-CAMS and SkySentinel (source \cite{Jenniskens2018}). The lower part of the figure shows a detail of the flare part of the light curve. }
\label{Fig:GLM_Arizona}
\end{figure*}

The Arizona fireball was observed on November 15, 2017. It entered the atmosphere with a velocity of $26$ km/s. It was recorded by the SkySentinel video network and by the LO-CAMS, Lowell part of the California All-sky Meteor Surveillance (CAMS) camera network. The peak magnitude was between $-16$ and $-17$ (for details, see \cite{Jenniskens2018}). Only two terminal
flares were recorded by GLM. The time resolution of video observations is low compared to GLM. Our light curve estimate is about one magnitude fainter than the light curve from the video observations, with a peak magnitude of $-16$. A comparison is shown in Figure \ref{Fig:GLM_Arizona}.

\subsubsection{Hamburg meteorite fall, January 17, 2018}
\begin{figure}[htb]
\centering
{ \includegraphics[width=\hsize]{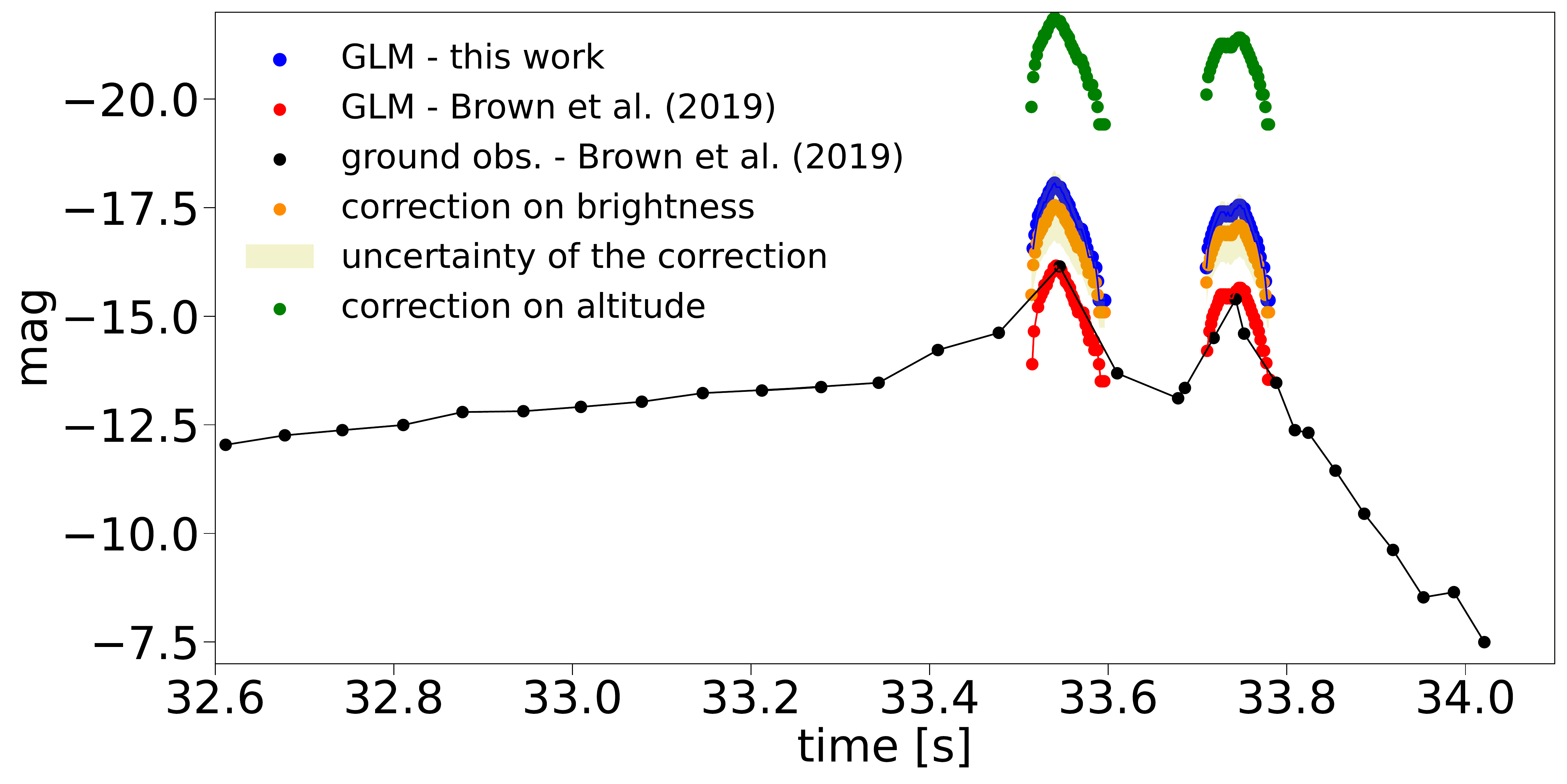}}
\caption{Light curve of the Hamburg fireball, computed from GLM data (red points) and video observations (black points). Time t=33s corresponds to Jan 17, 01:08:33 UT. }
\label{Fig:GLM_Hamburg}
\end{figure}

The Hamburg fireball from January 17, 2018 resulted in a meteorite fall in the area of Ann Arbor, Michigan. The fireball was observed by several security video cameras from the ground and the GLM--16 satellite recorded the two brightest peaks. The initial velocity was $15.83 \pm 0.05$ km/s and the two main flares occurred at altitudes of $24.1$ km and $21$ km \citep{Brown2019}. The light curve from the ground-based observations presented in \cite{Brown2019} is shown in Figure \ref{Fig:GLM_Hamburg}. In \cite{Brown2019}, the observed spectral energy density from GLM was converted into the visual absolute magnitude using assumptions from \cite{Jenniskens2018}, indicating that the limiting sensitivity for GLM is near the peak visual absolute magnitude of -$14$ and connecting this limiting magnitude with the floor of the observed spectral energy density. This GLM-converted light curve is also shown in Figure \ref{Fig:GLM_Hamburg}. The agreement between these two light curves is very good. We compared these light curves with the light curve of the two brightest peaks calculated from GLM data using our method. In this case, the GLM estimate is about two orders of magnitude brighter than the video observations as well as the GLM converted magnitude in \cite{Brown2019}. This brightest part of the light curve reconstructed in \cite{Brown2019} from the scattered light was calibrated to unsaturated parts of the light curve computed from direct fireball measurements in the video. Still, our GLM-derived light curve suggests that the peak magnitude was underestimated by \cite{Brown2019}.

\subsubsection{Alberta event, February 22, 2021}
The Alberta fireball occurred on February 22, 2021. It had a high velocity of $62.1$ km/s. It is the first directly observed decimeter-sized rocky meteoroid on a long-period comet orbit \citep{Vida2022}. Using a manual calibration of ground-based observations and GLM observations of three fast bolides, the empirical equation between absolute magnitude, $m,$ and energy, $E,$ in femto Joules observed by GLM, \begin{math}m = -9.2 - 2.5log_{10}(E)\end{math}, was derived in \cite{Vida2022}. This equation can be applied only to fireballs with velocities around $60$ km/s. Using Equations (\ref{eqnMAG}) and (\ref{eqnE_GLM}), and substituting energy $E$ in femto Joules, we get the following relation: \begin{math}m = -9.8 - 2.5log_{10}(E)\end{math}. This gives us a difference of $0.6$ magnitude between our calibration and that of Vida, which we consider to be a reasonable agreement.

\subsection{Comparison with USG sensors}
To test our calibration of GLM data we also used observations from space-based US Government (USG) sensors published by NASA \footnote{https://cneos.jpl.nasa.gov/fireballs/}. We compared $27$ bolides for which both GLM and USG are available. Velocities from $11$ km/s to $42$ km/s were reported for them by NASA. From the GLM narrowband energy radiated at $777$ nm, we computed the broadband energy radiated from the fireball in Joules using Eq. (\ref{eqnE_GLM}), integrating the reported light curve, and computing emission $I_{total}$ using Eq. (\ref{eqnO_vel}). When necessary, we corrected the light curve for gaps due to the overflow of the lightning detector. Although GLM can observe only the brightest part of the bolide in most cases, we assume that most of the energy is emitted in these bright parts, but some underestimation may be present, especially for meteors at the threshold detection level. The energy reported by USG is the energy radiated from the whole spectral range, assuming radiation of black body at temperature of $\approx 6000$K. The radiation $I_{total}$ computed from radiation at $777$ nm using Eq. (\ref{eqnO_vel}) is in fact only the radiation in the range of $380$ nm -- $850$ nm. To compare these two quantities, we divided the USG energy with the factor of $1.85,$ since the radiation in the whole spectral range of the black body is $1.85 \times $ the radiation of the black body in the spectral range between $380$ nm and $850$ nm.

\begin{figure}
\centering
\begin{subfigure}{0.5\textwidth}
\includegraphics[width=\textwidth]{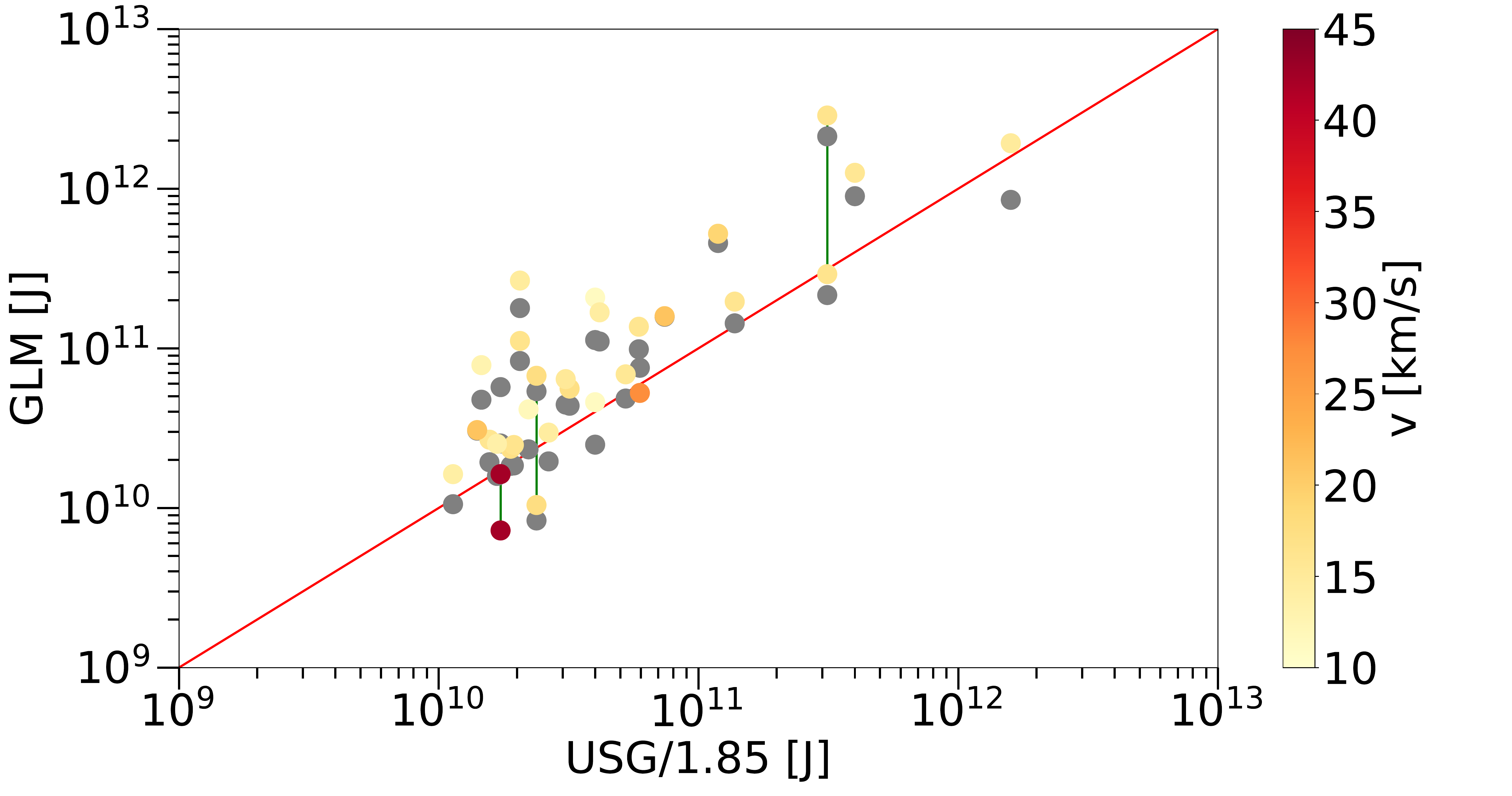}
\caption{}
\label{Fig:USGa}
\end{subfigure}
\hfill
\begin{subfigure}{0.5\textwidth}
\includegraphics[width=\textwidth]{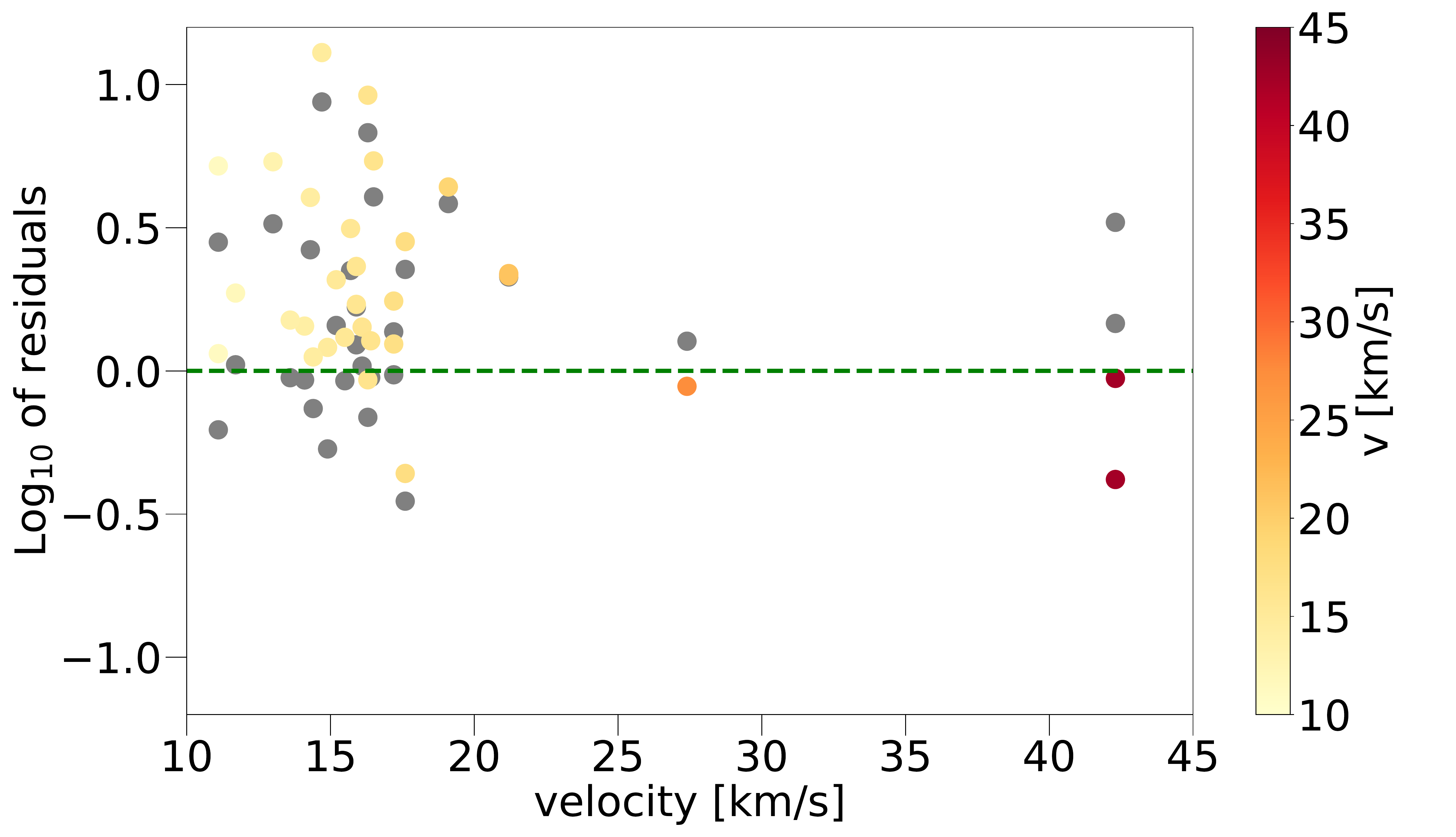}
\caption{}
\label{Fig:USGb}
\end{subfigure}
\caption{GLM radiated energy estimate and USG reported energy for selected fireballs in the spectral range from $380$ nm to $850$ nm. Gray points are computed assuming the blackbody spectrum. Colored points are computed assuming O I -- 1 radiation with correction on velocity. Figure \ref{Fig:USGa}: GLM and USG comparison of observations of same fireballs. Fireballs observed simultaneously with GLM 16 and GLM 17 are connected with the green line. Figure \ref{Fig:USGb}: Residuals ($J, log_{10}$ scale) from the GLM--USG matching curve.}
\label{Fig:USG}
\end{figure}

We can see the result of this comparison in Figure \ref{Fig:USGa}. Some fireballs have been detected from both GOES satellites. We computed the radiation for both detections. These points are connected by a green line. We also computed the radiated energy from GLM data assuming a $6000$ K black body spectrum. Following \cite{Jenniskens2018}, the GLM reported energy was multiplied by a factor of $1018$ to obtain the energy in the whole spectral range of the black body. Then it was also divided by the factor of $1.85$ to obtain radiation of the black body in the spectral range $380$ nm -- $850$ nm. These points are marked in gray in Figure \ref{Fig:USGa}.

In Figure \ref{Fig:USGb} we show the difference between the computed GLM energy and the energy reported by USG as a function of the velocity reported by USG. The difference is given in the units of order, namely a value of $1.0$ on the vertical axis means that the GLM energy was exactly one order larger than the USG energy. Here, again, color-coded points are computed using our calibration from Eq. (\ref{eqnO_vel}), and the gray points are energies computed assuming only blackbody radiation. According to Eq. (\ref{eqnO_vel}), for velocities of $22.6$ km/s, both methods give identical results. And while for velocities below $\approx 22.6$ km/s, the black body assumption gives the energy closer to the energy reported by USG -- for velocities higher than $22.6$ km/s, the energy computed assuming the oxygen triplet radiation is closer to the energy reported by USG. However, there are only two meteors with a velocity above $22.6$ km/s and, thus, the scatter of points at lower velocites is large.

\section{Extrapolation of the calibration}
\subsection{Brightness correction}
Since our sample of EN fireballs contains limited range of meteor magnitudes, we examined how the velocity calibration in Equations \ref{eqnO_vel} and \ref{eqnMAG} for the oxygen region radiation can be extrapolated for fireballs with magnitudes outside of the range of our sample. We investigated how the relative radiation in the oxygen region changed with meteor brightness. To eliminate the influence of velocity on this relative radiation, we stacked the EN fireballs into groups according to similar velocities. In this way, we can study how the relative radiation at $777$ nm $I_{777}/I_{total}$ depends on the absolute radiation $I_{777}$ at $777$ nm in given velocity group using customized Eq. \ref{eqnO_vel}:

\begin{multline}\label{eqnO_vel_extrapol}
log(I_{777}/I_{total}) - 0.026(\pm0.001) \times v + 3.294(\pm0.077) = \\ c(v) \times log(I_{777}) + d(v),
\end{multline}

where constants $c(v)$ and $d(v)$ depend on velocity and are thus different for different velocity groups. The dependence for each velocity group is shown in Figure \ref{AP:Fig:Fit_I777} in the appendix. To compute the uncertainties of each fit, we used the Monte-Carlo method with $10000$ points generated with normal distribution within errors for each point in Figure \ref{AP:Fig:Fit_I777}. Errors for each point were computed using the measured noise level for each spectrum. For each clone set, the least-squares fit was performed and then the final fit and its uncertainty were obtained by computing the mean and the standard deviation of these fits. Then, $c(v)$ is the slope of the fit and $d(v)$ is the intercept. To obtain how parameters $c(v)$ and $d(v)$ depend on velocity, we computed average velocity for each velocity group and plotted the dependence of $c(v)$ and $d(v)$ on this velocity. This is shown in Figure \ref{AP:Fig:parameterFit_I777} of the Appendix. Errors for the average velocity of each velocity group were computed using uncertainties of velocity measurements of each fireball. As an error of $c(v)$ and $d(v),$ the above-mentioned standard deviations of least-squares fits were used. To obtain the fit of these dependencies, we used the least-squares method. The uncertainty of the fit was obtained by once again using the Monte Carlo method, generating $10000$ clones with normal distribution and using means and standard deviations of fits of these clones.

The results of the fits were:

\begin{equation}\label{cv}
c(v) = -0.0053(\pm0.0041) \times v + 0.261(\pm0.224)
\end{equation}

and

\begin{equation}\label{dv}
d(v) = 0.021(\pm0.013) \times v - 1.00(\pm0.71).
\end{equation}

\begin{figure}
\centering
\begin{subfigure}{0.5\textwidth}
\includegraphics[width=\textwidth]{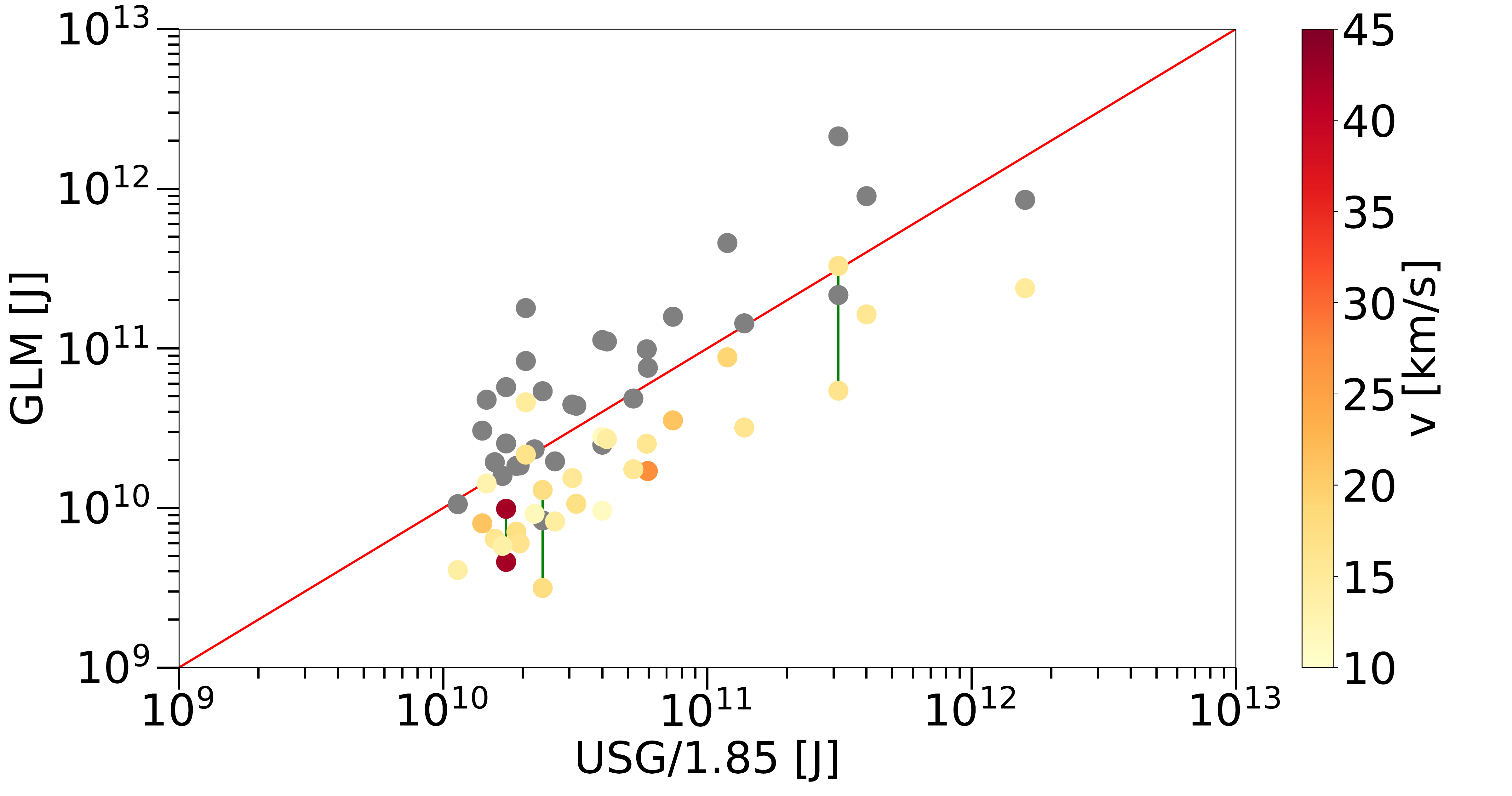}
\caption{}
\label{Fig:USGaCorr}
\end{subfigure}
\hfill
\begin{subfigure}{0.5\textwidth}
\includegraphics[width=\textwidth]{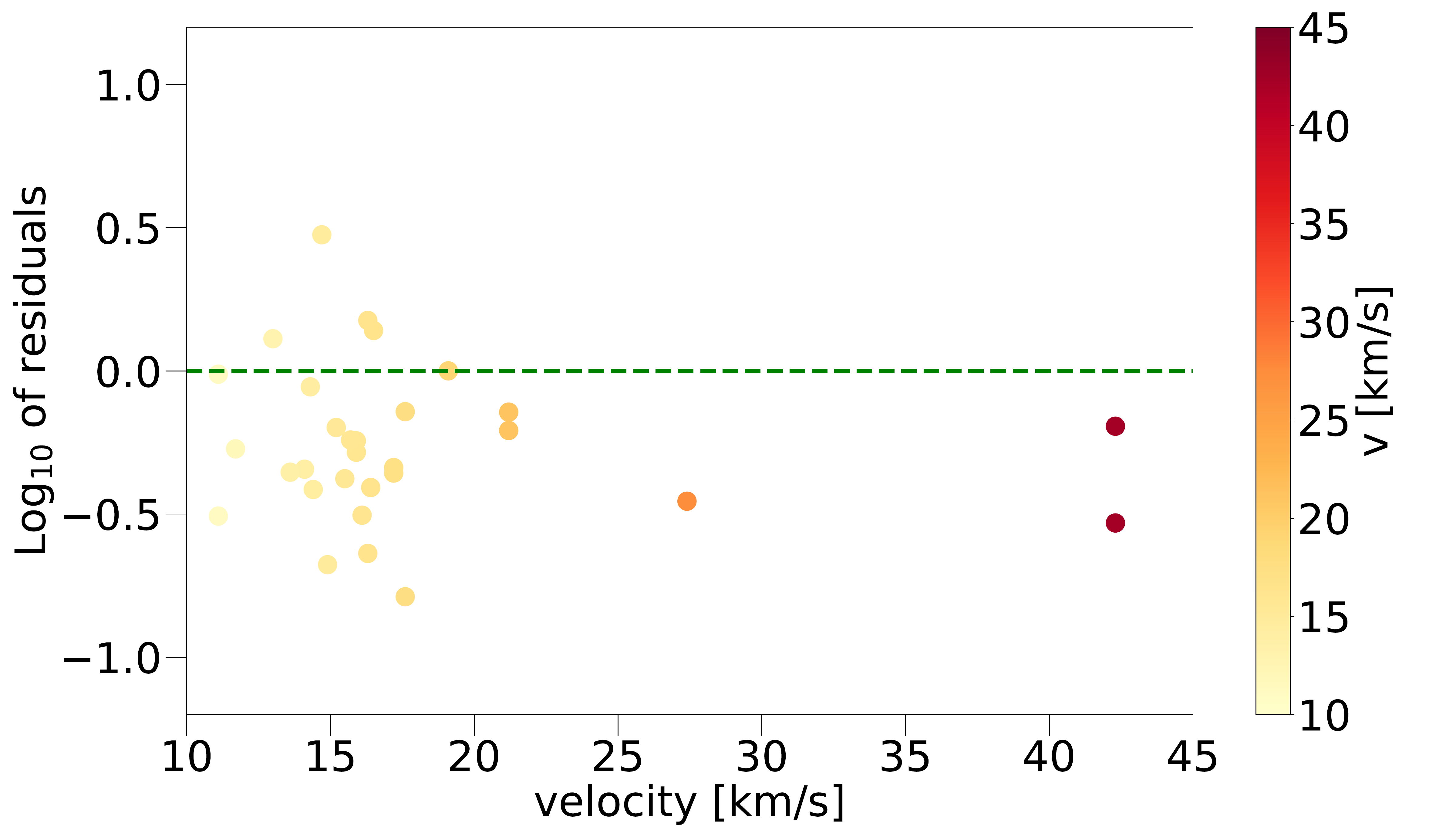}
\caption{}
\label{Fig:USGbCorr}
\end{subfigure}
\caption{GLM radiated energy estimate and USG reported energy for selected fireballs in the spectral range from $380$ nm to $850$ nm. Points are computed assuming O I -- 1 radiation with correction on velocity and after the correction on fireball brightness. Figure \ref{Fig:USGaCorr}: GLM and USG comparison. Figure \ref{Fig:USGbCorr}: Residuals ($J, log_{10}$ scale) from the GLM--USG matching curve.}
\label{Fig:USGCorr}
\end{figure}

With these parameters, we can compute corrected radiation $I_{total}$ (i. e., radiation between $380$ nm and $850$ nm) from known radiation at $777$ nm $I_{777}$ and known fireball velocity, $v$. When we assume that this correction can simply be extrapolated for fireballs outside of the magnitude range of the EN fireballs sample in this work (within a reasonable range), then this can be applied for radiation calibration of the GLM observations of bolides brighter than those in the EN sample. We applied this correction to the GLM and USG comparison in Figure \ref{Fig:USGCorr}. We can see that after this correction the GLM reported energies are in a bit better agreement with USG reported energies, compared to energies computed without this correction in Eq. \ref{Fig:USG}. In addition, within the uncertainty of this correction, they are in an agreement with energies computed from GLM assuming only blackbody radiation.

A similar correction as for total intensities can be applied to magnitudes computed from GLM observations, Eq. \ref{eqnMAG} is modified to:

\begin{multline}\label{eqnMag_vel_extrapol}
m_V + 2.5 \times log_{10}(I_{777}) - 0.0948 \times v + 3.45 = cm(v) \times log(I_{777}) + dm(v).
\end{multline}

Results of particular fit of each velocity group can be seen in Figure \ref{AP:Fig:Fit_mag} in the appendix and in the Figure \ref{AP:Fig:parameterFit_mag} the dependence of $cm(v)$ and $dm(v)$ on velocity can be seen. The result of the least-squares fitting of these parameters is:

\begin{equation}\label{cv}
cm(v) = -0.022(\pm0.005) \times v + 0.79 (\pm0.24)
\end{equation}

and

\begin{equation}\label{dv}
dm(v) =0.102(\pm0.016) \times v -3.31 (\pm0.77).
\end{equation}

If we again assume that this correction can be linearly extrapolated, we can apply it for fireballs in Section \ref{FireballsCompare}. For the British Columbia fireball, we can see in Figure \ref{Fig:GLM_British} that the corrected magnitude is within the uncertainty of the correction in agreement with the light curve of \cite{Jenniskens2018}, when they assumed only radiation in the continuum. The correction for the Arizona fireball was only minor and the corrected and uncorrected light curves were within the uncertainty of the correction (see Figure \ref{Fig:GLM_Arizona}). Also, the corrected light curve of the Hamburg fireball in Figure \ref{Fig:GLM_Hamburg} was only slightly different from the uncorrected light curve. This shows that the dependence of the relative radiation in the oxygen region on the brightness of the meteor is minor, but our method can be extrapolated for fireballs outside the magnitude range of our sample.

\subsection{Dependence on altitude for meteors with and without flare}

\begin{figure}[htb]
\centering
{ \includegraphics[width=\hsize]{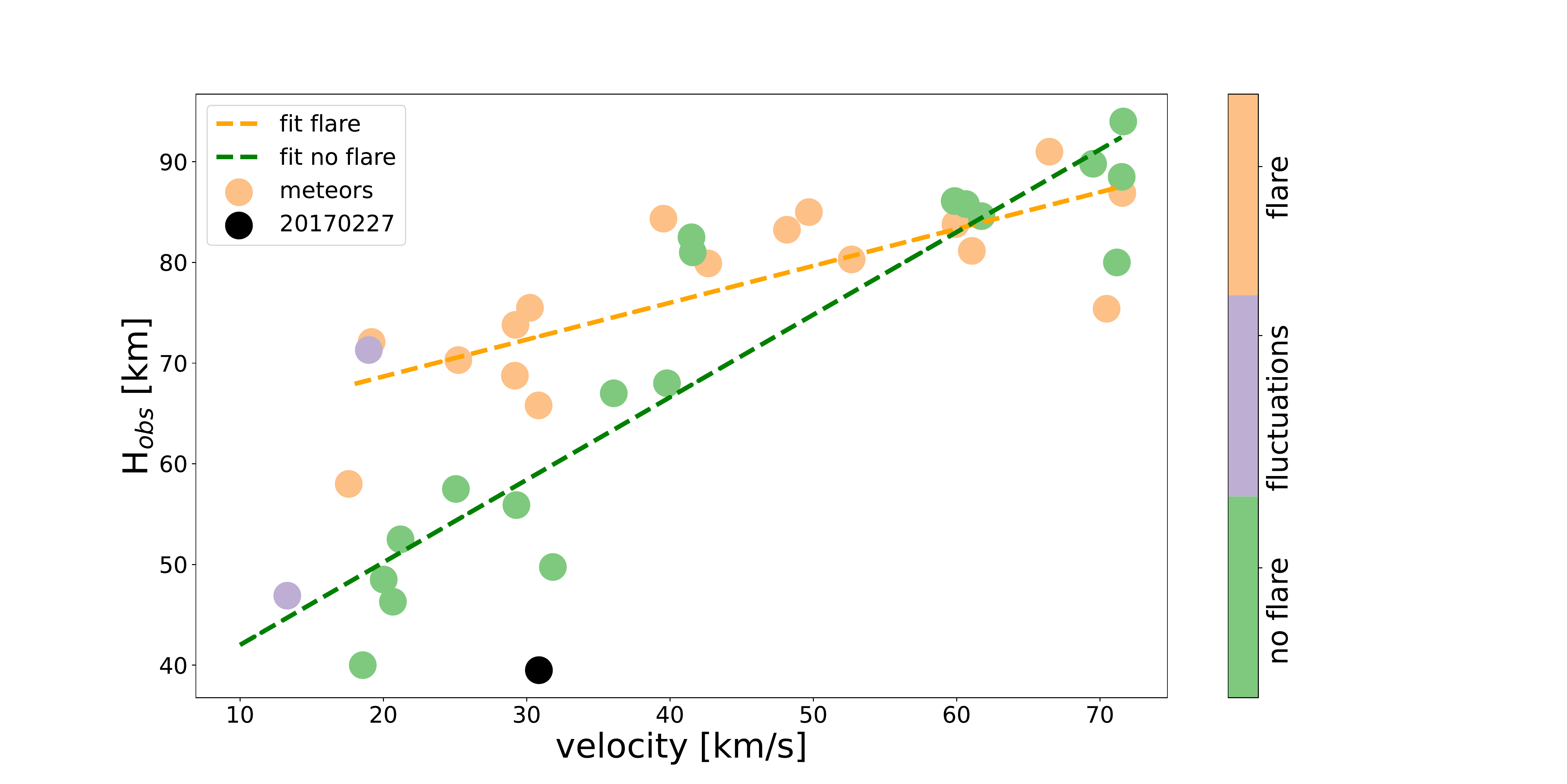}}
\caption{Altitude of scanned spectrum and velocities of meteors. Meteors with fluctuations, flares, or without flares are colored.}
\label{Fig:VelAlti}
\end{figure}

We observed that deviations of meteor altitude from the altitude typical for the given meteoroid velocity can affect the radiation at $777$ nm and, thus, the derived estimate of meteor magnitude. To examine this, we used the middle altitude of the spectrum scan and its dependence on the meteor velocity (see Figure \ref{Fig:VelAlti}). Naturally, the scan was correlated with the altitude of maximal brightness to maximize the spectrum S/N ratio. Meteors without flare showed a steeper dependency of altitude on velocity than those with flare. Fast meteors from both groups had the altitude more or less similar, slow meteors were lower in the atmosphere when they did not show any flare on their light curve. The least-squares fits of these two groups were used to determine the typical altitude for a given velocity. For meteors without flare in the spectrum, the typical altitude $H(v)_{N}$ in km was:

\begin{equation}
H(v)_{N} = 0.82 \times v + 34.0.
\label{eqn:HvNo}
\end{equation}

For meteors with the flare in the spectrum the typical altitude H(v)$_{F}$ was:

\begin{equation}
H(v)_{F} = 0.37 \times v + 61.4.
\label{eqn:HvF}
\end{equation}

\begin{figure}[htb]
\centering
\includegraphics[width=\hsize]{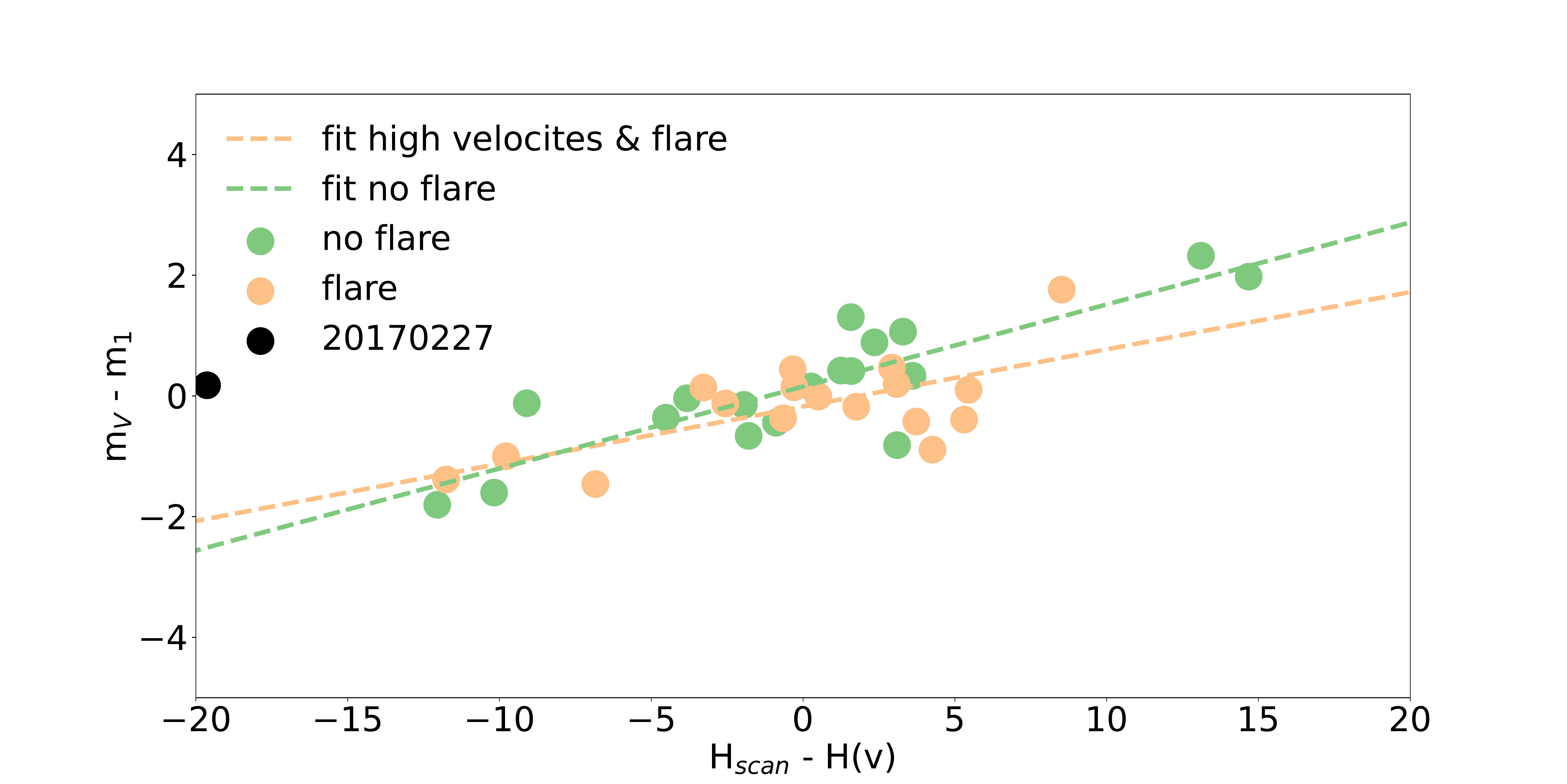}
\caption{Difference between visual magnitude, $m_V$, and magnitude, $m_1$, estimated from radiation at $777$ nm and its dependence on the deviation of the scanned spectrum altitude, $H_{obs}$, from the altitude typical for given velocity, $H(v)$.}
\label{Fig:magCorrHeight}
\end{figure}

Here, $v$ is the velocity of the meteor in km/s. The difference between this altitude and the actual altitude of the scanned spectrum ($H_{obs}$ - $H(v)$) was computed to determine correction on altitude for both groups. We refer to Figure \ref{Fig:magCorrHeight}, where the dependence of the difference between actual visual and estimated magnitude ($m_V - m_1$) on the altitude difference ($H_{obs}$ - $H(v)$) is shown. The least-squares fit slope can be used as the parameter of the magnitude correction on altitude. When no flare is in the light curve it can be:

\begin{equation}
m_H = m_1 + 0.14(H_{obs} - H(v)).
\label{eqn:corr3}
\end{equation}

Here, $m_H$ is the magnitude corrected for meteor altitude and H$_{obs}$ is the middle altitude at which the spectrum was scanned to estimate the magnitude $m_1$.
If there is a flare, we then have:

\begin{equation}
m_H = m_1 + 0.10(H_{obs} - H(v)).
\label{eqn:corr3Flare}
\end{equation}

In Figure \ref{Fig:magCorrHeight}, we marked meteor 20170227, with a magnitude estimate, m$_1$, very close to the visual magnitude, $m_V$, even though its spectrum scan altitude, $H_{obs}$, was lower than for any other meteor. This meteor is also marked in Figure \ref{Fig:Energy}, showing a low abundance of sodium and normal magnesium abundance for a given velocity. This indicates the high strength of the meteoroid material \citep{borovicka2005}. This is also in agreement with the parameter, $P_E$, computed from the terminal altitude of the luminous trajectory, the initial velocity, the initial mass, and the slope of the atmospheric trajectory \citep{CeplechaMcCrosky1976}. This $P_E$ can be used for the classification of meteoroid material \citep{Ceplecha1988}. For this fireball, we computed $P_E = -3.96$ and it was classified as type I (ordinary chondrite), with an origin in asteroids according to Ceplecha classification. This can explain why, for a given velocity, the fireball penetrated so deep in the atmosphere; on the other hand (as it can be seen in Figure \ref{Fig:Energy}), the normal relative radiation at $777$ nm for a given velocity caused an accurate magnitude estimate, computed from this radiation. We excluded this meteor from the correction on altitude estimation and did not apply this correction to this fireball.

\subsection{Applying both corrections}

\begin{figure*}
\centering
\begin{subfigure}[b]{0.475\textwidth}
\centering
\includegraphics[width=\textwidth]{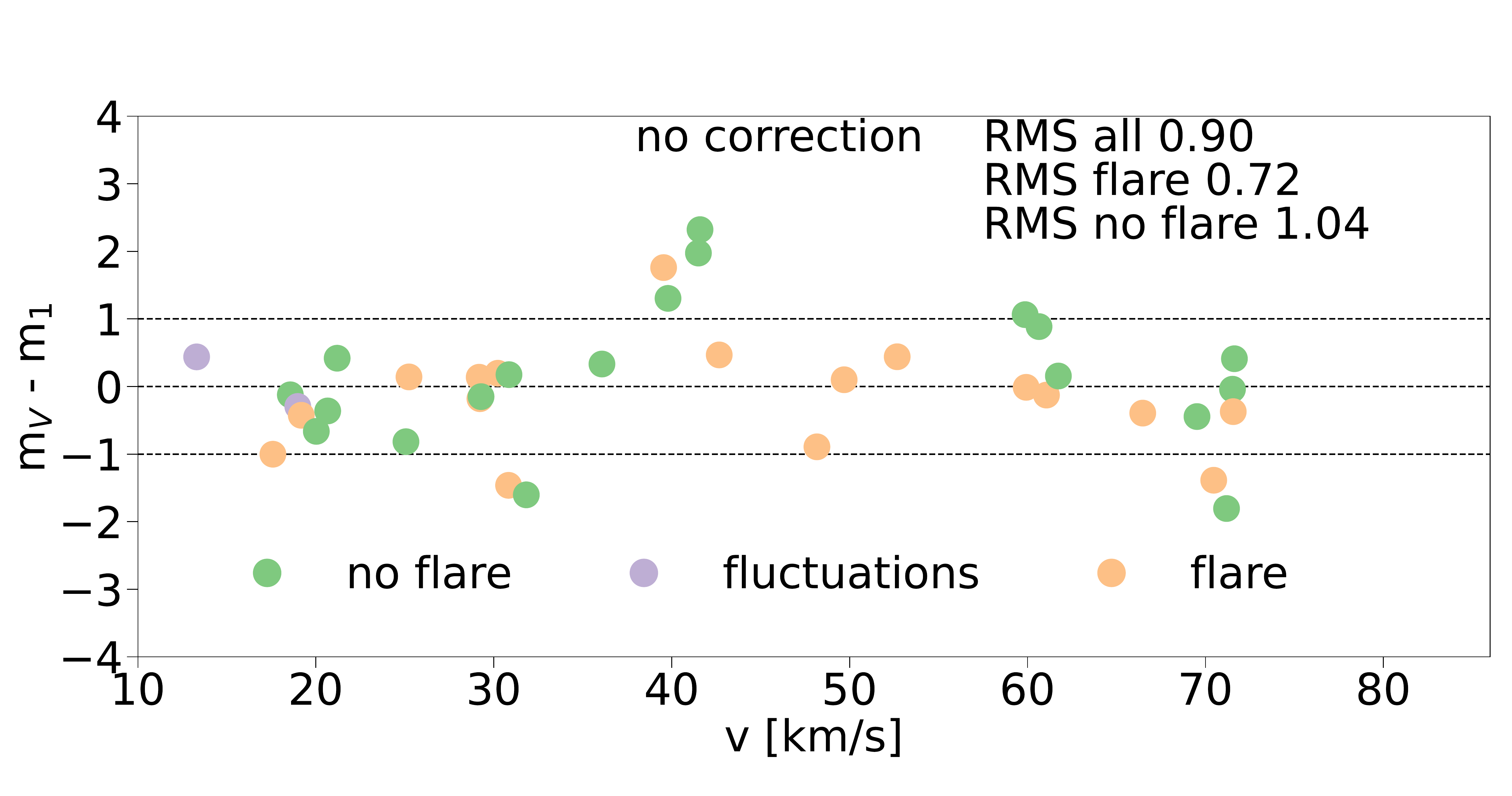}
\caption{}
\label{Fig:MagCorr2a}
\end{subfigure}
\hfill
\begin{subfigure}[b]{0.475\textwidth}
\centering
\includegraphics[width=\textwidth]{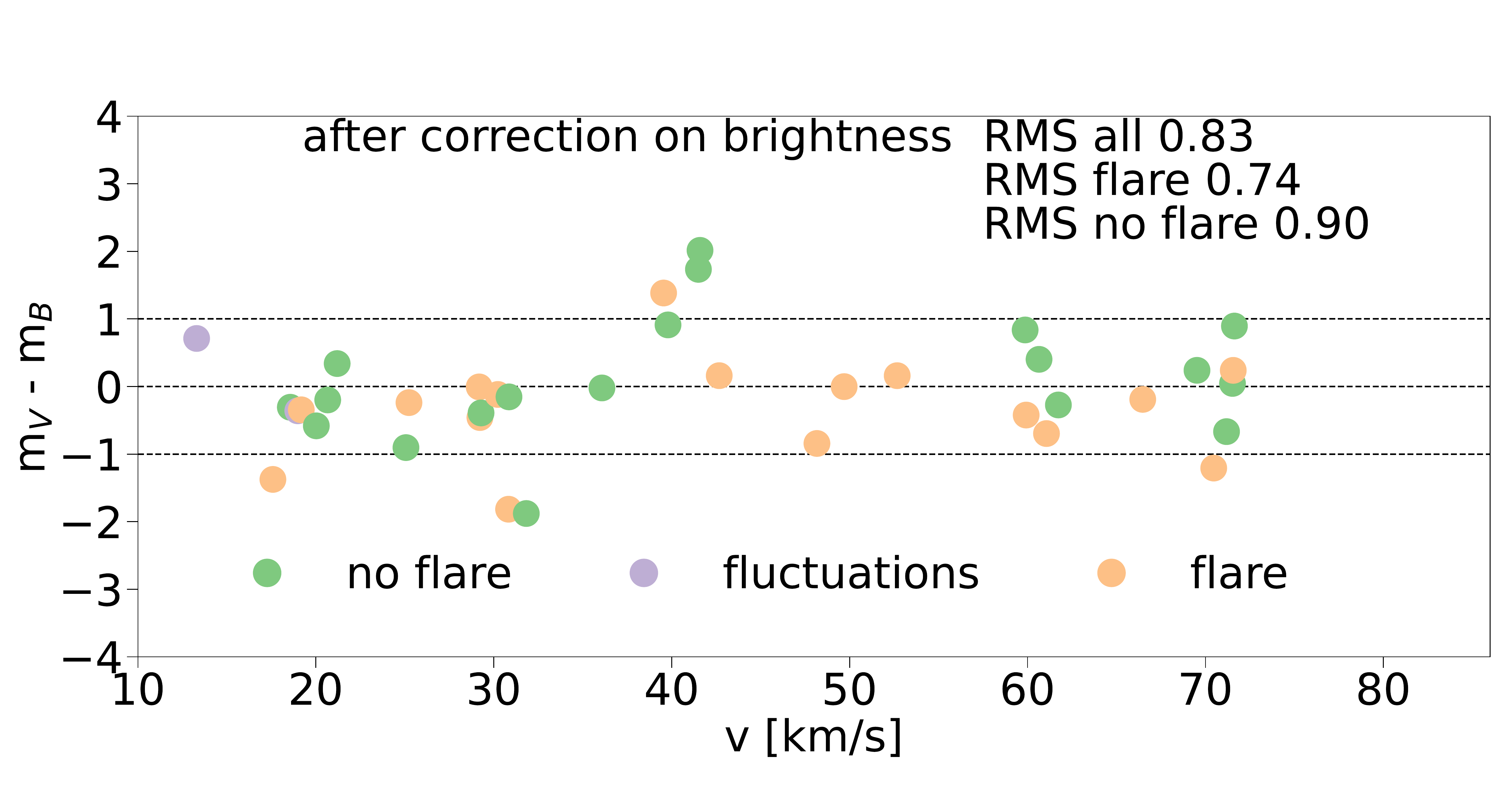}
\caption{}
\label{Fig:MagCorr2b}
\end{subfigure}
\vskip\baselineskip
\begin{subfigure}[b]{0.475\textwidth}
\centering
\includegraphics[width=\textwidth]{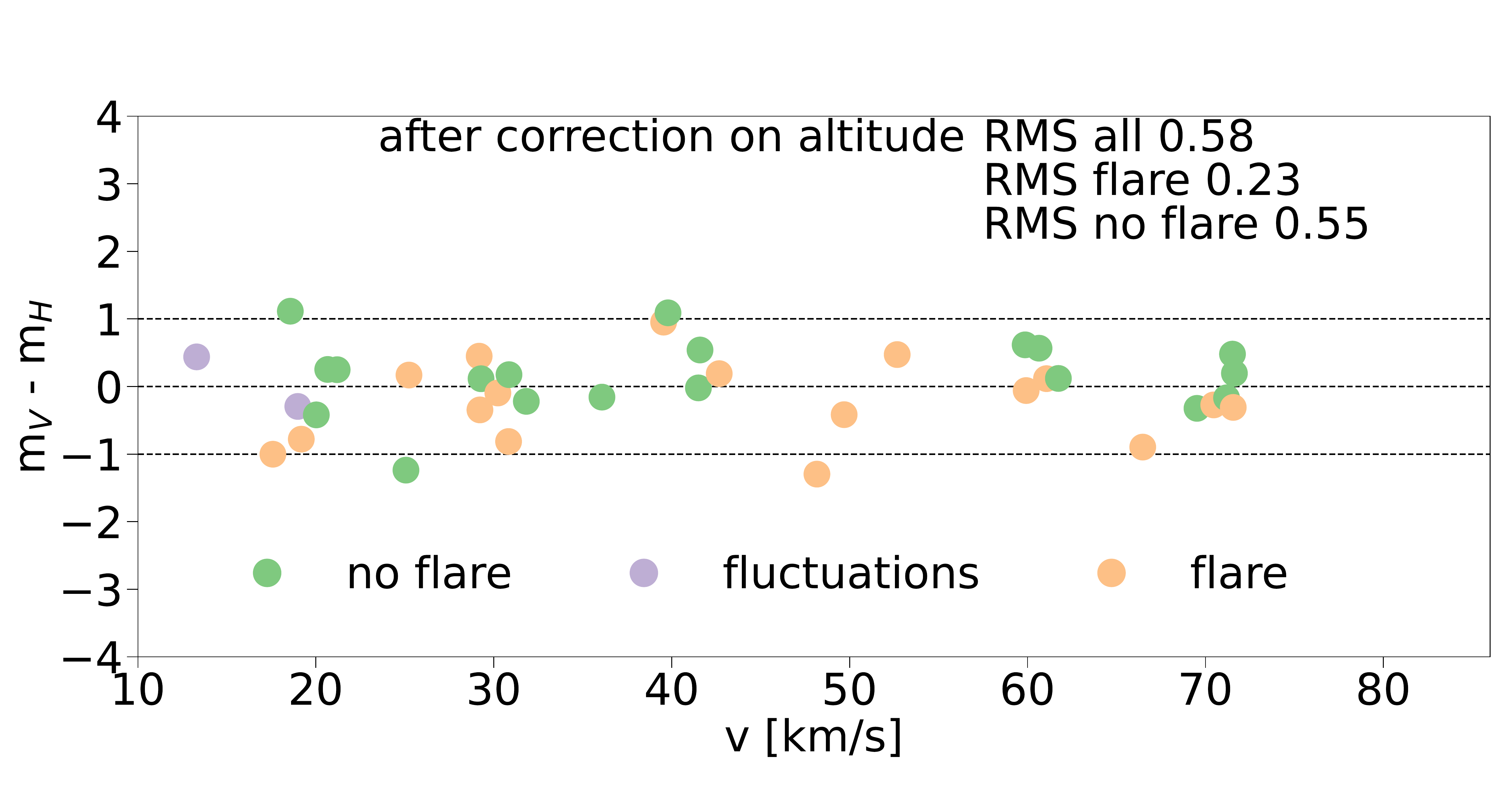}
\caption{}
\label{Fig:MagCorr2c}
\end{subfigure}
\begin{subfigure}[b]{0.475\textwidth}
\centering
\includegraphics[width=\textwidth]{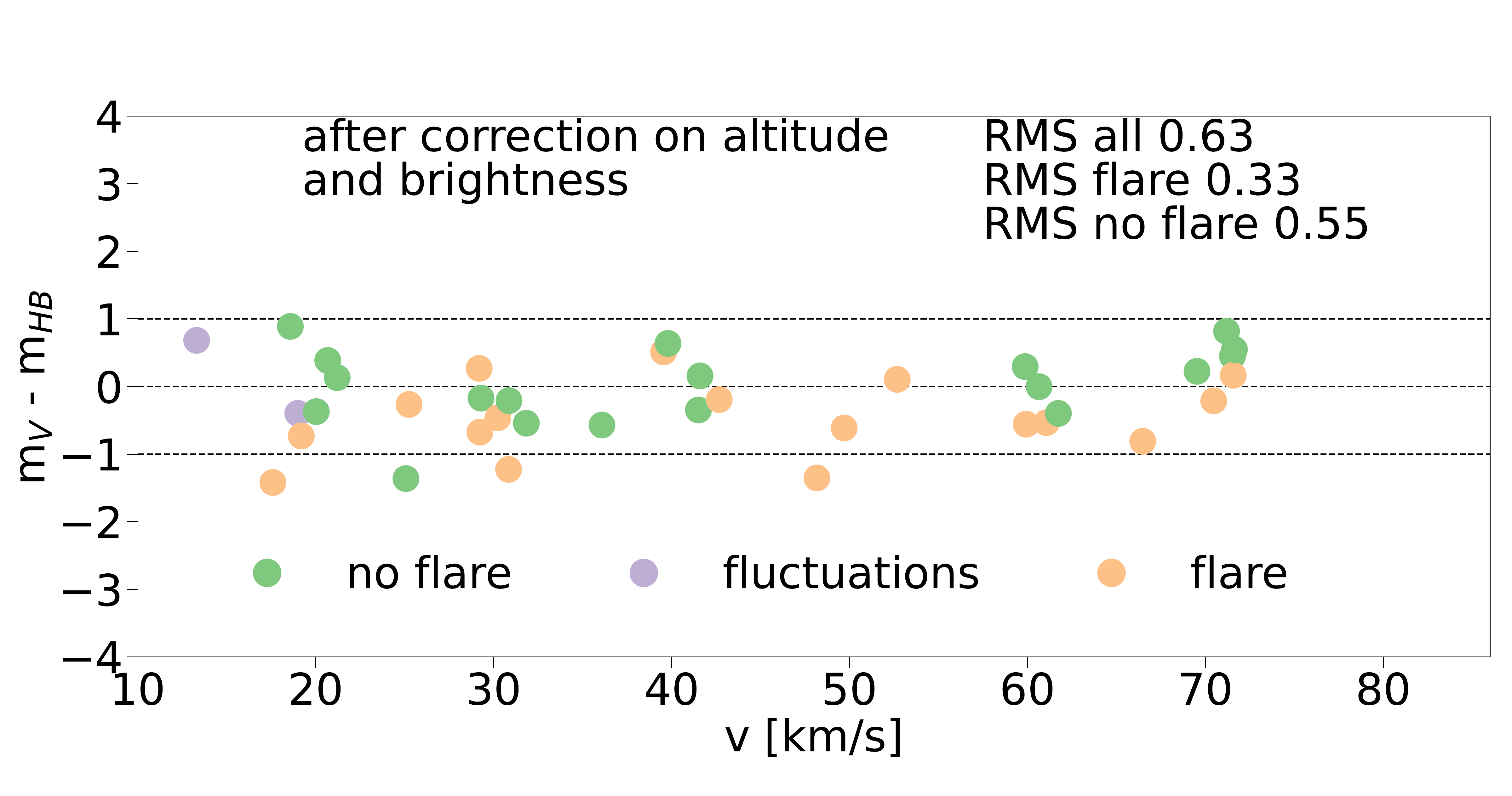}
\caption{}
\label{Fig:MagCorr2d}
\end{subfigure}
\caption[ Difference between visual magnitude $m_V$ and magnitude $m_1$ estimated from radiation at $777$nm and magnitudes corrected for brightness and altitude of a meteor. ]
{\small Difference between the visual magnitude, $m_V$, and the magnitude, $m_1$, estimated from radiation at $777$nm (Fig. \ref{Fig:MagCorr2a}), the magnitude $m_B$ corrected for brightness (Fig. \ref{Fig:MagCorr2b}), the magnitude, $m_H$, corrected for the altitude of a meteor (Fig. \ref{Fig:MagCorr2c}), and the magnitude, $m_{HB}$, corrected both for the brightness and the altitude of meteor (Fig. \ref{Fig:MagCorr2d}).}
\label{Fig:MagCorr2}
\end{figure*}

The brightness correction of the calibration was applied to our EN fireball sample in Figure \ref{Fig:MagCorr2b}, where observed visual magnitudes, $m_V$, are compared with magnitudes computed using the calibration in Eq. \ref{eqnMag_vel_extrapol}. When comparing the root mean square (rms) of this plot with the rms of the data in Figure \ref{Fig:MagCorr2a}, where no meteor brightness was applied, we can see a small improvement. The dependence of the relative radiation of oxygen is small but observable.

The altitude correction of the calibration was applied to the EN fireball sample. Result can be seen in Figure \ref{Fig:MagCorr2c}, where observed visual magnitudes, $m_V$, are compared with magnitudes computed using the calibration in Eq. \ref{eqnMAG} and these magnitudes were then corrected for the altitude using Equations \ref{eqn:corr3} and \ref{eqn:corr3Flare}. The rms of the sample improved overall when deviating fireballs with fast velocities and with velocities between $40$ km/s and $50$ km/s were corrected.
The altitude correction was also applied to magnitudes that have already been corrected with regard to brightness. This result can be seen in Figure \ref{Fig:MagCorr2d}.

\section{Discussion}

Using fireball observations of the Europan fireball network (EN) we developed a simple method to calibrate observations of bolides in the oxygen region at $777$ nm by GLM detectors on board the GOES satellites. This method uses the dependence of the oxygen radiation on the velocity of fireballs. The difference in radiation between slow and fast meteors in this region for similarly bright fireballs can be up to two orders (see Figure \ref{Fig:Energy}). The Europan fireball network is not overlapping with the coverage of the GLM detectors thus no simultaneous observations can be used. To test our method, we compared visual magnitudes of meteors observed by EN with magnitudes of the same meteors, computed from the radiation of the oxygen region using our calibration method and spectral cameras of EN. The difference between these two brightnesses was usually less than $2$ mag and most of the meteors showed a difference of less than $1$ mag. We also applied our method to fireballs that were simultaneously observed by GLM and from the ground. For slow meteors, our method was more or less in an agreement with calibration assuming only blackbody radiation in the whole spectrum (\cite{Jenniskens2018} or \cite{Brown2019}. The calibration was also in good agreement with the calibration of the fast Alberta fireball in the work of \cite{Vida2022}. What is crucial for our method is the knowledge of the fireball velocity, which is not always available for GLM observations. We also compared our calibration of GLM observations that were simultaneously observed with USG. The method was in good agreement with calibration that assumed only black body radiation since most of the fireballs were slower than $20$ km/s (see Figures \ref{Fig:USGa} and \ref{Fig:USGb}). In the future, when the Lightning Imager (LI) on the EUMETSAT's Meteosat Third Generation will be deployed, the overlap with our fireball network will allow the possibility to directly compare the same fireballs observed by lightning detectors and by the spectral cameras of our network (under the condition that it will be possible to filter the fireball data and those observations will be available).

Other influences that can affect oxygen radiation were studied. As it can be seen in Figure \ref{AP:Fig:Fit_I777} the radiation at $777$ nm relative to the broadband radiation, $I_{total}$, is increasing with absolute radiation at $777$ nm (i. e., with the brightness of meteor) for slow meteors. In slow meteors, we observed only weak or no oxygen lines. Due to the low excitation of atmospheric atoms and molecules in slow meteors the oxygen line is not present in the spectrum and only continuous radiation contributes to the 777 nm band. As the meteor brightness increases, the radiation becomes optically thick (approaching the black body spectrum) and the continuum becomes more important in relation to spectral lines in other parts of the spectrum. For meteors faster than $\approx 30$ km/s, the relative brightness at $777$ nm was constant, with increasing meteor brightness and for the fastest meteors, it was decreasing. In fast and bright meteors, the oxygen line can be already optically thick and thus its relative brightness can be smaller than in fainter meteors with similar velocity. When we applied this correction to our sample with a limited magnitude range, the result was quite negligible. However, when we linearly extrapolated this correction to brighter fireballs (in Section \ref{FireballsCompare}), the results in general improved.

The altitude calibration showed that spectra observed at an altitude lower than the altitude typical for a given velocity showed lower relative brightness of radiation at the oxygen region compared to spectra observed at a higher altitude. As the meteor penetrates lower in the atmosphere, the ablation rate increases, and the relative brightness of oxygen is smaller. This behavior was similar for spectra observed during the flare, but the typical altitude for a given velocity was different when observed during the flare compared to observation without the flare.

When we applied the altitude correction to the sample of EN fireballs and compared the corrected magnitudes, $m_H$, with visual magnitudes $m_V$, we were able to correct some of the most deviating meteors (see Figure \ref{Fig:MagCorr2c}). Without the altitude correction, some meteors with velocity $\approx 40$ km/s and higher altitudes than expected had the oxygen line brighter than expected for their velocity. Two very fast meteors were at lower altitudes with fainter oxygen lines. The meteor that deviated most from the typical height had high $P_E$ parameter (see Figure \ref{Fig:magCorrHeight}). Thus, we investigated if there is some relation between the meteoroid strength (represented with the $P_E$ parameter) and the difference $m_V - m_1$, where $m_1$ is the first, uncorrected, estimate of the magnitude computed from the radiation at $777$ nm. As it can be seen in Figure \ref{AP:Fig:Hscan_Pe_vel}, there is no clear relation between the $P_E$ parameter, the altitude of the spectrum, and the velocity of the fireball. Most of the meteors deviated less than $5$ km from the expected altitude for a given velocity. Those meteors for which spectra were scanned at a lower altitude than expected had a wide range of velocities and also they had different values of the $P_E$ parameter. Only three meteors had spectrum scanned higher than $5$ km above the expected altitude. They had velocities between $40$ and $50$ km/s with similar $P_E$ values between $-5.0$ and $-5.6,$ corresponding to classes II and IIIA according to the classification of \cite{CeplechaMcCrosky1976} as carbonaceous chondrites and regular cometary material, respectively. It seems that apart from the velocity of the meteor, the altitude where the spectrum is observed can affect the relative radiation of the oxygen. In other words, the altitude and flare correction show that at higher altitudes and outside flares the oxygen line is relatively brighter since the ablation rate for meteoritic lines is lower compared to observations of the spectrum during the flare or at lower meteor altitudes. When we applied the altitude correction on magnitudes that have already been corrected for brightness, the overall rms was a bit worse than the rms of magnitudes corrected only for altitude (without previous correction on brightness). This shows that the brightness correction is minor for the magnitude range of fireballs in our sample. Moreover, we can see in Figure \ref{Fig:MagCorr2} that meteors with medium and fast velocities showed the best improvements when both brightness and altitude corrections were applied. On the other hand, the slowest meteors did not improve (in general). While using correction on altitude and brightness, we were able to improve the difference between the magnitude estimated from the radiation at $777$ nm and the magnitude in the visible spectral range, $m_V$, within one magnitude in brightness for most fireballs observed by SDAFO.

The application of the altitude-flare correction to the observations from GLM is not trivial. The altitude calibration is based on a sample of cm-sized meteoroids, larger meteoroids can, in general, penetrate deeper in the atmosphere and the typical altitude for a given velocity is lower, thus we assume that this calibration is limited for meteoroids of brightness between $\approx -8$ and $-15 $ mag. As an example, we used the altitude correction for the Hamburg fireball. We used the correction for meteors with flares (Equations \ref{eqn:HvF} and \ref{eqn:corr3Flare}). As an observed altitude, we used the altitude reported for the first flare $H_{obs}$: $24.1$ km. The typical altitude for the reported velocity is $H(v)_{F} = 67.3 km$. This large difference between $H_{obs}$ and $H(v)_{F}$ caused a probable over-correction of the magnitude in Figure \ref{Fig:GLM_Hamburg}. For the given velocity, we did not observe any EN fireball with a flare at such a low altitude. Unfortunately only for a small fraction of GLM fireballs is the altitude estimated and if it is not known from ground observation, it can be unreliable.

\subsection{Threshold magnitude for GLM observations}

Based on the comparison of GLM events observed with simultaneous ground observations \cite{Jenniskens2018} estimated the threshold magnitude for slow fireballs observed by GLM detectors to be $-14$. Since we observed that radiation in the oxygen region is $\approx 1/1000$ of overall radiation in the visible spectrum and $> 1/100$ for meteors faster than $\approx 50$ km/s, the threshold brightness for fast meteors observed by GLM has to be lower. Using the relation in Eq. \ref{eqnMAG}, we can estimate the threshold brightness for fast fireballs. If according to \cite{Jenniskens2018} some threshold radiation in oxygen region $I_{777}$ corresponds to a magnitude of $m_v = -14$ for slow meteors, we can use Eq. \ref{eqnMAG} and as a given slow velocity we can use for example $v = 15$ km/s. The same threshold value for radiation in the oxygen region, the term $-2.5 \times log_{10}(I_{777})$, will correspond to a different threshold magnitude if we use different fireball velocities. For example, for velocity of $70$ km/s, the threshold magnitude, $mv$, is $\approx -8.8$. Although only three meteors in our sample are brighter than $-14$ magnitude, if we count the velocity dependence of oxygen radiation, then $11$ fireballs from our sample were above the threshold sensitivity of the GLM detectors. Our GLM data calibration method is independent on this threshold value. When using only one threshold value for any given meteor velocity, inaccuracy can arise. For example, the threshold value of $-14$ mag used by \cite{Brown2019} for the absolute calibration of the Hamburg fireball GLM observation corresponds to the fireball's velocity of $15.83$ km/s in this case - but if the same threshold value was used for faster meteors, it would overestimate the derived magnitudes.

\subsection{Accuracy of the GLM observations}

\begin{figure}[htb]
\centering
{ \includegraphics[width=\hsize]{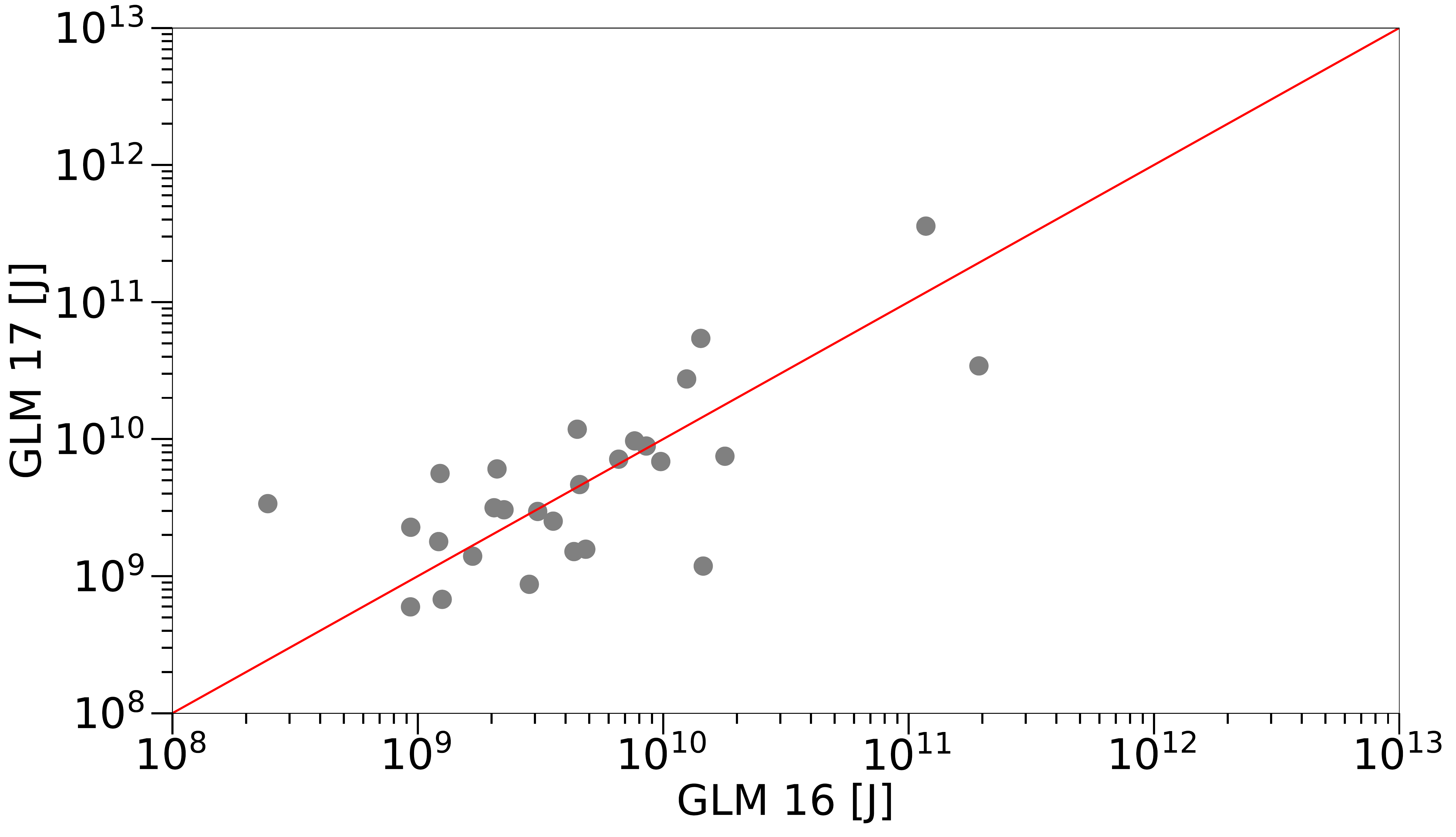}}
\caption{Comparison of radiation for stereo GLM meteors observed by both GLM 16 and GLM 17 detectors for an assumed velocity $15$ km/s.}
\label{Fig:stereo}
\end{figure}

When comparing the calibrated radiation observed by GLM with observations reported by USG in Figure \ref{Fig:USGa}, we can notice that fireballs observed simultaneously by two GOES satellites can differ in calibrated radiation up to one order of magnitude. These stereo bolide detections are connected by a green line in Figure \ref{Fig:USGa}. Since there were only three stereo cases we compared a larger number of stereo GLM observations in Figure \ref{Fig:stereo}. The problem, namely, that for most of these meteors the velocity is unknown, was solved using an assumed velocity of $15$ km/s for all meteors since we are not interested in absolute values of radiation, but we are comparing relative radiation of the same fireball observed by two detectors. Thus, the absolute values in the Figure \ref{Fig:stereo} are artificial, but the ratio between calibrated radiation from both detectors is velocity-independent. We can see that for some cases, the difference was more than one order and the dispersion is in an agreement with the one observed in Figure \ref{Fig:USGa}. This is another uncertainty in the GLM data that has to be taken into account. The source of this discrepancy is not in our calibration, but it is in the reported data and thus it is unknown to the authors. Considering this uncertainty, the difference between the radiation estimated with the velocity calibration and the radiation estimated only by the blackbody radiation at $777$ nm is well within this uncertainty.

\subsection{Sodium and magnesium in cm-sized meteoroids}

As an additional part of the study, the radiation of sodium and magnesium in cm-sized meteors were studied.
As expected, the relative sodium contribution to the spectrum was greater in slower meteors than in faster meteors (Figure \ref{Fig:Energy_Na}). Slow meteors with colder plasma are more favorable for the radiation of sodium with low-excitation potential. Moreover, meteors slower than $\approx 30$ km/s showed a large scatter in Na radiation in cases with a weak relative intensity of Na. This can be explained by variations in the actual amount of sodium in each meteoroid. The variation of sodium abundance in meteoroids is well known \citep{borovicka2005, Vojacek2015, Vojacek2019, Matlovic2019} and can be explained by either close approaches to the Sun or by space weathering.

For sodium, the three brightness fits implied that for bright meteors the slope of the relative brightness of Na to the velocity was smaller than for fainter meteors. This suggests that the high-temperature component is not as dominant in fast bright meteors as it is in weaker meteors of the same velocity. Moreover, the sodium line can be optically thick in bright slow meteors, and therefore not as dominant as would correspond to its actual abundance.


Due to the relatively high excitation potential of magnesium Mg I -- 2, the relative brightness of magnesium in (Figure \ref{Fig:Energy_Mg}) increased with increasing meteor velocity, but for meteors faster than $\approx 25$ km/s, the high-temperature spectra components started to contribute to the spectrum, thus lowering the magnesium relative contribution. Moreover, above $25$ km/s, the temperature probably does not increase anymore, as evidenced by the nearly constant Na/Mg ratio, shown in Figure \ref{Fig:NaMg}.

Similarly, as for sodium, we observed different slopes of three brightness bins for magnesium. The decrease is slowest for the brightest meteors. We offer the same explanation as for the case of sodium: the reduced dominance of the high-temperature component in bright fireballs is due to their higher optical thickness.

The ratio of Na/Mg radiation is in an agreement with the work of \cite{borovicka2005} and \cite{Vojacek2019} for millimeter-sized meteoroids. It increases for slow meteors, with a ratio is more or less constant for velocities faster than $25$ km/s. We did not observe a shift in absolute values of ratio $I_{589} { / } I_{517}$ for larger meteoroids, as reported in \cite{Matlovic2019}. These authors explained the shift by the presence of larger bodies in their sample and thus the reduced level of space weathering with regard to the volatile sodium from the meteoroid body. The sample of \cite{Matlovic2019} overlapped with our work in terms of absolute brightness of meteors ($-1$ to $-14$ mag, compared to the range from $-8$ to $-15$ mag in our work) and, thus, also sizes of meteoroids. Therefore, we cannot confirm this absolute Na/Mg ratio for this meteoroid size.

\section{Conclusions}
We studied the spectra of bright meteors with a focus on the oxygen triplet O I -- 1 at $777$ nm.
The intensity of the oxygen line steeply increases with meteor velocity. The line is invisible in slow meteors, but it is one of the brightest lines in the spectra of fast meteors. The radiant intensity in the narrow $1.1$ nm spectral window around $777$ nm used by the GOES satellites amounts to only $1/1000$ of the radiant intensity of the $380$ - $850$ nm window in meteors of $\approx 11$ km.s$^{-1}$. The radiation is mostly due to a continuum. However, this share increases to $1/30$ at $70$ km.s$^{-1}$, when the oxygen line dominates. As a consequence, the GLM limiting magnitude depends on meteor velocity. For slow meteors, it is $\approx -14$ mag, but according to our spectral observations of photographic fireballs, the dependence of relative radiation at $777$ nm on the fireball velocity suggests that meteors as faint as $-9$ mag with speeds up to $70$ km/s can be observed by the GLM.

We investigated the discrepancy between GLM--16 and GLM--17 on more stereo observations and we found that they can differ by about one order of magnitude when compared.
We also investigated the discrepancy between GLM–16 and GLM–17 on multiple stereo observations and found that they can vary by about one order of magnitude.

We have provided an empirical formula for converting the energy observed in the $777$ nm band into a meteor V-band magnitude. The formula can be used if the meteor velocity is known. Our data also suggest that the oxygen line intensity depends on the meteor altitude as well as on whether the observation is carried out during a meteor flare. The brightness of the fireball also influences the relative radiation in the oxygen region. Second-order refinements are therefore provided to the conversion formula. The altitude and flare refinement reflect the fact that for a given velocity, the oxygen line is more important when the ablation rate is low, namely, outside flares and at higher altitudes. The brightness correction reflects the fact that the dependence of the $777$ nm band intensity on meteor velocity is less steep for very bright meteors, where the radiation is optically thicker. The typical altitude for a given velocity used in the altitude correction formula was derived from the sample of fireballs observed by the European Fireball Network with a magnitude range from $-8$ to $-15$ mag. This correction was determined using this range of meteor brightnesses only and it is likely that it cannot be applied to brighter events penetrating deeper in the atmosphere.

We note that as a byproduct of this work, we also studied spectral regions at $517$ nm and $589$ nm, where magnesium and sodium (respectively) dominate. We did not observe a shift in absolute values of ratio $I_{589} { / } I_{517}$ for larger meteoroids reported in \cite{Matlovic2019}.

\section*{Acknowledgment}

This work was supported by the grant 19-26232X of the Grant Agency of the Czech Republic, GA \v CR and by the Praemium Academiae of the CAS. We would like to thank E. Sansom for her valuable comments in the review that helped to significantly improve the manuscript.


\bibliography{44217.bbl}

\begin{appendix}

\begin{figure*}[p]
\section{Brightness correction for total intensity}
\centering
\begin{subfigure}{0.45\textwidth}
\includegraphics[width=\hsize]{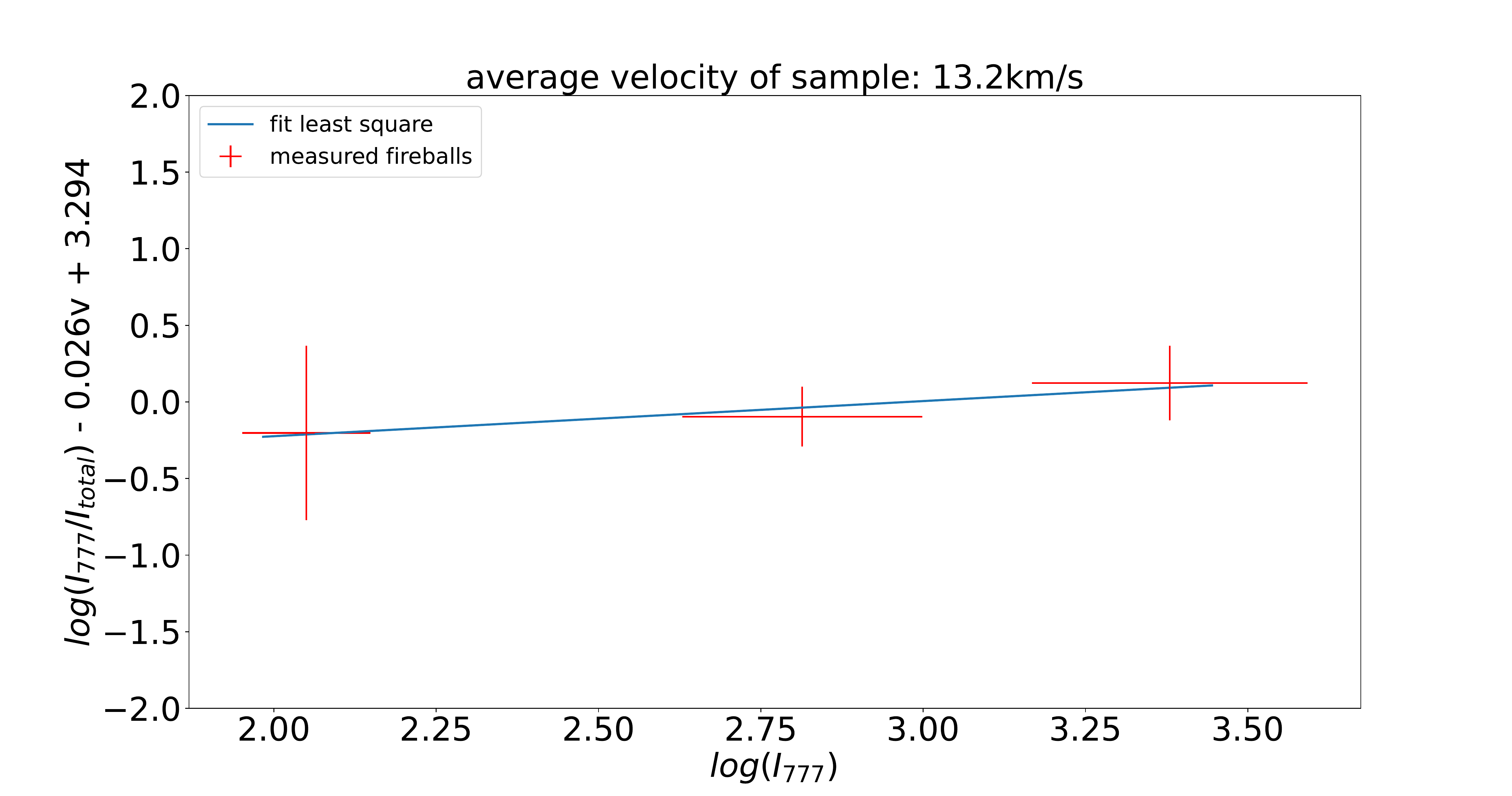}
\label{AP:Fig:13}
\end{subfigure}
\begin{subfigure}{0.45\textwidth}
\includegraphics[width=\hsize]{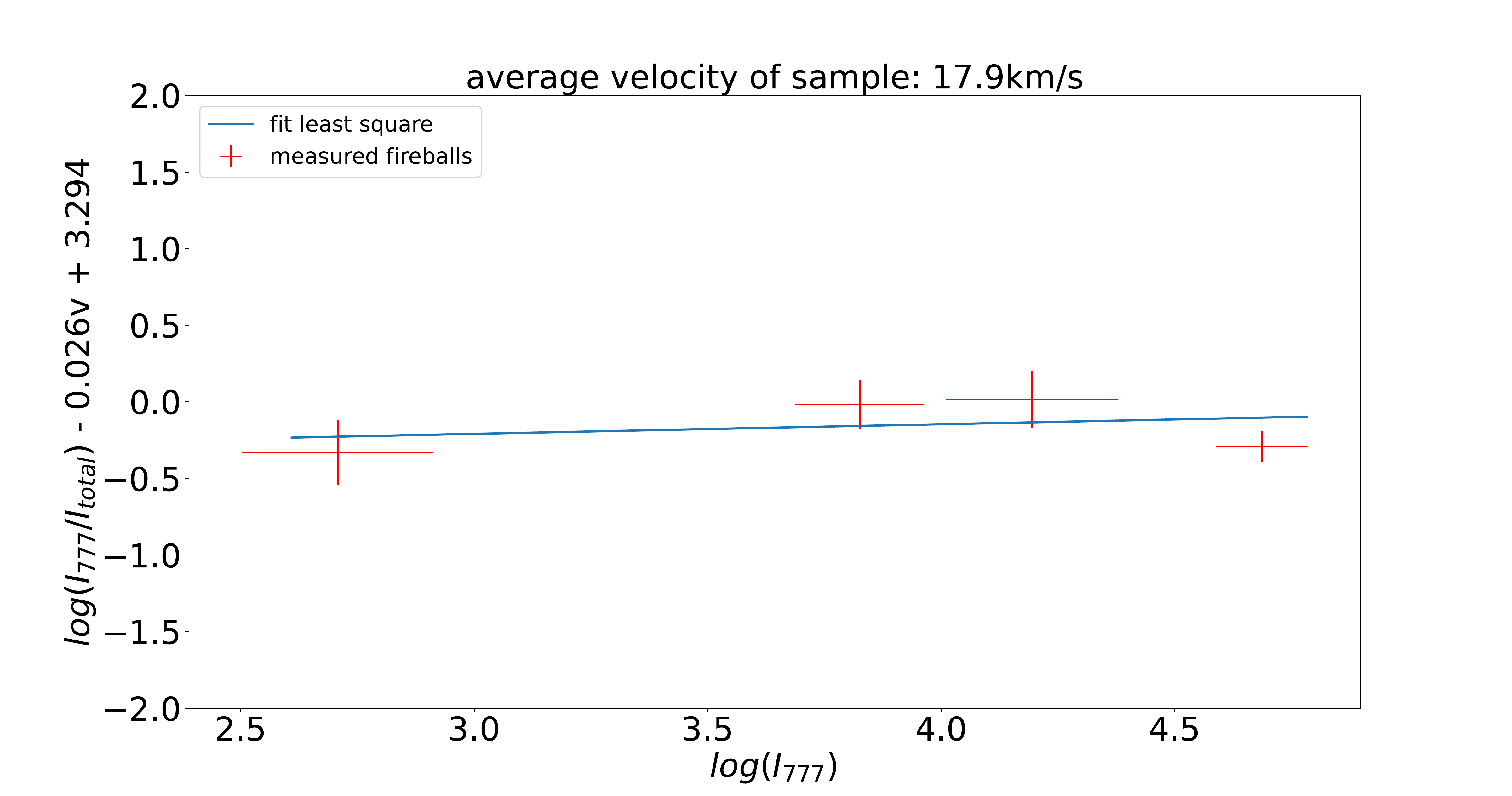}
\label{AP:Fig:18}
\end{subfigure}
\begin{subfigure}{0.45\textwidth}
\includegraphics[width=\hsize]{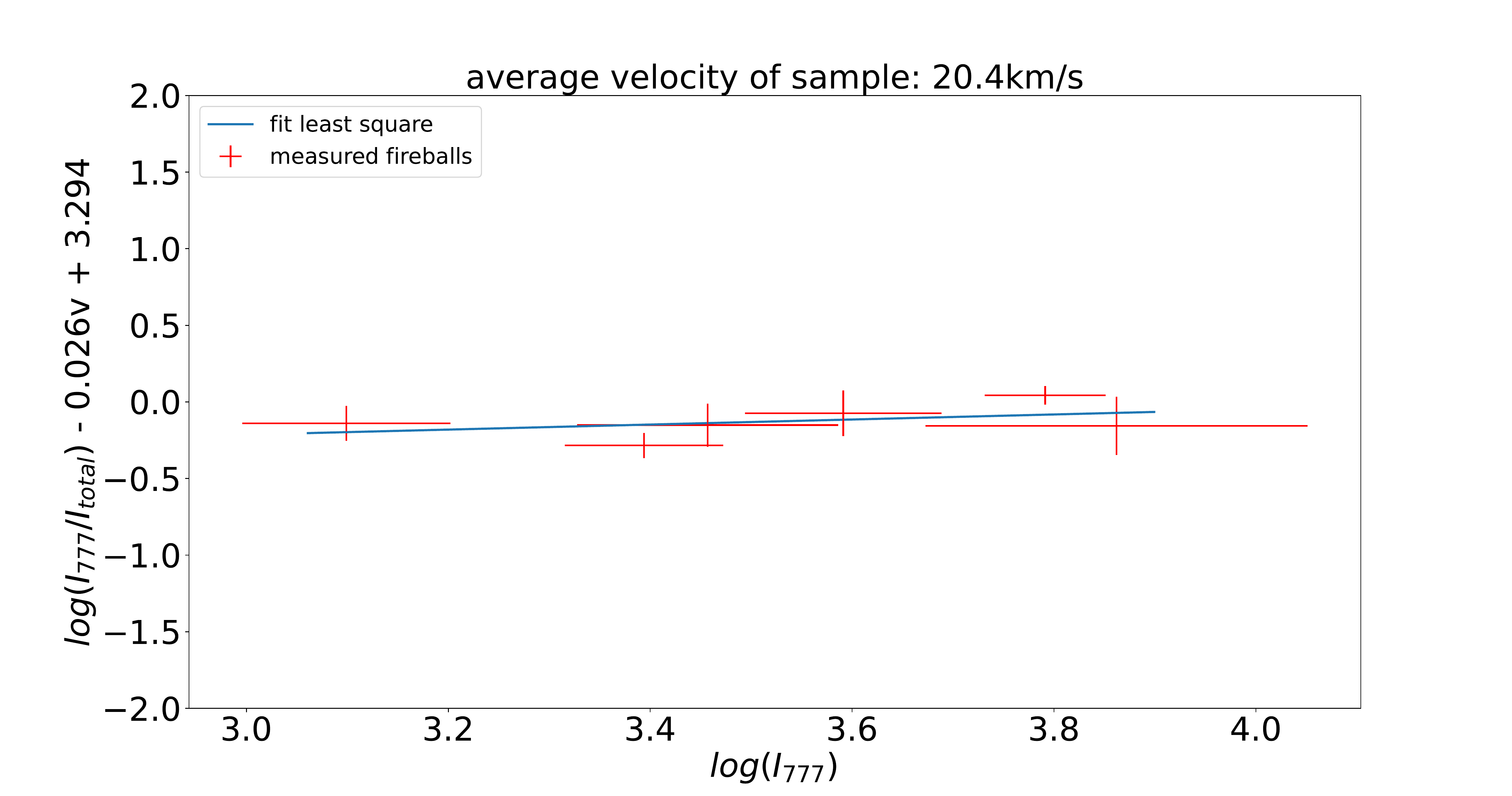}
\label{AP:Fig:20}
\end{subfigure}
\begin{subfigure}{0.45\textwidth}
\includegraphics[width=\hsize]{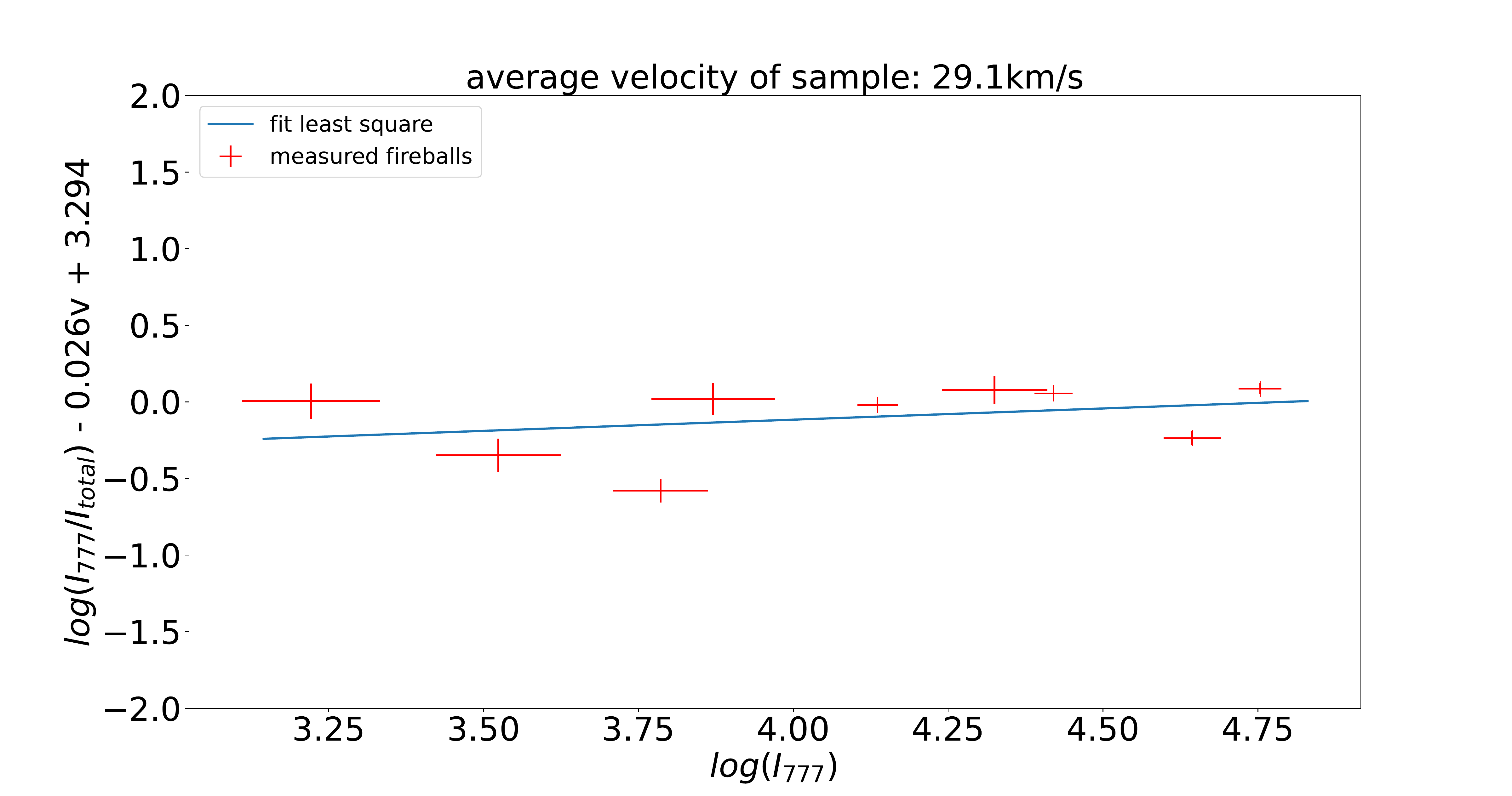}
\label{AP:Fig:30}
\end{subfigure}
\begin{subfigure}{0.45\textwidth}
\includegraphics[width=\hsize]{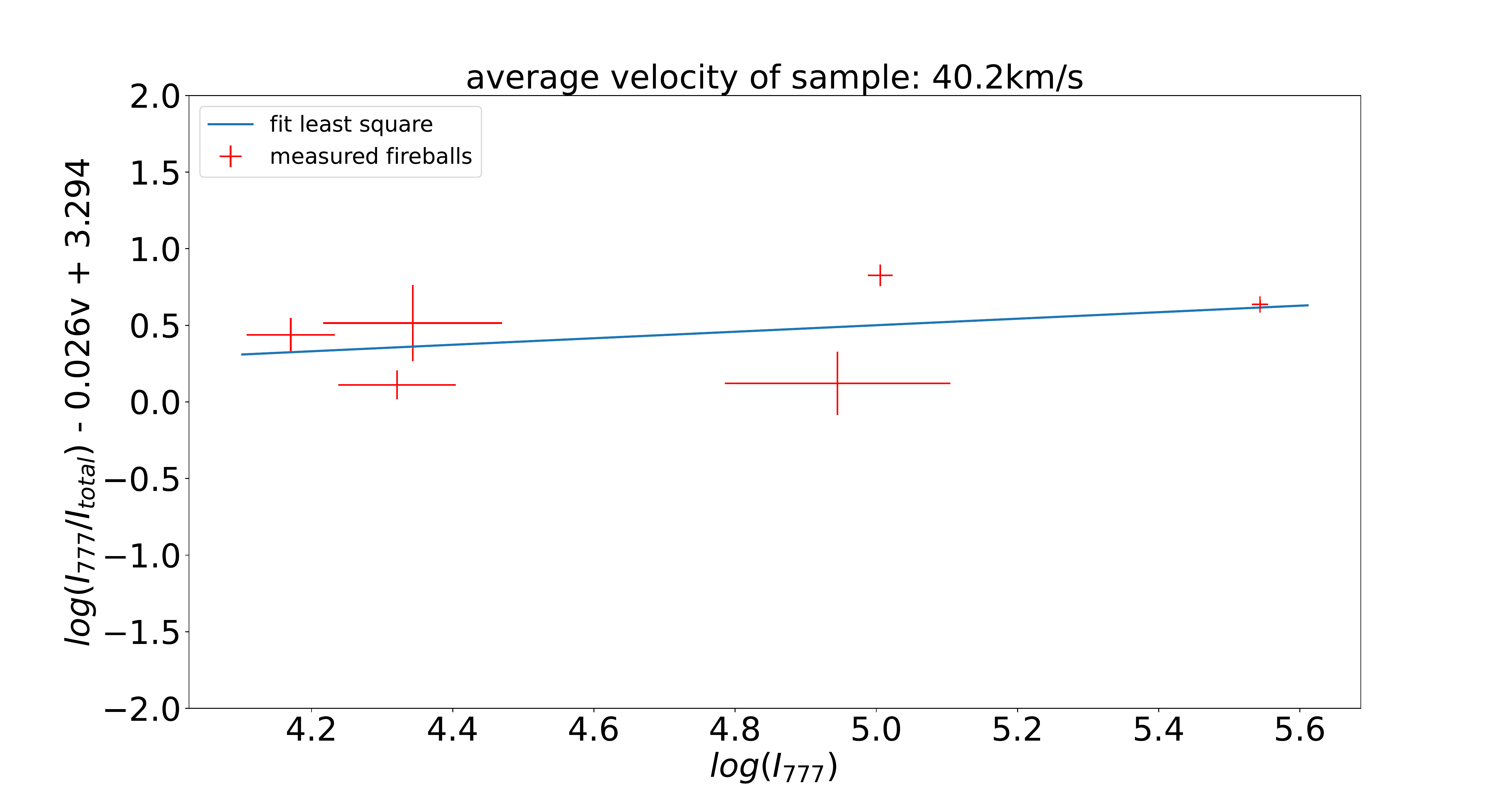}
\label{AP:Fig:40}
\end{subfigure}
\begin{subfigure}{0.45\textwidth}
\includegraphics[width=\hsize]{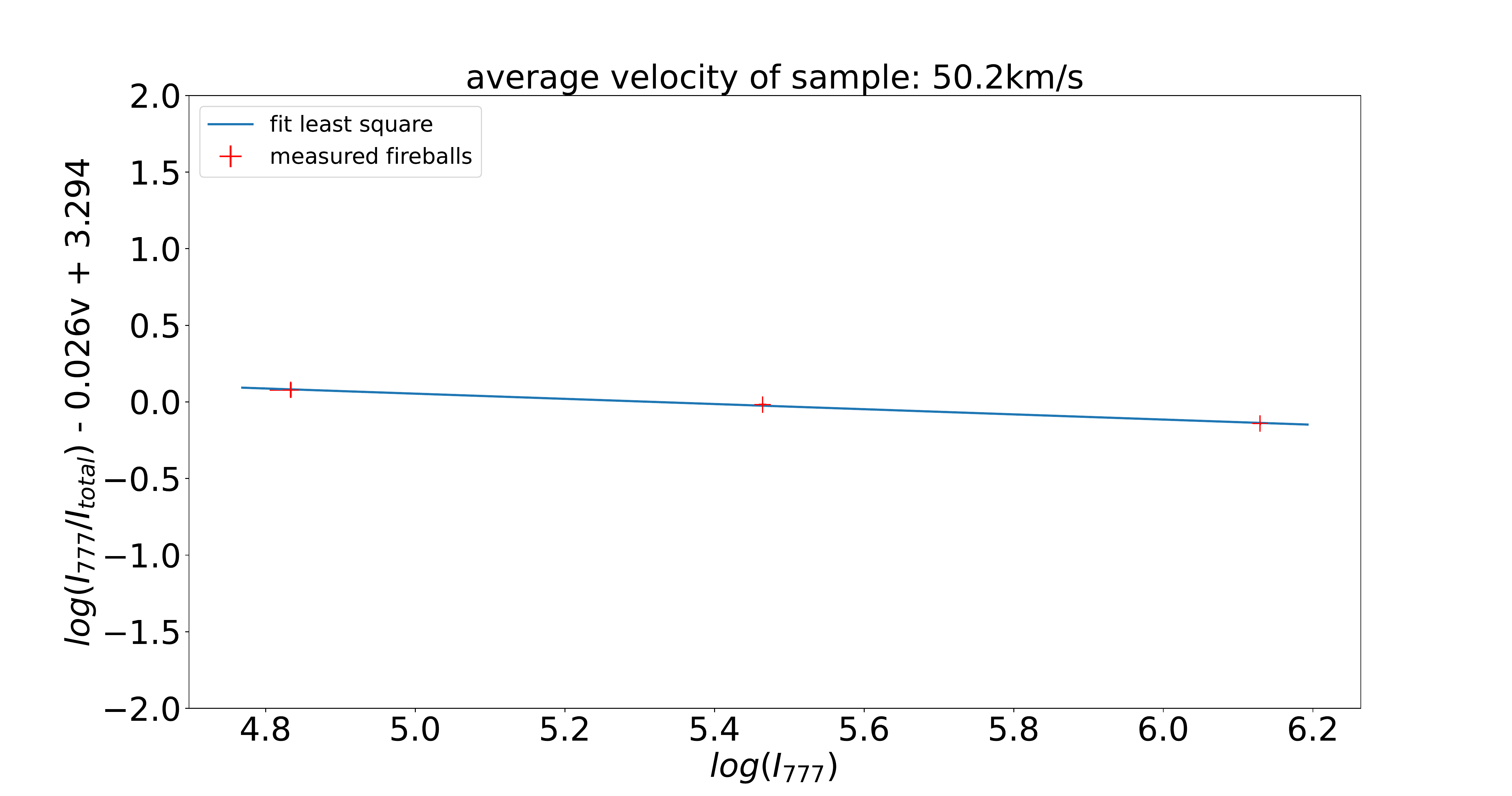}
\label{AP:Fig:50}
\end{subfigure}
\begin{subfigure}{0.45\textwidth}
\includegraphics[width=\hsize]{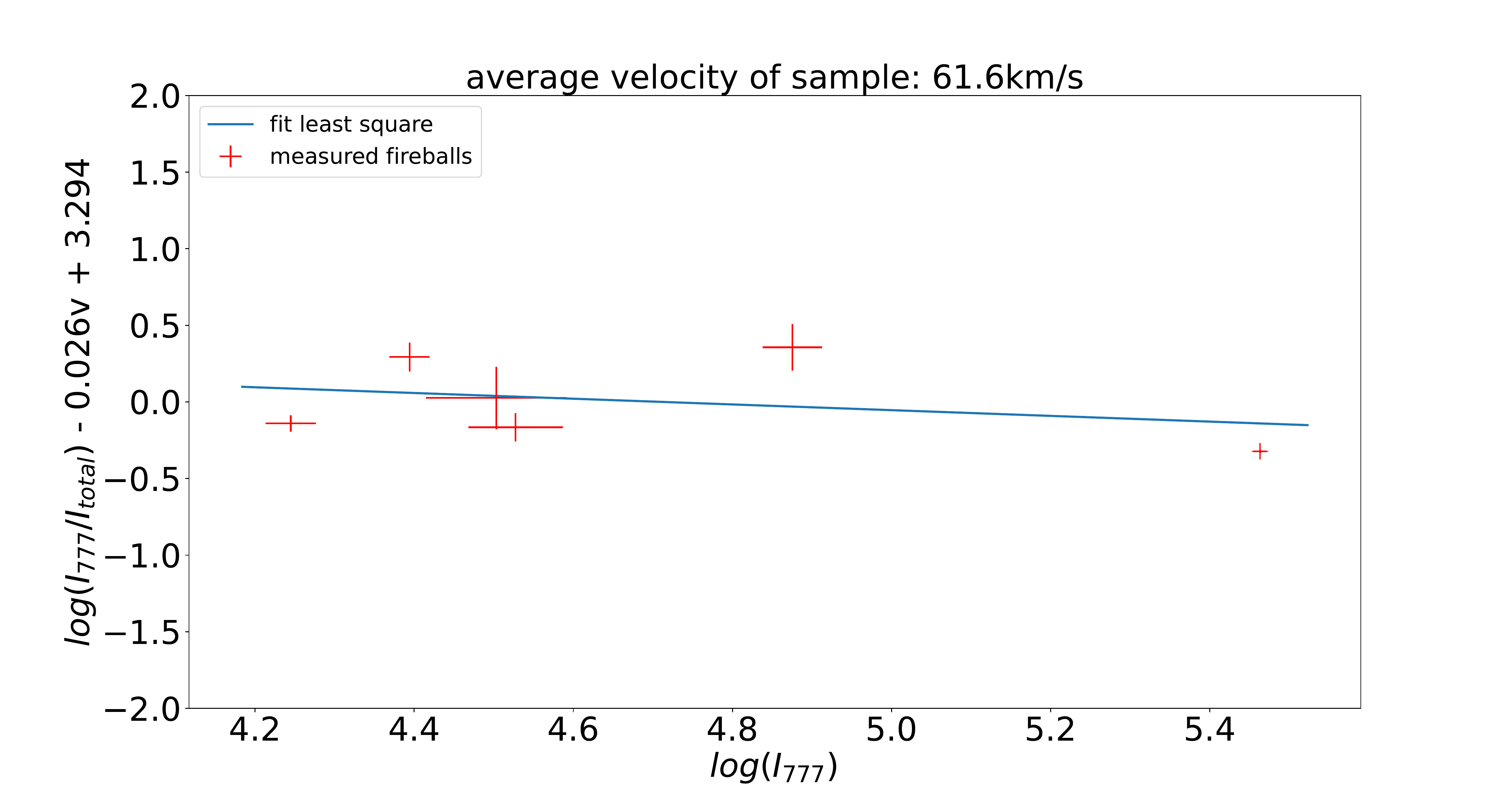}
\label{AP:Fig:60}
\end{subfigure}
\begin{subfigure}{0.45\textwidth}
\includegraphics[width=\hsize]{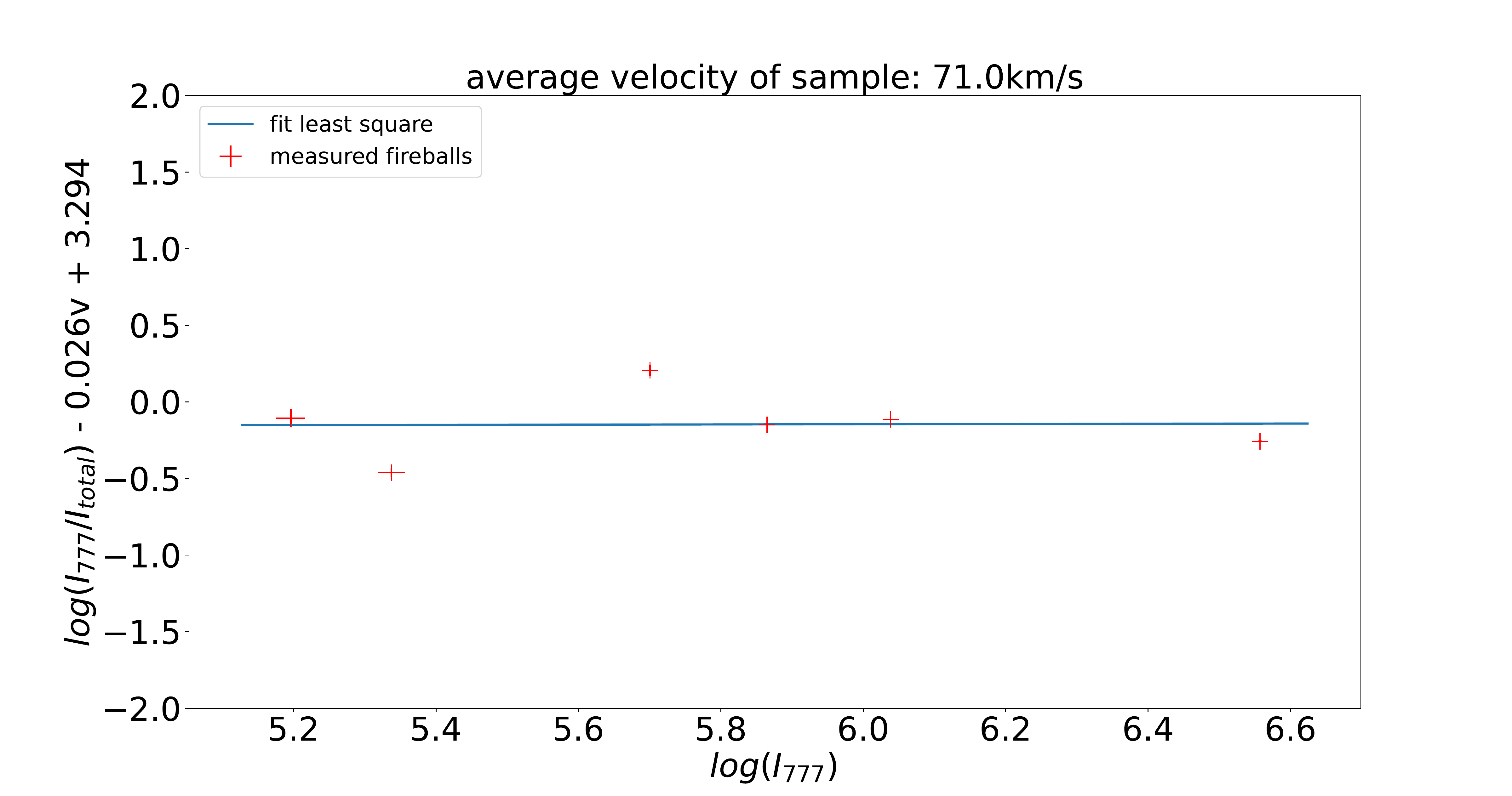}
\label{AP:Fig:70}
\end{subfigure}
\caption{Dependece of the relative radiation $log(I_{777}/I_{total})$ - 0.026v + 3.294 on the radiation in oxygen region $I_{777}$. Least-squares fits are given.}
\label{AP:Fig:Fit_I777}
\end{figure*}

\begin{figure*}[h]
\centering
\begin{subfigure}{0.45\textwidth}
\includegraphics[width=\hsize]{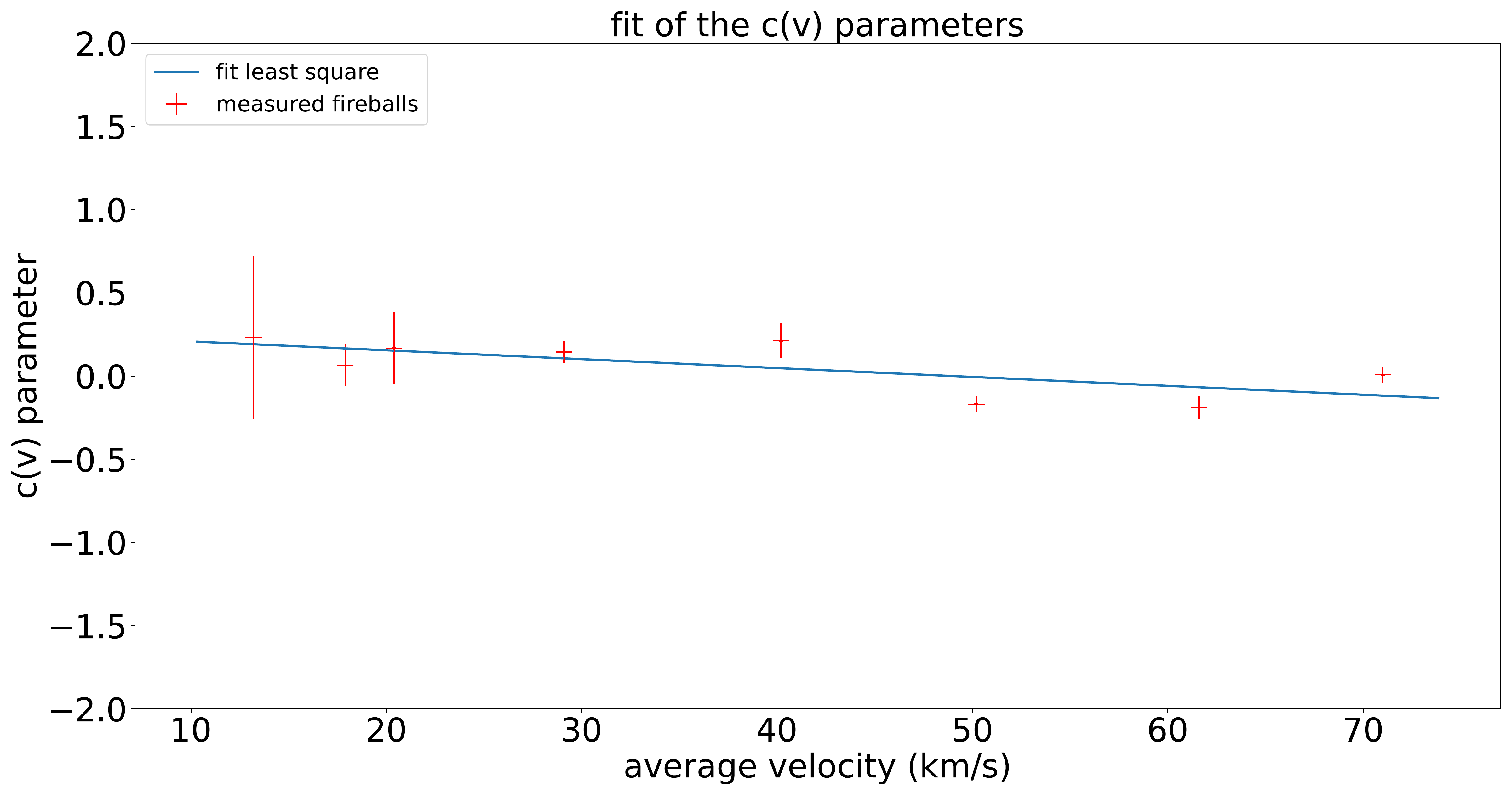}
\label{AP:Fig:ap_fit}
\end{subfigure}
\begin{subfigure}{0.45\textwidth}
\includegraphics[width=\hsize]{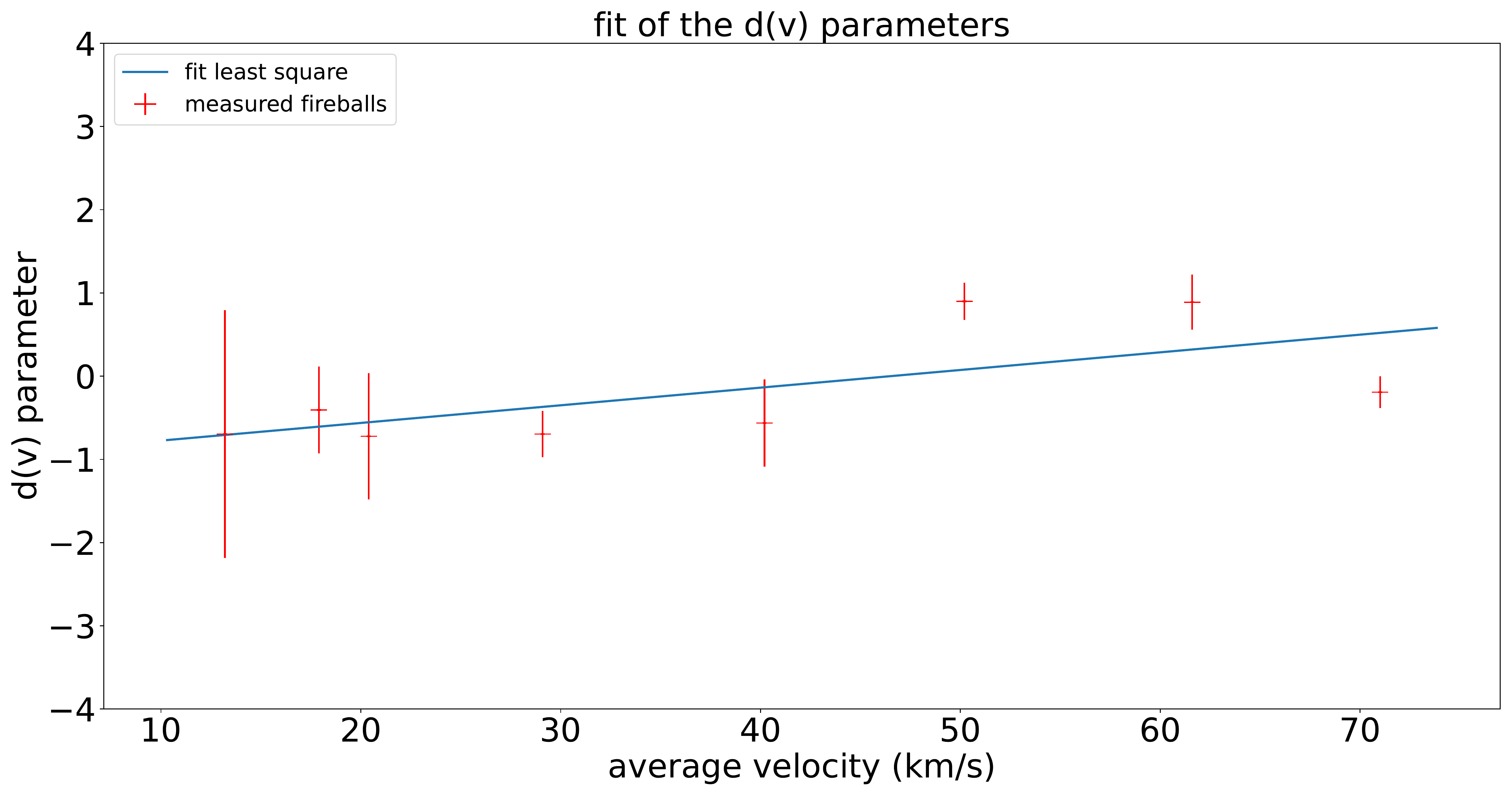}
\label{AP:Fig:bp_fit}
\end{subfigure}

\caption{Parameters of fits from Figure \ref{AP:Fig:Fit_I777} and their dependence on the meteor velocity.}
\label{AP:Fig:parameterFit_I777}
\end{figure*}

\begin{figure*}[p]
\section{Brightness correction for magnitude}
\centering
\begin{subfigure}{0.45\textwidth}
\includegraphics[width=\hsize]{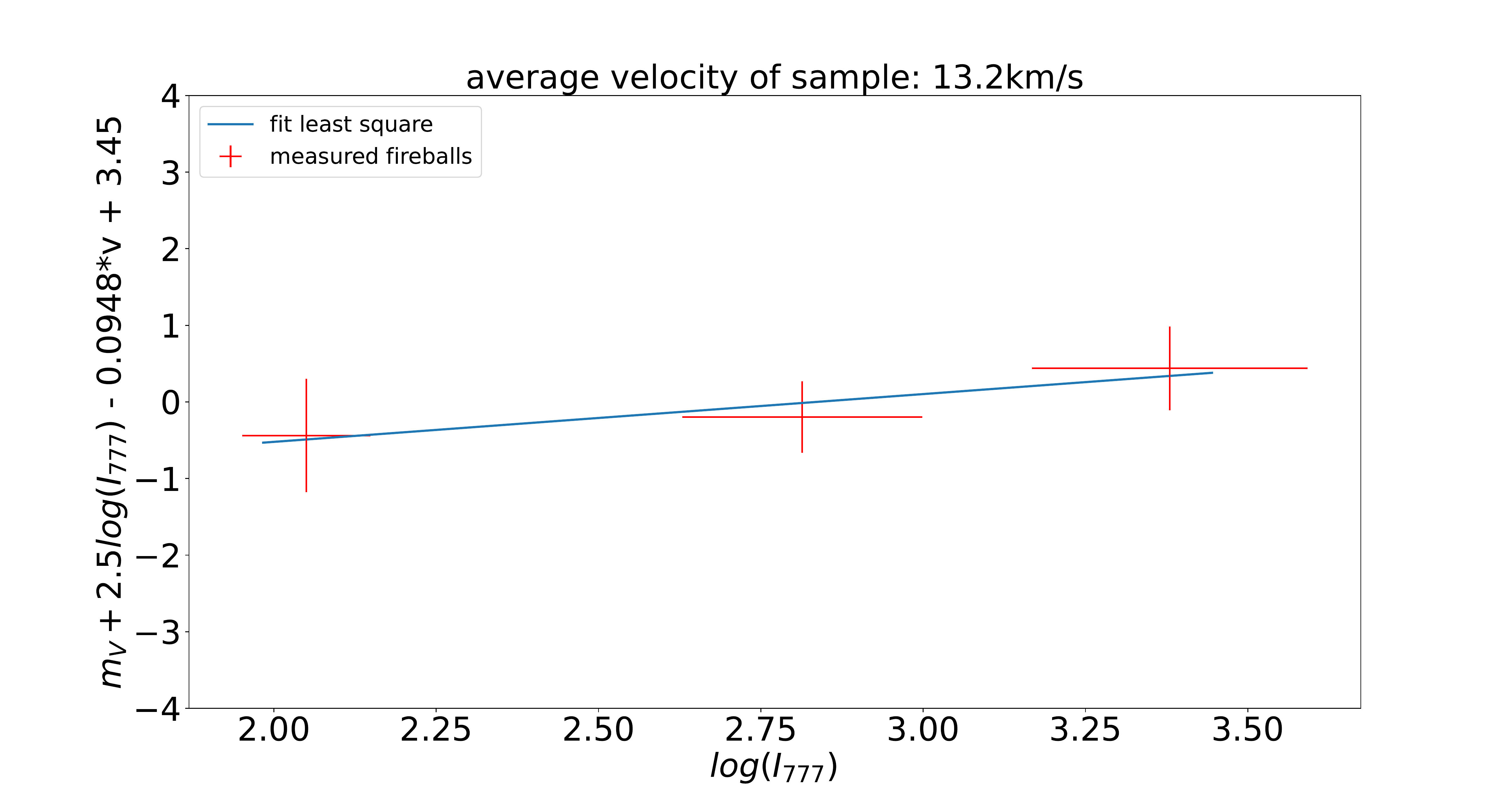}
\label{AP:Fig:13mag}
\end{subfigure}
\begin{subfigure}{0.45\textwidth}
\includegraphics[width=\hsize]{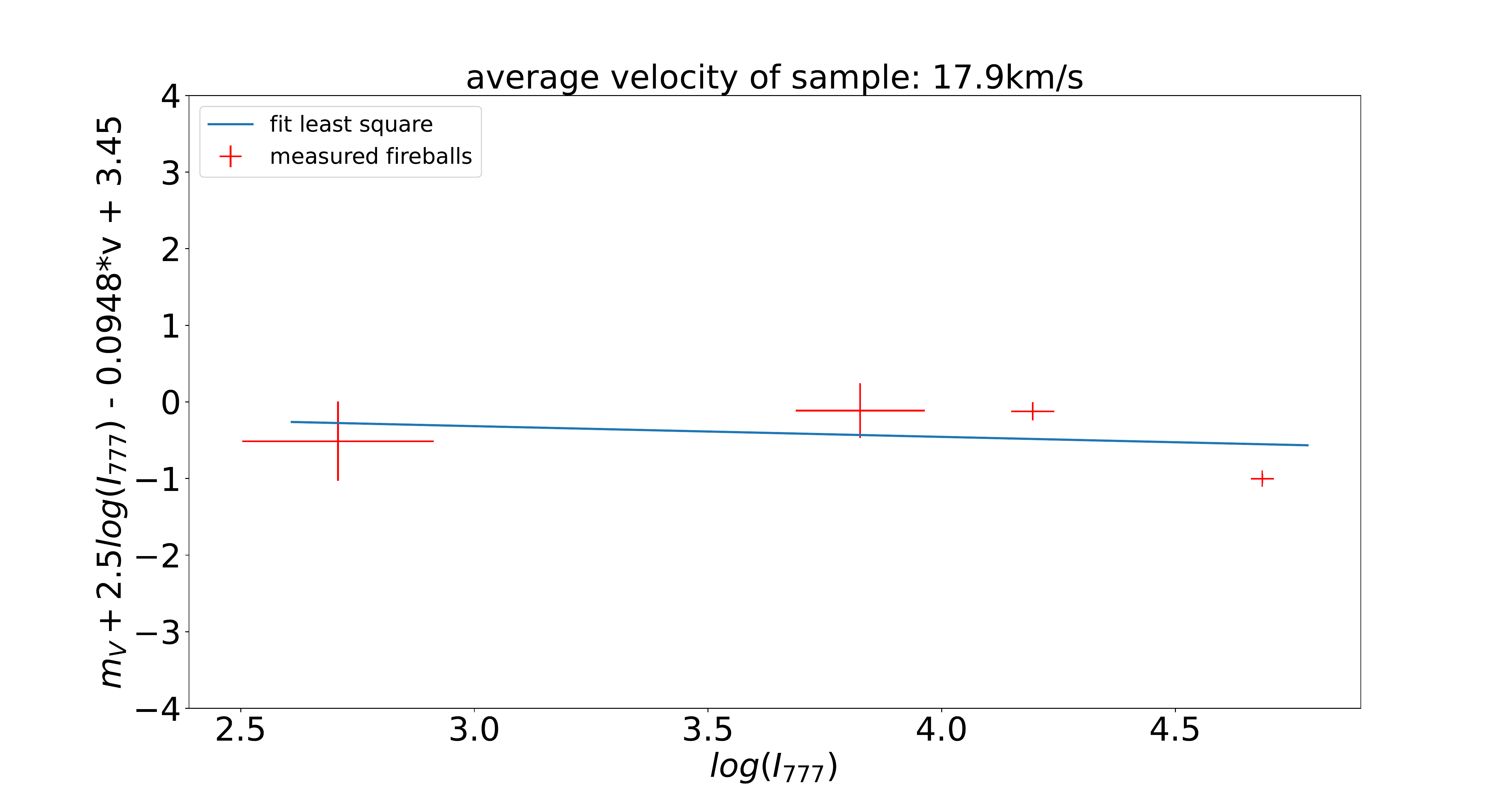}
\label{AP:Fig:18mag}
\end{subfigure}
\begin{subfigure}{0.45\textwidth}
\includegraphics[width=\hsize]{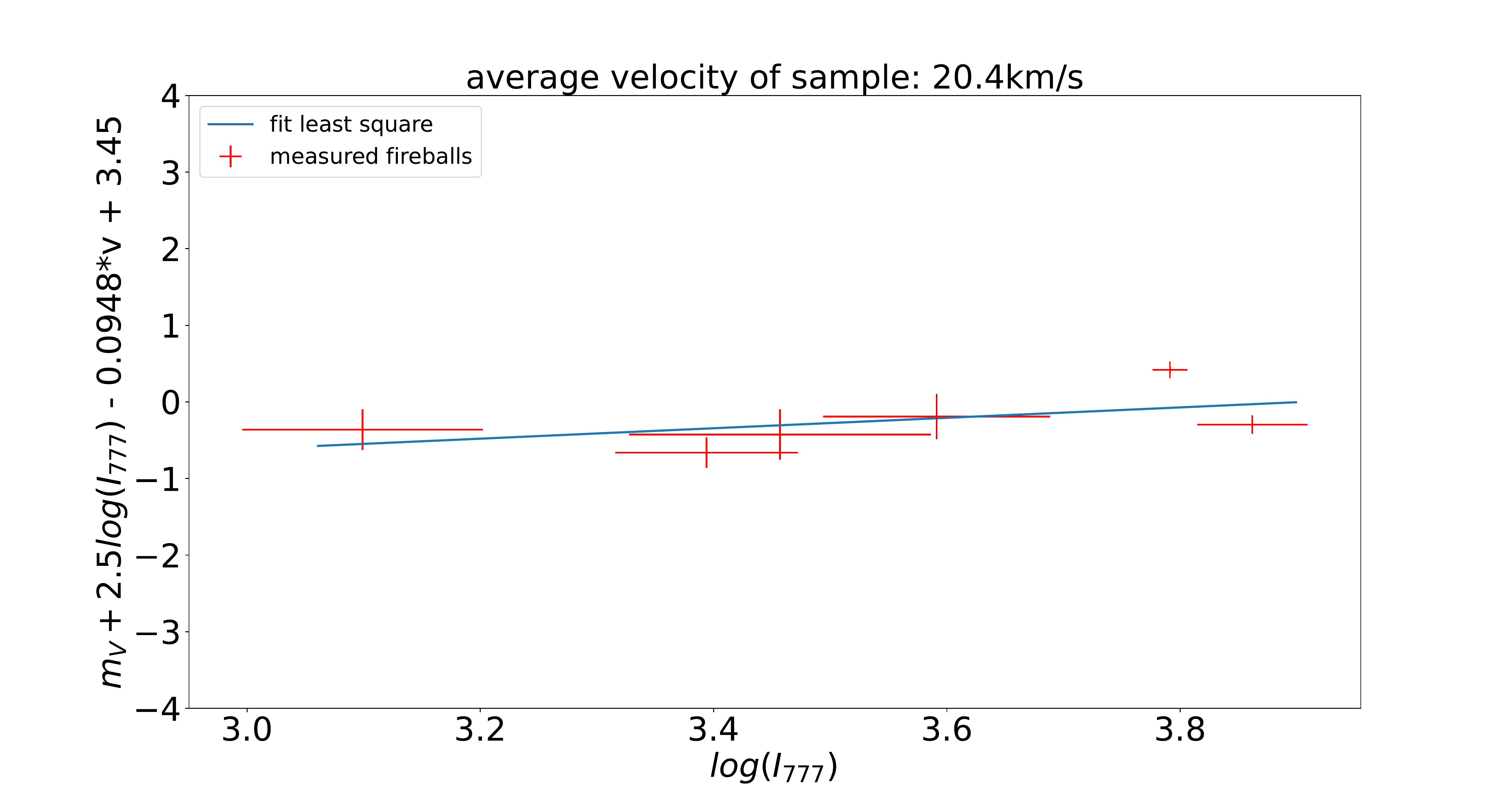}
\label{AP:Fig:20mag}
\end{subfigure}
\begin{subfigure}{0.45\textwidth}
\includegraphics[width=\hsize]{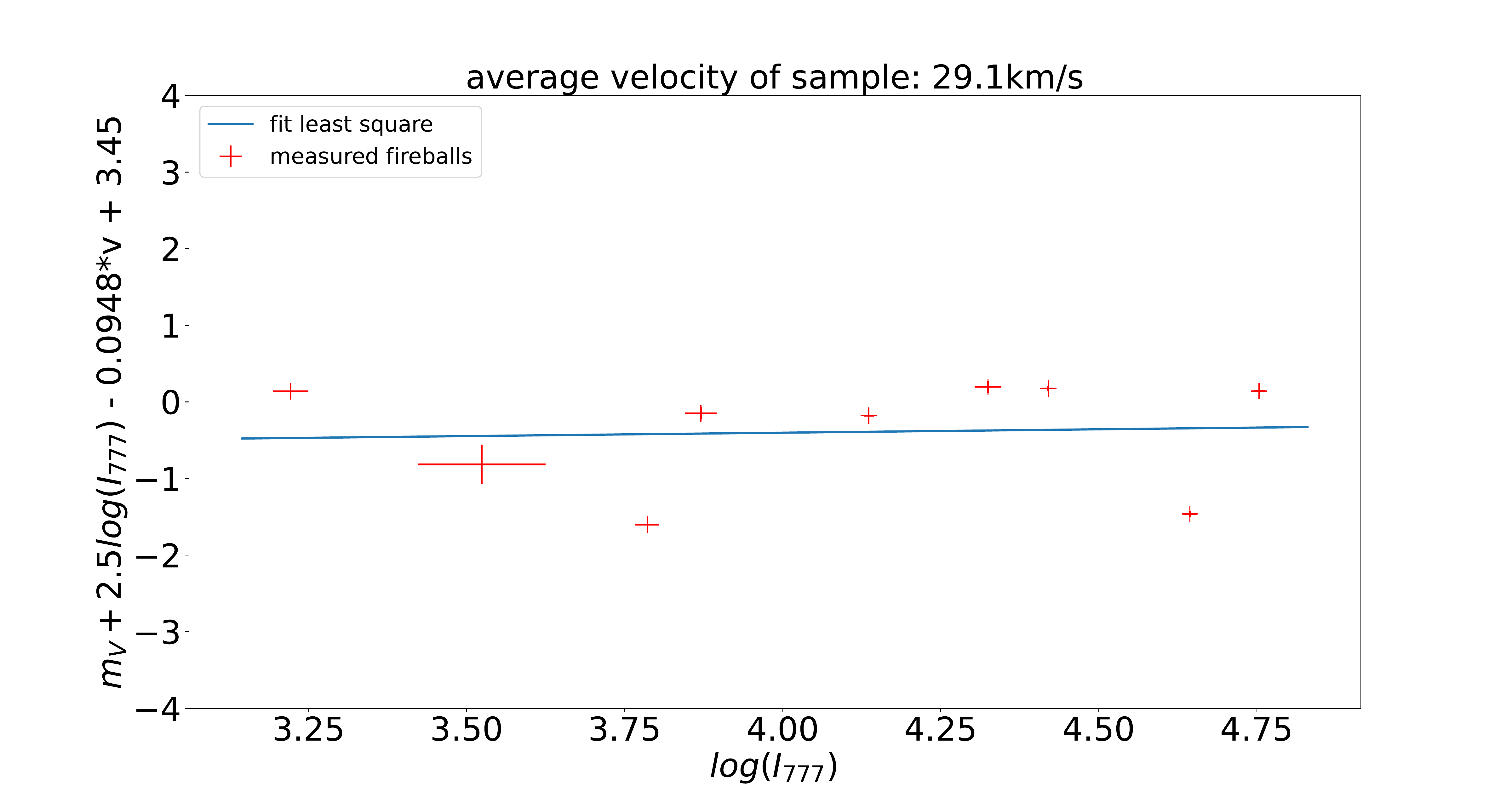}
\label{AP:Fig:30mag}
\end{subfigure}
\begin{subfigure}{0.45\textwidth}
\includegraphics[width=\hsize]{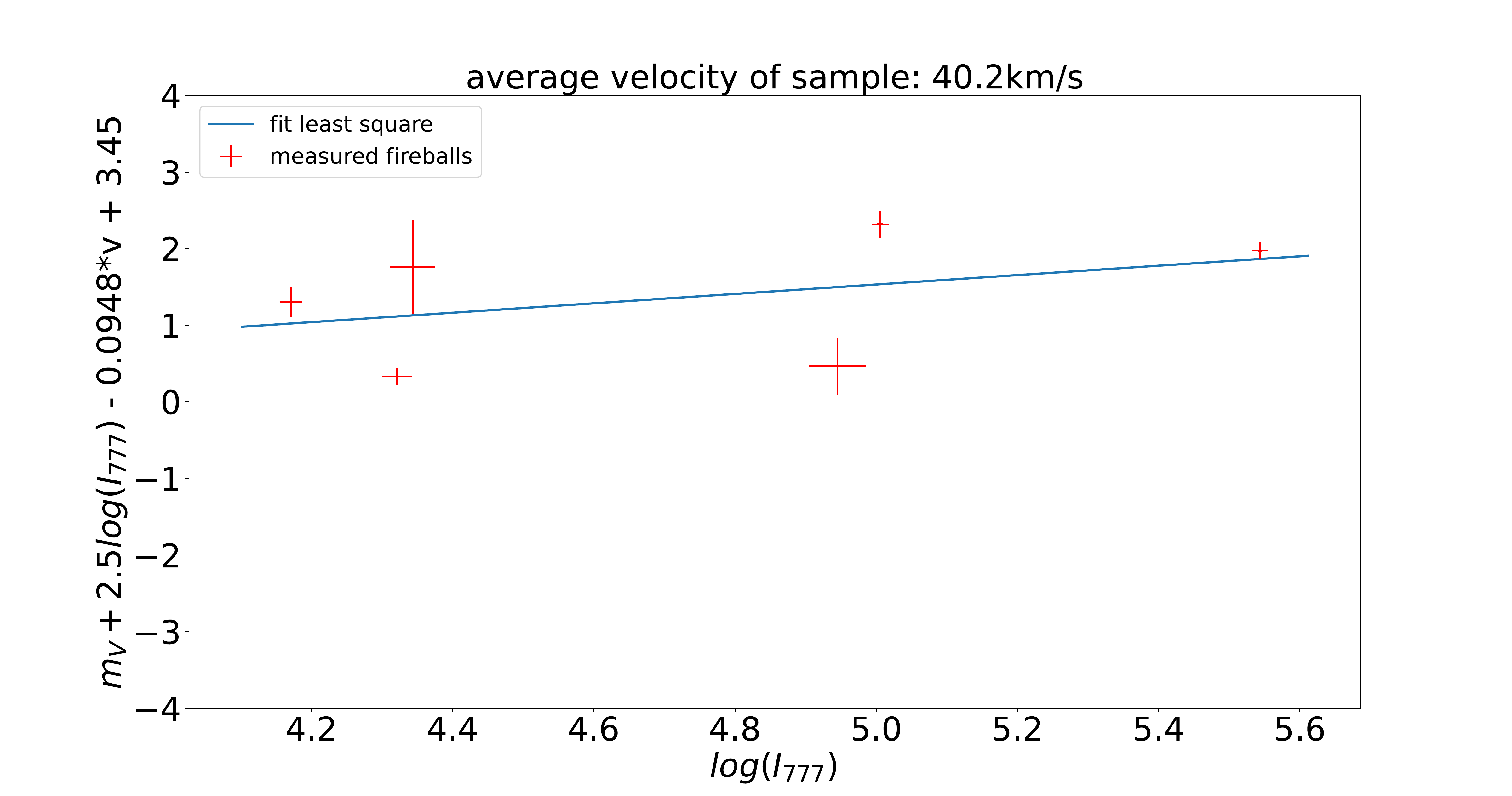}
\label{AP:Fig:40mag}
\end{subfigure}
\begin{subfigure}{0.45\textwidth}
\includegraphics[width=\hsize]{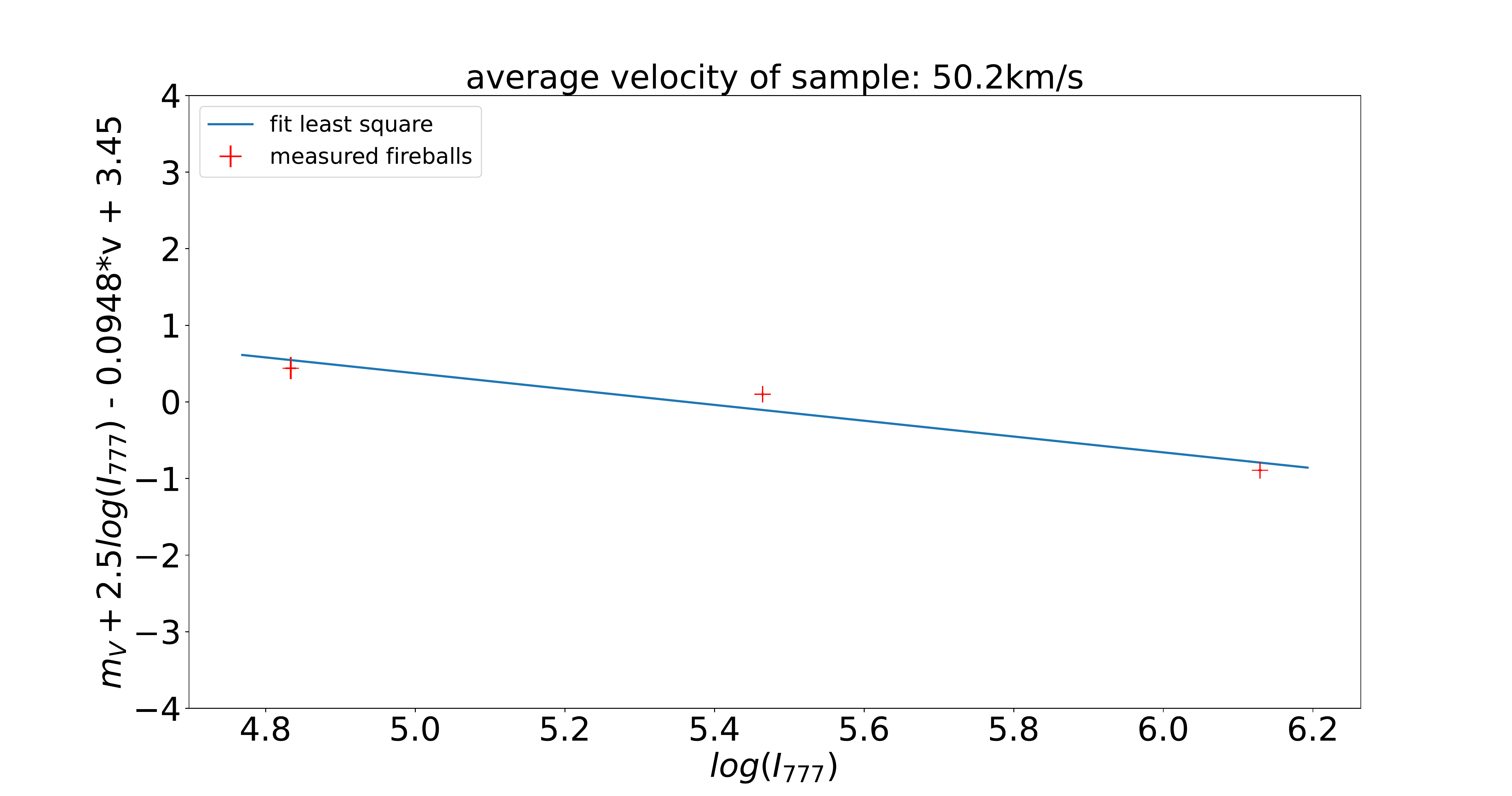}
\label{AP:Fig:50mag}
\end{subfigure}
\begin{subfigure}{0.45\textwidth}
\includegraphics[width=\hsize]{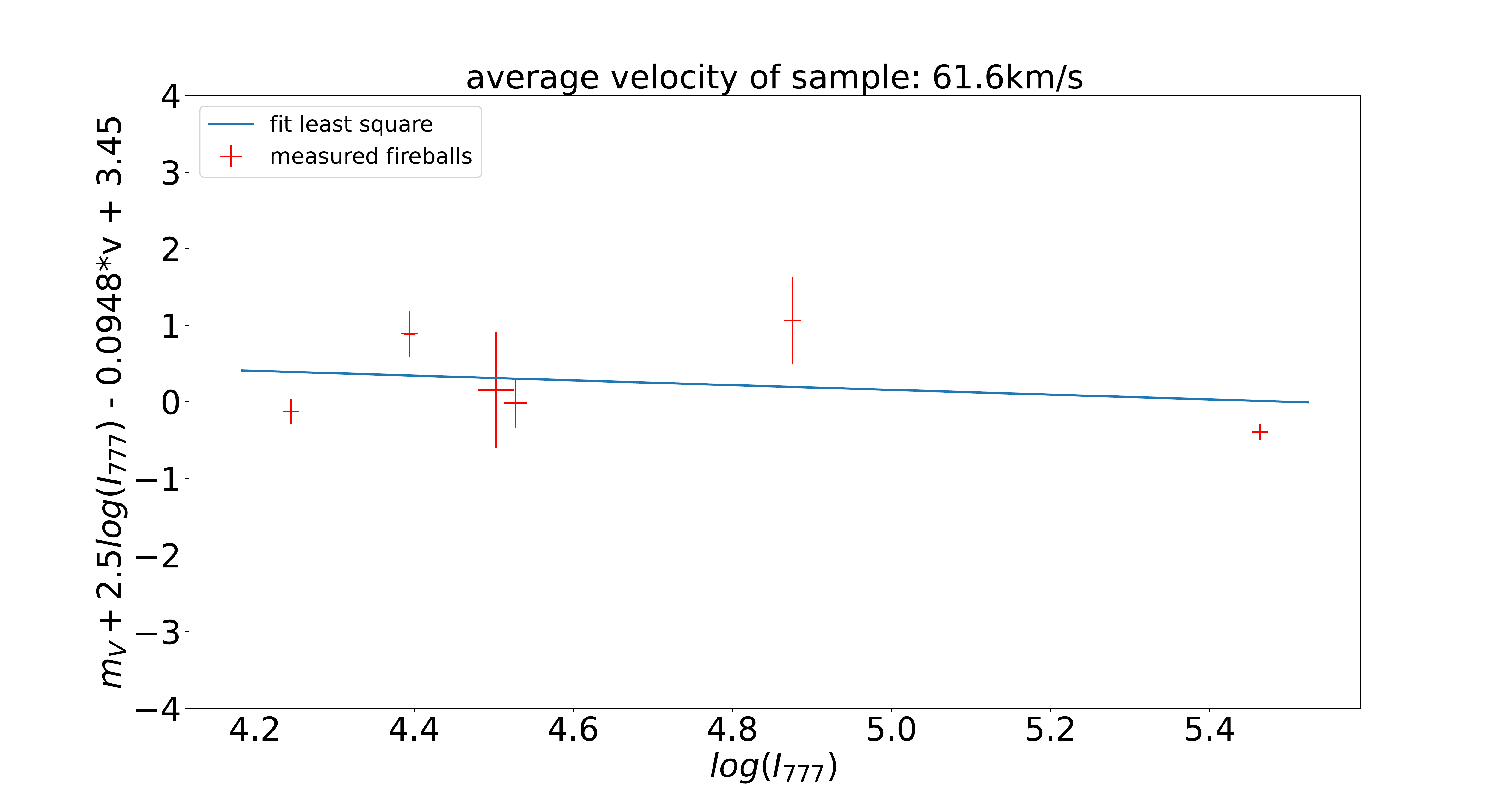}
\label{AP:Fig:60mag}
\end{subfigure}
\begin{subfigure}{0.45\textwidth}
\includegraphics[width=\hsize]{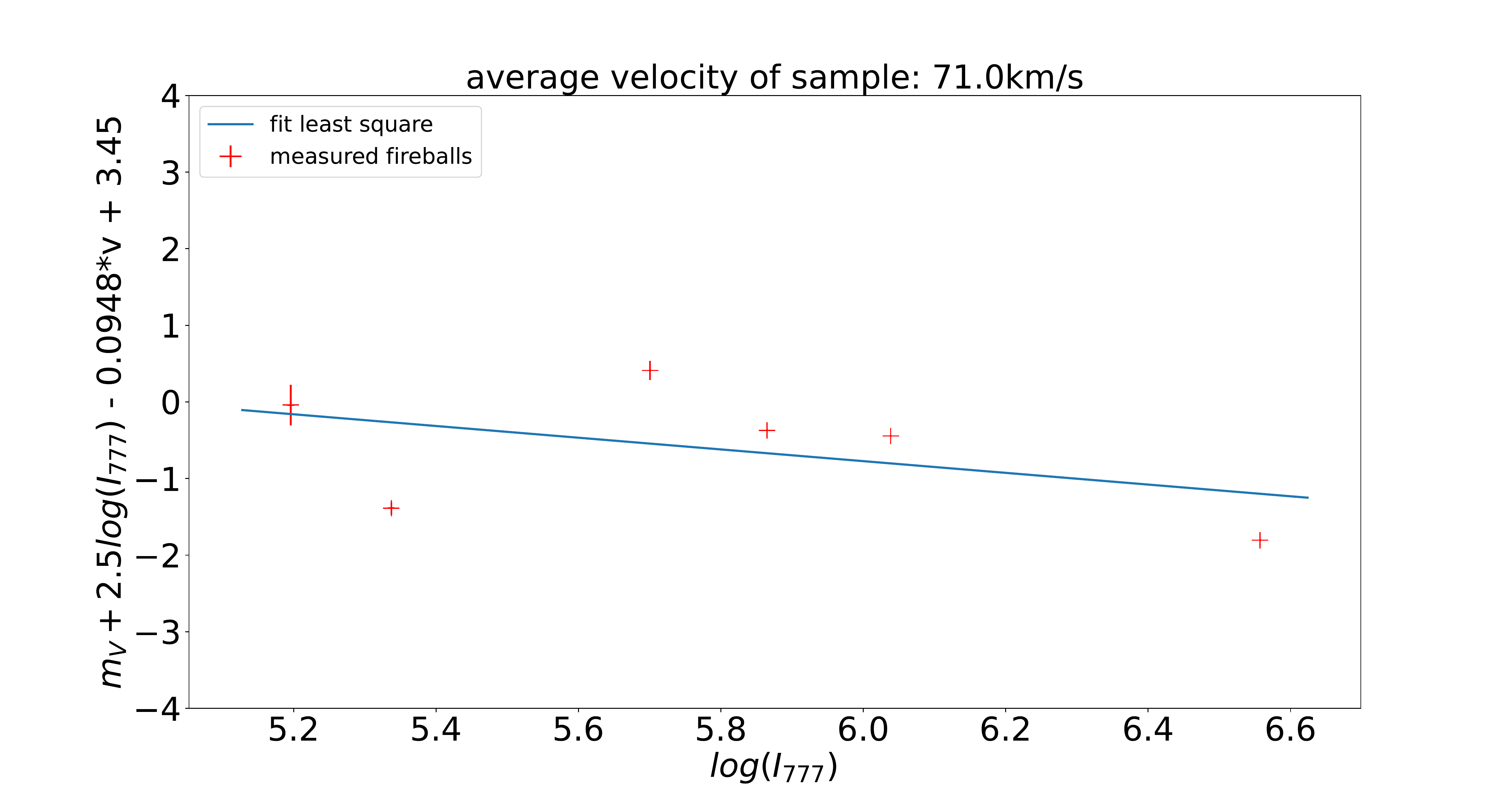}
\label{AP:Fig:70mag}
\end{subfigure}
\caption{Dependence of $m_V + 2.5 \times log(I_{777})$ - 0.0948 $\times$ v + 3.45 on the radiation in oxygen region $I_{777}$. Least-squares fits are given.}
\label{AP:Fig:Fit_mag}
\end{figure*}

\begin{figure*}[h]
\centering
\begin{subfigure}{0.45\textwidth}
\includegraphics[width=\hsize]{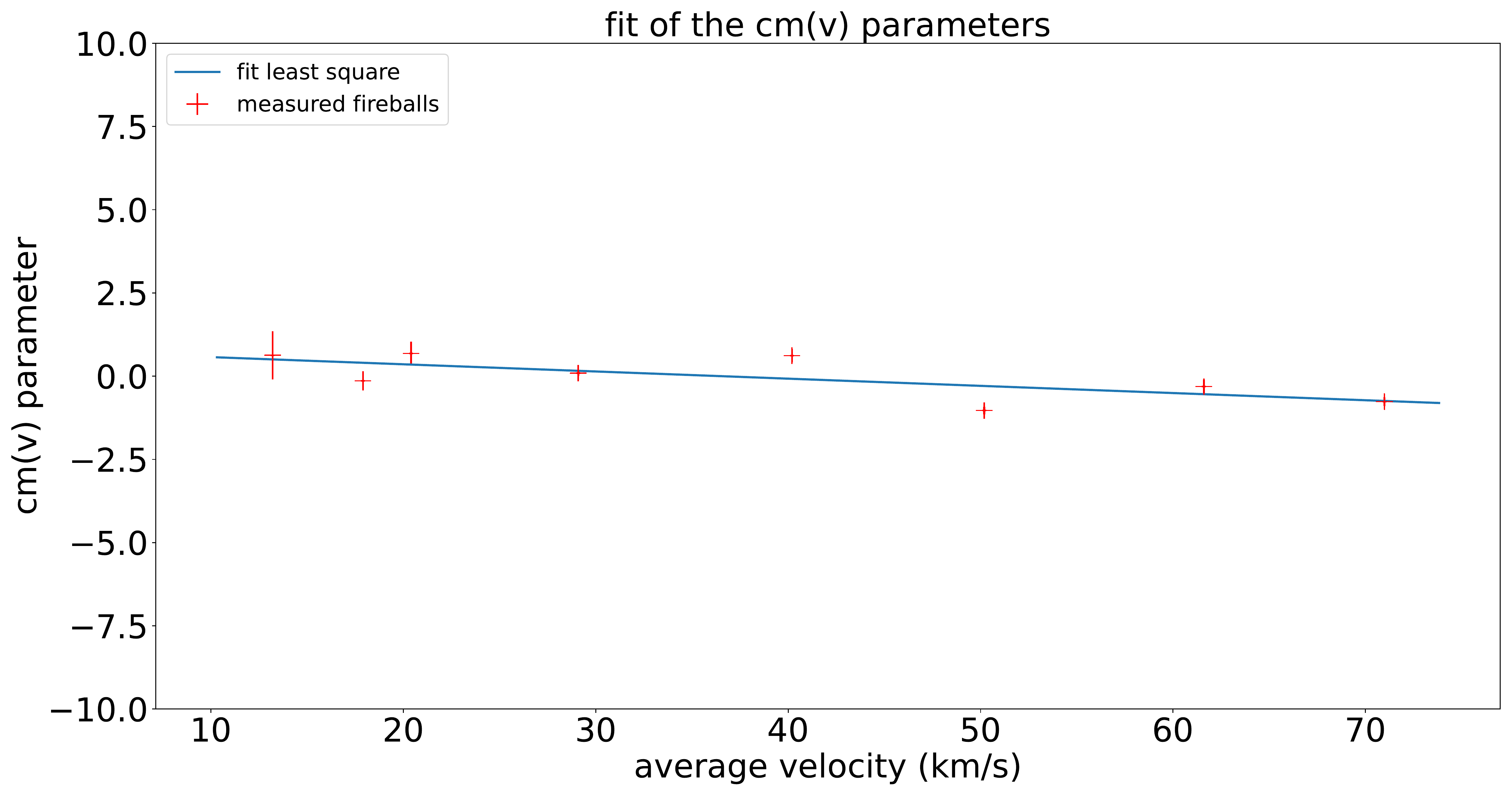}
\label{AP:Fig:ap_mag_fit}
\end{subfigure}
\begin{subfigure}{0.45\textwidth}
\includegraphics[width=\hsize]{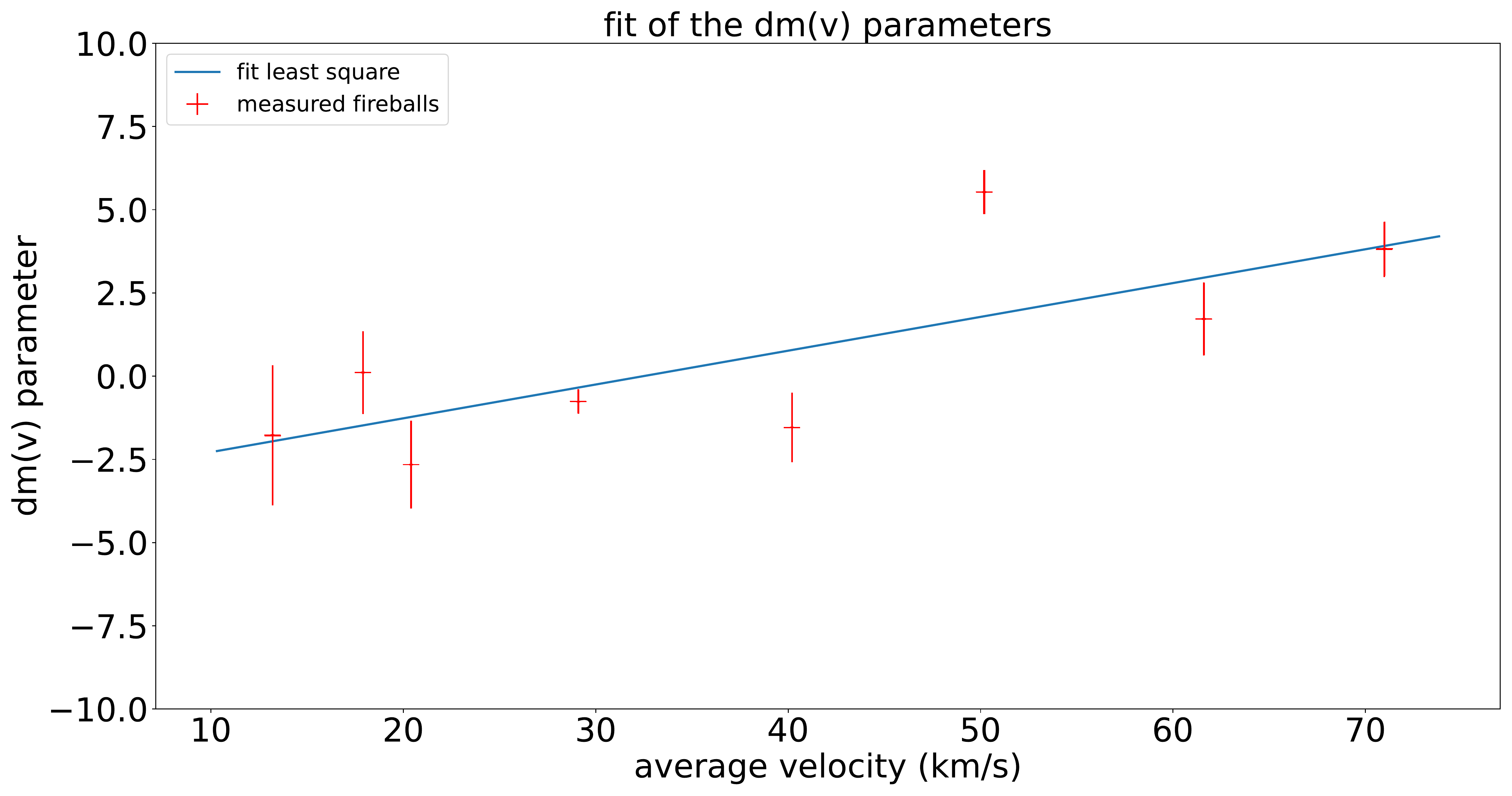}
\label{AP:Fig:bp_mag_fit}
\end{subfigure}

\caption{Parameters of the fits given in Figure \ref{AP:Fig:Fit_mag} and their dependence on the meteor velocity.}
\label{AP:Fig:parameterFit_mag}
\end{figure*}

\begin{figure*}[h]
\centering
\begin{subfigure}{0.45\textwidth}
\includegraphics[width=\hsize]{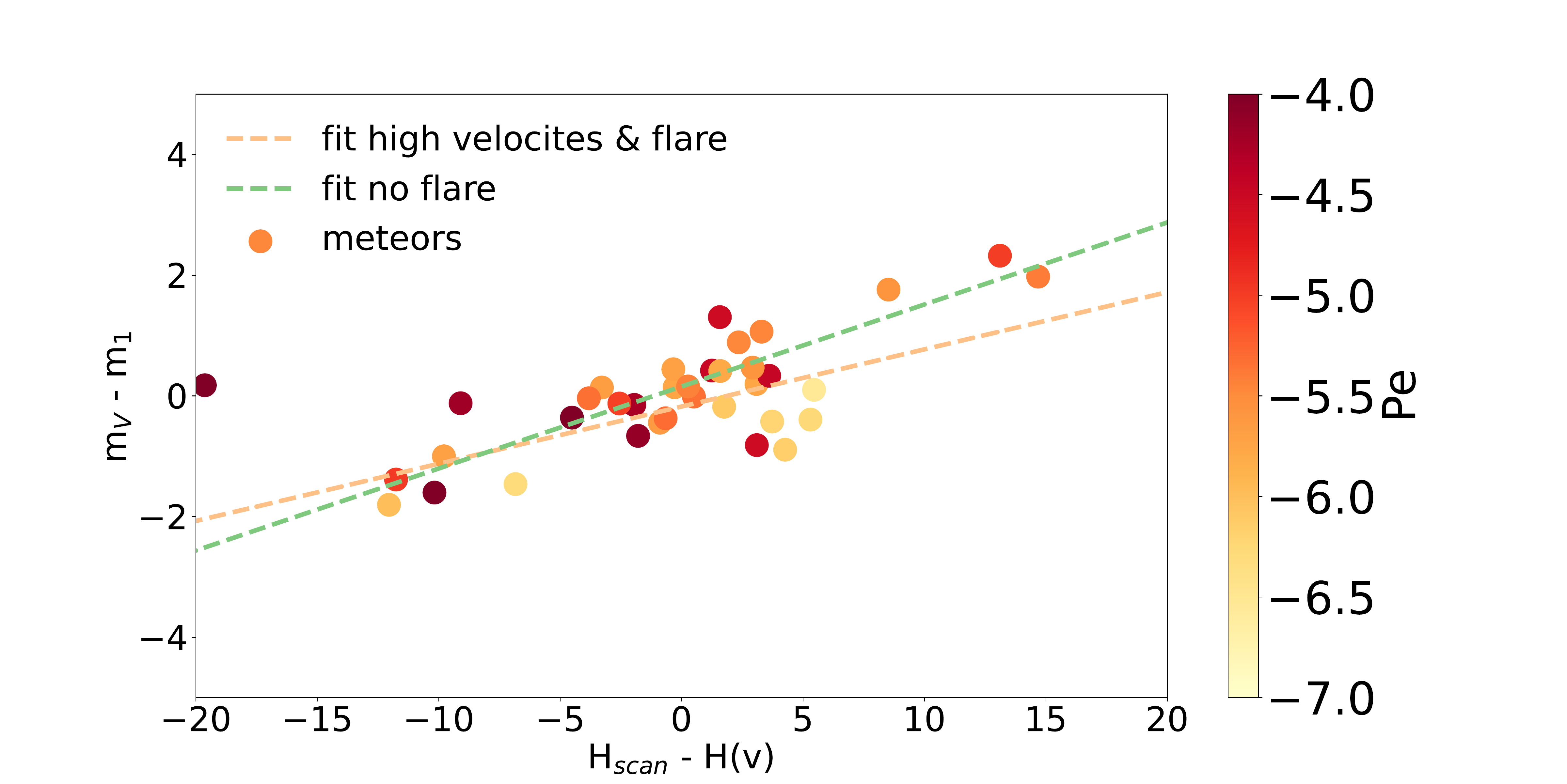}
\caption{}
\label{AP:Fig:Hscan_Pe}
\end{subfigure}
\begin{subfigure}{0.45\textwidth}
\includegraphics[width=\hsize]{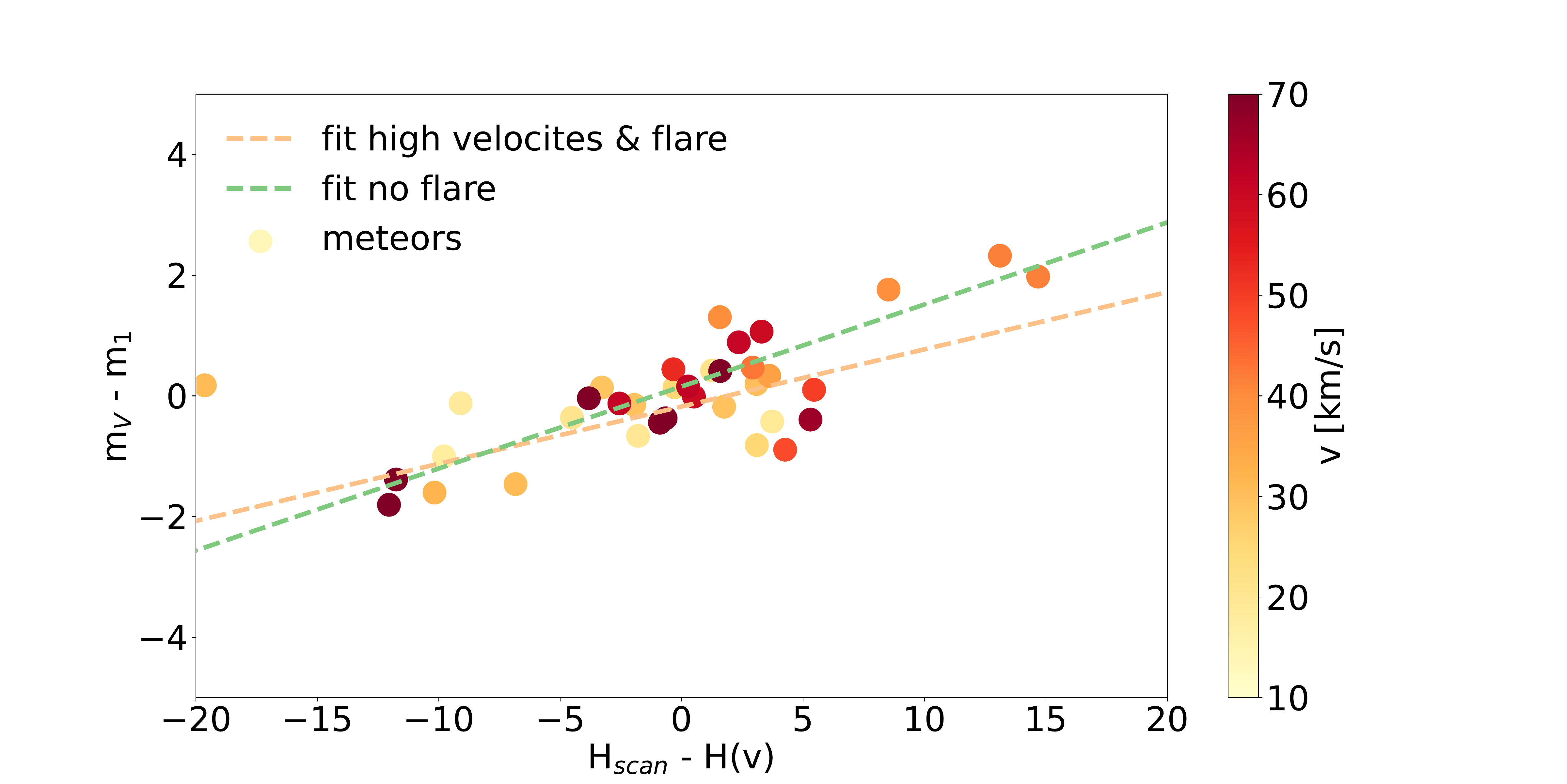}
\caption{}
\label{AP:Fig:Hscan_vel}
\end{subfigure}

\caption{Difference between the visual magnitude, $m_V$, and the magnitude, $m_1$, estimated from the radiation at $777$ nm and its dependence on the deviation of the scanned spectrum altitude, $H_{obs}$, from the altitude typical for given velocity, $H(v)$. Points are colored according to the $P_E$ parameter of the meteoroids given in Figure \ref{AP:Fig:Hscan_Pe} and according to the velocity in Figure \ref{AP:Fig:Hscan_vel}.}
\label{AP:Fig:Hscan_Pe_vel}
\end{figure*}

\end{appendix}

\end{document}